\documentclass{aa}
\usepackage{graphicx}
\usepackage{color}
\usepackage{txfonts}
\usepackage{xspace}
\usepackage{lscape}
\usepackage{multirow}
\usepackage{morefloats}
\usepackage{hyperref}
\usepackage[all]{hypcap}
\usepackage{morefloats}
\usepackage{caption}
\usepackage{subfigure}  
\usepackage{breakurl} % to break links in footnote

\newcommand{\llin}{$\ell_{\rm{lin}}$\xspace}

\newcommand{\tbeam}{$\theta_{\rm{mb}}$\xspace}

\newcommand{\tex}{$T_{\rm{ex}}$\xspace}
\newcommand{\tmb}{$T_{\rm{mb}}$\xspace}
\newcommand{\tant}{$T_{\rm{A}}^*$\xspace}
\newcommand{\tsys}{$T_{\rm{sys}}$\xspace}
\newcommand{\tpeak}{$T_{\rm{peak}}$\xspace}
\newcommand{\vlsr}{$V_{\rm{lsr}}$\xspace}
\newcommand{\vpeak}{$V_{\rm{peak}}$\xspace}

\newcommand{\sint}{$S_{\rm{int}}$\xspace}
\newcommand{\lco}{$L_{\rm{CO}}$\xspace}
\newcommand{\lbol}{$L_{\rm{bol}}$\xspace}
\newcommand{\mclump}{$M_{\rm{clump}}$\xspace}
\newcommand{\lmratio}{$L/M$\xspace}
\newcommand{\atgtop}{$\rm{TOP100}$\xspace}
\newcommand{\champ}{$\rm{CHAMP}^+$\xspace}
\newcommand{\flash}{$\rm{FLASH}^+$\xspace}

\newcommand{\sanjose}{$\rm{SJG13}$\xspace}
\newcommand{\vankempen}{$\rm{vK09}$\xspace}

\newcommand{\kms}{$\rm{km\,s^{-1}}$\xspace}
\newcommand{\um}{$\mu\rm{m}$\xspace}

\newcommand{\lsun}{L$_\odot$\xspace}
\newcommand{\msun}{M$_\odot$\xspace}

\newcommand{\fiq}{$\rm{70w}$\xspace}
\newcommand{\irq}{$\rm{IRw}$\xspace}
\newcommand{\irb}{$\rm{IRb}$\xspace}
\newcommand{\hii}{$\rm{H\,\textsc{ii}}$\xspace}

\begin{document}

   \title{ATLASGAL-selected high-mass clumps in the inner Galaxy. \\ VII. Characterisation of mid-$J$ CO emission}
   \author{F.~Navarete\inst{1,2}
           \and
           S.~Leurini\inst{3,1}
           \and
           A.~Giannetti\inst{4,1}
           \and
           F.~Wyrowski\inst{1}
           \and
           J.~S.~Urquhart\inst{5,1}
           \and \\
           C.~K\"onig\inst{1}
           \and
           T.~Csengeri\inst{1}
           \and
           R.~G\"usten\inst{1}
           \and
           A.~Damineli\inst{2}
           \and
           K.~M.~Menten\inst{1}
          }

   \institute{Max-Planck-Institut f\"ur Radioastronomie, Auf dem H\"ugel 69, 53121 Bonn, Germany\\
              \email{fnavarete@mpifr-bonn.mpg.de} % \email{navarete@gmail.com}
         \and
             Universidade de S\~ao Paulo, 
             Instituto de Astronomia, Geof\'isica e Ci\^encias Atmosf\'ericas,
             Departamento de Astronomia, \\
             Rua do Mat\~ao 1226, Cidade Universit\'aria S\~ao Paulo-SP, 05508-090, Brazil % \email{navarete@usp.br}
         \and 
         INAF - Osservatorio Astronomico di Cagliari,
Via della Scienza 5, I-09047, Selargius (CA)
         \and
             INAF - Istituto di Radioastronomia \& Italian ALMA Regional Centre, Via P. Gobetti 101, I-40129 Bologna, Italy 
         \and
             Centre for Astrophysics and Planetary Science, The University of Kent, Canterbury, Kent CT2 7NH, UK\\
             }

   \authorrunning{Navarete et al.}
   \titlerunning{Characterisation of mid-$J$ CO emission towards \atgtop clumps}
   \date{}
%	\pagerange{\pageref{firstpage}--\pageref{lastpage}} \pubyear{2018}

  \abstract
  {High-mass stars are formed within massive molecular clumps, where a large number of stars form close together.  The evolution of the clumps with different masses and luminosities is mainly regulated by its high-mass stellar content and the formation of such objects is still not well understood.}
{In this work, we characterise the mid-$J$ CO emission in a statistical sample of 99 clumps (\atgtop) selected from the ATLASGAL survey that are representative of the Galactic proto-cluster population.}
{High-spatial resolution APEX-CHAMP$^+$ maps of the CO\,(6--5) and CO\,(7--6) transitions were obtained and combined with additional single-pointing APEX-FLASH$^+$ spectra of the CO\,(4--3) line. The data were convolved to a common angular resolution of 13\farcs4. We analysed the line profiles by fitting the spectra with up to three Gaussian components, classified as narrow or broad, and computed CO line luminosities for each transition. Additionally,  we  defined a distance-limited sample of 72 sources within 5\,kpc to check the robustness of our analysis against beam dilution effects. We have studied the correlations of the line luminosities and profiles for the three CO transitions with the clump properties and investigate if and how they change as a function of the evolution.}
{All  sources were detected above 3-$\sigma$ in all  three CO transitions and most of the sources exhibit broad CO emission likely associated with molecular outflows. We find that the extension of the mid-$J$ CO emission is correlated with the size of the dust emission traced by the Herschel-PACS 70\,\um maps.
    The CO line luminosity (\lco) is correlated with the luminosity and mass of the clumps. However, it does not correlate with the luminosity-to-mass ratio.}
{The dependency of the CO luminosity with the properties of the clumps is steeper for higher-$J$ transitions. Our data seem to exclude that this trend is biased by self-absorption features in the CO emission, but rather suggest that different $J$ transitions arise from different regions of the inner envelope.
Moreover, high-mass clumps show similar trends in CO luminosity as lower mass clumps, but are systematically offset towards larger values, suggesting that higher column density and (or) temperature (of unresolved) CO emitters are found inside  high-mass clumps.}


   \maketitle

\label{firstpage}

\section{Introduction}
\label{introduction}

High-mass stars are responsible for the dynamical and chemical evolution of the interstellar medium and of their host galaxies by injecting heavier elements and energy in their surrounding environment by means of their strong UV emission and winds. Despite their importance, the processes that lead to the formation of high-mass stars are still not well understood \citep{Zinnecker07}.
    
    Observations at high-angular resolution have confirmed a high degree of multiplicity for high-mass stars, suggesting these objects are not formed in isolated systems \citep{Grellmann13}. The same scenario is supported by three-dimensional simulations of high-mass star formation \citep{Krumholz09,Rosen16}. These objects are formed on a relatively short timescale ($\sim$10$^5$\,yr), requiring large accretion rates \citep[$\sim$10$^{-4}$\,M$_\odot$\,yr$^{-1}$,][]{Hosokawa09}.
     Such conditions can only be achieved in the densest clumps in molecular clouds, with sizes of $\lesssim$\,1\,pc and masses of order 100-1000\,M$_\odot$ \citep{Bergin07}. These clumps are associated with large visual extinctions, thus observations at long wavelengths are required to study their properties and the star formation process.

After molecular hydrogen (H$_2$), which is difficult to observe directly in dense cold gas, carbon monoxide (CO) is the most abundant molecular species.
    Thus, rotational transitions of CO are commonly used to investigate the physics and kinematics of star-forming regions (SFRs).
    Traditionally, observations of CO transitions with low angular momentum quantum number $J$ from $J$\,=\,1--0 to 4--3 (here defined as low-$J$ transitions) have been used for this purpose (e.g. see \citealt{Schulz95}, \citealt{Zhang01} and \citealt{Beuther02}).
    These lines have upper level energies, $E_{\rm{u}}$, lower than 55\,K and are easily excited at relatively low temperatures and moderate densities.
    Therefore, low-$J$ CO lines are not selective tracers of the densest regions of SFRs, but are contaminated by emission from the ambient molecular cloud.
    On the other hand, higher-$J$ CO transitions  are less contaminated by ambient gas emission and likely probe the warm gas directly associated with embedded young stellar objects (YSOs).
In this paper we make use of the $J$\,=\,6--5 and 7--6 lines of CO, with  $E_{\rm{u}}$\,$\sim$\,116\,K and 155\,K, respectively, and in the following we refer to them simply as  mid-$J$ CO transitions.
    Over the past decade the Atacama Pathfinder Experiment telescope \citep[APEX\footnote{Based on observations with the APEX telescope under programme IDs M-087.F-0030-2011, M-093.F-0026-2014 and M-096.F-0005-2015. APEX is a collaboration between the Max-Planck-Institut f\"ur Radioastronomie, the European Southern Observatory, and the Onsala Space Observatory.},][]{Gusten06} has enabled routine observations of mid-$J$ CO lines, while {\it Herschel} and SOFIA have opened the possibility of spectroscopically resolved observations of even higher-$J$ transitions ($J$=10--9 and higher, e.g. \citealt{Gomez12}; \citealt[][hereafter, \sanjose]{SanJose13};  \citealt{Leurini15,Mottram17}).
    \citet[][hereafter, \vankempen]{vanKempen09} and \citet{vanKempen09b} have shown the importance of mid-$J$ CO transitions in tracing warm gas in the envelopes and outflows of low-mass protostars. More recently, \sanjose used the High Frequency Instrument for the Far Infrared \citep[HIFI,][]{deGraauw10} on board of Herschel to study a sample of low- and high-mass star-forming regions in high-$J$ transitions of several CO isotopologues (e.g. CO, $^{13}$CO and C$^{18}$O $J$\,=\,10--9), finding that the link between entrained outflowing gas and envelope motions is independent of the source mass.

In this paper, we present CO\,(6--5) and CO\,(7--6) maps towards a sample of 99 high-mass clumps selected from the APEX Telescope Large Area Survey of the Galaxy (ATLASGAL), which has provided an unbiased coverage of the plane of the inner Milky Way in the continuum emission at 870\,$\mu$m \citep{Schuller09}. Complementary single-pointing observations of the CO\,(4--3) line are also included in the analysis in order to characterise the CO emission towards the clumps.
    Section\,\ref{sec_obs_sample} describes the sample and Sect.\,\ref{observations} presents the observations and data reduction. In Sect.\,\ref{sec_COprofile} we present the distribution and extent of the mid-$J$ CO lines and their line profiles, compute the CO line luminosities and the excitation temperature of the gas, and compare them with the clump properties. In Sect.\,\ref{sec_discussion} we discuss our results in the context of previous works. 
    Finally, the conclusions are summarised in Sect.\,\ref{sec_summary}.

\section{Sample}
\label{sec_obs_sample}

    ATLASGAL detected the  vast  majority  of  all  current  and  future  high-mass  star  forming clumps ($M_{\rm{clump}}>1000 {\rm{M}}_\odot$) in the inner Galaxy. Recently, \citet{2018MNRAS.473.1059U} completed the distance assignment for $\sim$ 97 per cent of the ATLASGAL sources and analysed their masses, luminosities and temperatures based on thermal dust emission, and discussed how these properties evolve.
    Despite the statistical relevance of the ATLASGAL sample, detailed spectroscopic observations are not feasible on the whole sample.
    Therefore, we defined the ATLASGAL Top 100 \citep[hereafter, \atgtop,][]{Giannetti14,Koenig15}, a flux-limited sample of clumps selected from this survey with additional infrared (IR) selection criteria to ensure it encompasses a full range of luminosities and evolutionary stages (from 70\,$\mu$m-weak quiescent clumps to \hii regions).
    The 99 sources analysed in this paper are a sub-sample of the original \atgtop  \citep{Koenig15} and are classified as follows:
\begin{itemize}
   \item Clumps which either do not display any point-like emission in the  Hi-GAL Survey \citep{Molinari10} 70\,$\mu$m images and (or) only show weak, diffuse emission at this wavelength (hereafter, \fiq, 14 sources);
   \item Mid-IR weak sources that are either not associated with any point-like counterparts or the associated compact emission is weaker than 2.6\,Jy in the MIPSGAL survey \citep{Carey09} 24\,$\mu$m images (hereafter \irq, 31 sources);
   \item Mid-IR bright sources in an active phase of the high-mass star formation process, with strong compact emission seen in 8\,$\mu$m and 24\,$\mu$m images, but still not associated with radio continuum emission (hereafter \irb, 33 sources); 
   \item Sources in a later phase of the high-mass star formation process that are still deeply embedded in their envelope, but are bright in the mid-IR and associated with radio continuum emission (\hii regions, 21 sources).
\end{itemize}

    \citet{Koenig15} analysed the physical properties of the \atgtop sample in terms of distance, mass and luminosity. They found that at least 85\% of the sources have the ability to form high-mass stars and that most of them are likely gravitationally unstable and would collapse without the presence of a significant magnetic field.
    These authors showed that the \atgtop represents a statistically significant sample of high-mass star-forming clumps covering a range of evolutionary phases, from the coldest and quiescent 70\,$\mu$m-weak to the most evolved clumps hosting \hii regions, with no bias in terms of distance, luminosity and mass among the different classes. 
    The masses and bolometric luminosities of the clumps range from $\sim$20 to 5.2$\times$10$^5$\,\msun and from $\sim$60 to 3.6$\times$10$^6$\,\lsun, respectively. The distance of the clumps ranges between 0.86 and 12.6\,kpc, and  72 of the 99 clumps have distances below 5\,kpc. This implies that observations of the \atgtop at the same angular resolution  sample  quite different linear scales.
 In Appendix\,\ref{appendix_co_tables}, Table\,\ref{tbl_observations_short}, we list the main properties of the observed sources. 
    We adopted the Compact Source Catalogue (CSC) names from \citet{Contreras13} for the \atgtop sample although the centre of the maps may not exactly coincide with those positions (the average offset is $\sim$5\farcs4, with values ranging between $\sim$0\farcs5--25\farcs8, see Table\,\ref{tbl_observations_short}).

    In this paper, we investigate the properties of mid-$J$ CO lines for a sub-sample of the original \atgtop as part of our effort to observationally establish a solid evolutionary sequence for high-mass star formation.
    In addition to the dust continuum analysis of \citet{Koenig15}, we further characterised the \atgtop in terms of the content of the shocked gas in outflows traced by SiO emission \citep{Csengeri16} and the ionised gas content \citep{Kim17}, the CO depletion \citep{Giannetti14}, and the progressive heating of gas due to feedback of the central objects \citep{Giannetti17,Tang17}.
    These studies confirm an evolution of the targeted properties with the original selection criteria and strengthen our initial idea that the  \atgtop sample  constitutes a valuable inventory of high-mass clumps in different evolutionary stages.
    
\section{Observations and data reduction}
\label{observations} 

\subsection{\champ observations}
\label{sec_obs_champ}

    Observations of the \atgtop sample were performed with the APEX 12-m telescope on the following dates of 2014 May 17-20, July 10, 15-19, September 9-11 and 20. The \champ \citep{Kasemann06, Gusten08} multi-beam heterodyne receiver was used to map the sources simultaneously in the CO\,(6--5) and CO\,(7--6) transitions. Information about the instrument setup configuration is given in Table\,\ref{table:champ_setup}.

\begin{table*}[ht!]
\setlength{\tabcolsep}{5pt}	
\caption{\label{table:champ_setup}Summary of the observations.}
\centering
\begin{tabular}{cccccccccccc}
\hline\hline
Trans.	& $E_{\rm{u}}$	& Freq.	& Instr.	& $\eta_{\rm{mb}}$	& Beam		& $\Delta V$	& \tsys	& \multicolumn{2}{c}{rms (K)} &  Observed \\
		& (K)			& (GHz)		&			&				& size (\arcsec)	& (\kms)& (K)	& median	& range	& sources	\\        
\hline
CO\,(4--3)	& 55	& 461.04 & FLASH$^+$ & 0.60	& 13.4	& 0.953	& 1398\,$\pm$\,761	& 0.35 & 0.12--1.50 & 	98	\\
CO\,(6--5)	& 116	& 691.47 & \champ & 0.41	& 9.6	& 0.318 & 1300\,$\pm$\,250	& 0.21 & 0.07--0.75 & 	99	\\	
CO\,(7--6)	& 155	& 806.65 & \champ & 0.34	& 8.2	& 0.273	& 5000\,$\pm$\,1500	& 0.91 & 0.29--2.10 & 	99	\\
\hline
	\end{tabular}
\tablefoot{
    The columns are as follows:
    (1) observed transition;
    (2) upper-level energy of the transition;
    (3) rest frequency;
    (4) instrument;
    (5) main-beam efficiency ($\eta_{\rm{mb}}$);
    (6) beam size at the rest frequency;
    (7) spectral resolution;
    (8) mean systemic temperature of the observations;
    (9)-(10) median and range of the rms of the data of the single-pointing CO\,(4--3) spectra and the spectra extracted at central position of the CO\,(6--5) and CO\,(7--6) maps at their original resolution;
    (11) number of observed sources per transition (AGAL301.136$+$00.226 was not observed in CO\,(4--3)).}
\end{table*}
\setlength{\tabcolsep}{6pt}

    The \champ array has 2\,$\times$\,7 pixels that operate simultaneously in the radio frequency tuning ranges 620-720\,GHz in the low frequency array (LFA) and the other half in the range 780-950\,GHz in the high frequency array (HFA), respectively.
    The half-power beam widths (\tbeam) are 9\farcs0 (at 691\,GHz) and 7\farcs7 (807\,GHz), and the beam-spacing is $\sim$2.15\,\tbeam for both sub-arrays.
    The observations were performed in continuous on-the-fly (OTF) mode and maps of 80\arcsec\,$\times$\,80\arcsec size, centred on the coordinates given in Table\,\ref{tbl_observations_short}, were obtained for each source.
%%%%%%
The area outside of the central 60\arcsec$\times$60\arcsec\xspace region of each map is covered by only one pixel of the instrument, resulting in a larger rms near the edges of the map.
    The sky subtraction was performed by observing a blank sky field, offset from the central positions of the sources by {600\arcsec} in right ascension.
    The average precipitable water vapour (PWV) of the observations varied from 0.28 to 0.68\,mm per day, having a median value of 0.50\,mm.
    The average system temperatures (\tsys) ranged from 1050 to 1550\,K and 3500 to 6500\,K, at 691 and 807\,GHz, respectively.
    Pointing and focus were checked on planets at the beginning of each observing session. The pointing was also  checked every hour on Saturn and Mars, and on hot cores (G10.47+0.03~B1, G34.26, G327.3$-$0.6, and NGC6334I) during the observations.
    
    Each spectrum was rest-frequency corrected and baseline subtracted using the ''Continuum and Line Analysis Single Dish Software'' (\texttt{CLASS}), which is part of the \texttt{GILDAS} software\footnote{\url{http://www.iram.fr/IRAMFR/GILDAS}}. 
    The data were binned to a final spectral resolution of 2.0\,\kms in order to improve the signal-to-noise ratio of the spectra.
    The baseline subtraction was performed using a first-order fit to the line-free channels outside a window of $\pm$\,100\,\kms wide, centred on the systemic velocity, V$_{\rm lsr}$, of each source. We used a  broader window for sources exhibiting wings broader than $\sim$\,80\,\kms (AGAL034.2572+00.1535, AGAL301.136$-$00.226, AGAL327.393+00.199, AGAL337.406$-$00.402, AGAL351.244+00.669 and AGAL351.774$-$00.537, see  Table\,\ref{tbl_intpropCO_fixbeam}).
    Antenna temperatures (\tant) were converted to main-beam temperatures (\tmb) using beam efficiencies of 0.41 at 691\,GHz and 0.34 at 809\,GHz\footnote{\url{www3.mpifr-bonn.mpg.de/div/submmtech/heterodyne/champplus/champ\_efficiencies.16-09-14.html}}.
    Forward efficiencies are 0.95 in all observations.
    The gridding routine \texttt{XY\_MAP} in \texttt{CLASS} was used to construct the final datacubes. This routine convolves the gridded data with  a Gaussian of one third of the beam telescope size, yielding a final angular resolution slightly coarser (9\farcs6 for CO\,(6--5) and 8\farcs2 for CO\,(7--6)) than the original beam size (9\farcs0 and 7\farcs7, respectively). The final spectra at the central position of the maps have an average rms noise of 0.20 and 0.87\,K for CO\,(6--5) and CO\,(7--6) data, respectively.

\begin{figure}
    \centering
    \includegraphics[width=0.9\linewidth]{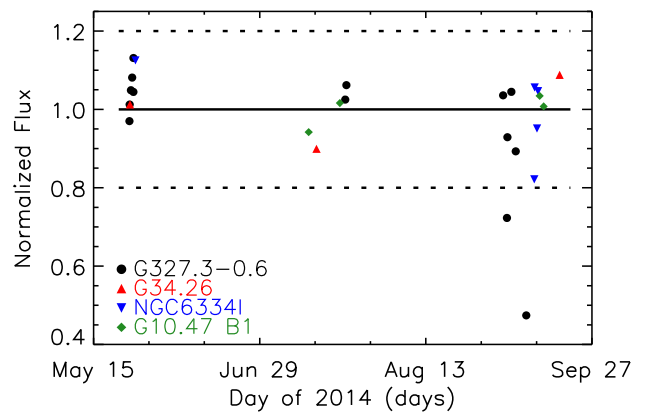}\\[-1.0ex]
    \caption{Ratio of the daily integrated flux to the average flux of the CO\,(6--5) transition for the calibration sources observed during the campaign.
             A solid horizontal line is placed at 1.0 and the dashed lines indicate a deviation of 20\% from unity.}
    \label{fig:calibration_sources}
\end{figure}

    Figure\,\ref{fig:calibration_sources} presents the ratio of the daily integrated flux to the corresponding average flux for the CO\,(6--5) transition of each hot core used as calibrator as a function of the observing day.
    The deviation of the majority of the points with respect to their average value is consistent within a $\pm$20\% limit; thus, this value was adopted as the uncertainty on the integrated flux for both mid-$J$ CO transitions.
On September 10, the observations of G327.3$-$0.6 showed the largest deviation from the average flux of the source (points at $y$\,$\sim$\,0.7 and $y$\,$\sim$\,0.5 in Fig.\,\ref{fig:calibration_sources}). For this reason, we associate an uncertainty of 30\% on the integrated flux of the sources AGAL320.881$-$00.397, AGAL326.661+00.519 and AGAL327.119+00.509, and of 50\% for sources AGAL329.066$-$00.307 and AGAL342.484+00.182, observed immediately after these two scans on G327.3$-$0.6.

\subsection{\flash observations}
\label{sec_obs_flash}

    The \flash \citep{Klein14} heterodyne receiver on the APEX telescope was used to observe the central positions of the CHAMP$^+$ maps in CO\,(4--3) on 2011 June 15 and 24, August 11 and 12. Table\,\ref{table:champ_setup} summarises the observational setup. The observations were performed in position switching mode with an offset position of  {600\arcsec} in right ascension  for sky-subtraction. Pointing and focus were checked on planets at the beginning of each observing session. The pointing was also regularly checked during the observations on Saturn and on hot cores (G10.62, G34.26, G327.3$-$0.6, NGC6334I and SGRB2(N)).
    The average PWV varied from 1.10 to 1.55\,mm per day with a median value of 1.29\,mm.
The system temperatures of the observations ranged from 650 to 2150\,K.

    The single-pointing observations were processed using \texttt{GILDAS/CLASS} software.
    The data were binned to a final spectral resolution of 2.0\,\kms and a fitted line was subtracted to stablish a straight baseline.
    The antenna temperatures were converted to \tmb by assuming beam and forward efficiencies of 0.60 and 0.95, respectively. 
    The resulting CO\,(4--3) spectra have an average rms noise of 0.36\,K.
    The uncertainty on the integrated flux of \flash data was estimated to be $\sim$20\% based on the continuum flux of the sources observed during the pointing scans.

\subsection{Spatial convolution of the mid-$J$ CO data}
 \label{sec_CO_convolution}
 
    The CO\,(6--5) and CO\,(7--6) data were convolved to a common angular resolution of 13\farcs4, matching the beam size of the single-pointing CO\,(4--3) observations. The resulting spectra are shown in Appendix\,\ref{appendix_co_fixbeam}.

The median rms of the convolved spectra are 0.35\,K, 0.17\,K and 0.87\,K for the  CO\,(4--3), CO\,(6--5) and CO\,(7--6) transitions, respectively. These values differ from those reported in Table\,\ref{table:champ_setup} for the \champ data where the rms at the original resolution of the dataset is given.
    Since our sources are not homogeneously distributed in distance (see Sect.\,\ref{sec_obs_sample}), spectra convolved to the same angular resolution of 13\farcs4 sample linear scales between 0.06 and 0.84\,pc.
    In order to study the effect of any bias introduced by sampling different linear scales within the clumps, the CO\,(6--5) and CO\,(7--6) data were also convolved to the same linear scale, \llin, of $\sim$0.24\,pc, which corresponds to an angular size $\theta$ (in radians) of:
\begin{equation}
 \theta = \tan^{-1}\left( \frac{\ell_{\mathrm{lin}}}{d} \right)
 \label{eq_beamsize}
\end{equation}
\noindent that depends on the distance of the source. The choice of \llin is driven by the nearest source, AGAL353.066+00.452, for which the part of the map with a relatively uniform rms (see Sect.\,\ref{sec_obs_champ}) corresponds to a linear scale of $\sim$0.24\,pc.
    Since we are limited by the beam size of the CO\,(6--5) observations ($\sim$10\arcsec), the same projected length can be obtained only for sources located at distances up to $\sim$5.0\,kpc. 
    This limit defines a sub-sample of 72 clumps (ten \fiq, 20 \irq, 26 \irb and 16 \hii regions). 

    The rest of the paper focuses on the properties of the full \atgtop sample based on the spectra convolved to 13\farcs4.
    The properties of the distance-limited sub-sample differ from those of the 13\farcs4 data only for the line profile (see Sect.\,\ref{sec_intpropCO}). A detailed comparison between the CO line luminosity and the properties of the clumps for the distance-limited sample is presented in Appendix\,\ref{appendix_distlim}.

\subsection{Self-absorption and multiple velocity components}
\label{sec_selfabs}

    The CO spectra of several clumps show a double-peak profile close to the ambient velocity (e.g. AGAL12.804$-$00.199, AGAL14.632$-$00.577, and AGAL333.134$-$00.431, see Fig.\,\ref{fig_fixbeam_gaussian_fit}).
    These complex profiles could arise from different velocity components in the beam or could be due to self-absorption given the likely high opacity of CO transitions close to the systemic velocity.
    To distinguish between these two scenarios, the 13\farcs4 CO spectra obtained in Sect.\,\ref{sec_CO_convolution} were compared to the C$^{17}$O\,(3--2) data from \citet{Giannetti14} observed with a similar angular resolution (19\arcsec). In the absence of C$^{17}$O observations (AGAL305.192$-$00.006, AGAL305.209$+$00.206 and AGAL353.066$+$00.452), the C$^{18}$O\,(2--1) profiles were used.
    Since the isotopologue line emission is usually optically thin \citep[cf.][]{Giannetti14}, it provides an accurate determination of the systemic velocity of the sources and, therefore, can be used to distinguish between the presence of multiple components or self-absorption in the optically thick $^{12}$CO lines. 
    Thus, when  C$^{17}$O or C$^{18}$O show a single peak corresponding in velocity to a dip in CO, we consider the CO spectra to be affected by self-absorption.
    Otherwise, if also the isotopologue data show a double-peak profile, the emission is likely due to two different velocity components within the beam.
    From the comparison with the CO isotopologues, we found 83 clumps with self-absorption features in the CO\,(4--3) line, 79 in the CO\,(6--5), 70 in the CO\,(7--6) transition.
    These numbers indicate that higher-$J$ CO transitions tend to be less affected by self-absorption features when compared to the lower-$J$ CO lines.
    Finally, only 15 objects do not display self-absorption features in any transitions.
    The CO spectra affected by self-absorption features are flagged with an asterisk symbol in Table\,\ref{tbl_intpropCO_fixbeam}.

To assess the impact of self-absorption on the analysis presented in Sect.\,\ref{sec_CO_correlations}, in particular on the properties derived from the integrated flux of the CO lines, we compared the observed integrated intensity of each CO transition  with the corresponding values obtained from the Gaussian fit presented in Sect.\,\ref{sec_gaussfit}. 
    This comparison indicated that self-absorption changes the offsets and the scatter of the data but not the slopes of the relations between the CO emission and the clump properties.
    Then, we investigated the ratio between the observed and the Gaussian integrated intensity values as a function of the evolutionary classes of the \atgtop sample. We found that 95\% of the sources exhibit ratios between 0.7 and 1.0 for all three lines.
    We also note a marginal decrease on the ratios from the earliest \fiq class ($\sim$1.0) to \hii regions ($\sim$0.8), indicating that self-absorption does not significantly affect the results presented in the following sections.
    We further investigated the effects of self-absorption by studying the sub-sample of $\sim$15 sources not affected by self-absorption (that is, the sources that are not flagged with an asterisk symbol in Table\,\ref{tbl_intpropCO_fixbeam}) and verified that the results presented in the following sections for the full sample are consistent with those of this sub-sample, although spanning a much broader range of clump masses and luminosities. More details on the analysis of the robustness of the relations reported in Sect.\,\ref{sec_CO_correlations} are provided in Appendices\,\ref{appendix_distlim} and \ref{appendix_gaussfit}.
    Five sources (see Appendix\,\ref{appendix_secondary}) show a  second  spectral feature in the  $^{12}$CO transitions and in the isotopologue data of \citet{Giannetti14} shifted in velocity from the rest velocity of the source.
%%%%%%%%%%%%%%%%%%%%%%%%
    We compared the  spatial distribution of the integrated intensity CO\,(6--5) emission with the corresponding ATLASGAL 870\,\um images (see Fig.\,\ref{fig_laboca_secondary_components})  for these five clumps. We found that in all sources the morphology of the integrated emission of one of the two peaks (labelled as P2 in Tables\,\ref{tbl_gauss_fitting_co43_fixbeam} to \ref{tbl_gauss_fitting_co76_fixbeam}) has a different spatial distribution than the  dust emission at 870\,\um and, thus, is likely not associated with the \atgtop clumps.
    These components are excluded from any further analysis in this paper.
    
\subsection{Gaussian decomposition of the CO profiles}
\label{sec_gaussfit}

\begin{figure*}
 \centering
 \includegraphics[width=\linewidth]{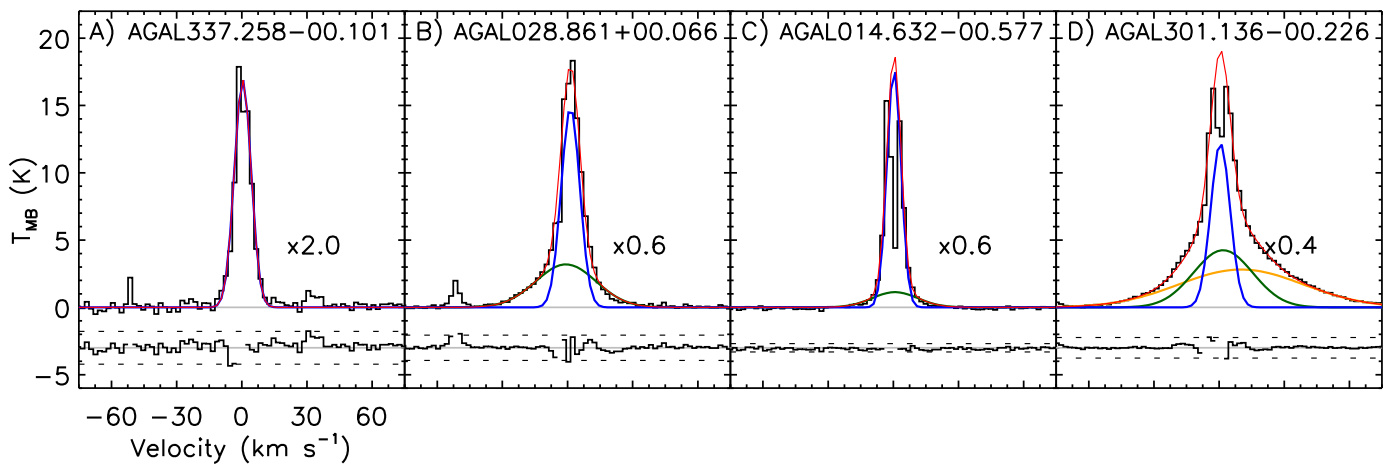}\\[-2.0ex]
 \caption{Gaussian decomposition for the CO\,(6--5) line using up to 3 components.
 Panel A: single Gaussian fit;
 Panel B: two-components fit;
 Panel C: two-components fit (channels affected by self absorption are masked);
 Panel D: three-components fit (channels affected by self absorption are masked).
 The spectra, fits and residuals were multiplied by the factor shown in each panel.
 The Gaussian fits are shown in blue, green and yellow, ordered by their line width; the sum of all components is shown in red. The grey lines indicate the baseline and the dashed horizontal lines placed in the residuals correspond to the 3-$\sigma$ level. Self-absorption features larger than 5-$\sigma$ were masked out from the residuals. 
 }
 \label{fitting_examples}
\end{figure*}

    The convolved CO spectra were fitted using multiple Gaussian components.
    The fits were performed interactively using the {\sc minimize} task in \texttt{CLASS/GILDAS}. A maximum number of three Gaussian components per spectrum was adopted.
    Each spectrum was initially fitted with one Gaussian component: if the residuals had sub-structures larger than 3-$\sigma$, a second or even a third component was added. 
    In case of self-absorption (see Sect.\,\ref{sec_selfabs}), the affected channels were masked before performing the fit.
    Any residual as narrow as the final velocity resolution of the data (2.0\,\kms) was ignored.
    In particular for CO\,(4--3), absorption features shifted in velocity from the main line are detected in several sources. These features are likely due to emission in the reference position, and were also masked before fitting the data. Examples  of the  line profile decomposition are given in Fig.\,\ref{fitting_examples}.

\begin{table}
 \centering
\setlength{\tabcolsep}{3pt}
 \caption{\label{table_gauss_ncomp}Results of the Gaussian fit of the CO spectra convolved to a common angular resolution of 13\farcs4.}
\begin{tabular}{l|ccc}
 \hline
 \hline
Transition		&	One comp.			&	Two Comp. 			&	Three comp.	\\
 \hline
CO\,(4--3)	&	27 (7,8,3,4)		&	68 (4,13,21,4)			&	3 (0,0,1,1)		\\
CO\,(6--5)	&	12 (6,4,1,1)     	&	58 (8,18,20,7)			&	29 (0,7,12,7)		\\
CO\,(7--6)	&	35 (14,14,6,1)		&	53 (0,16,24,10)			&	10 (0,1,3,6)		\\
 \hline
 \end{tabular}
 \tablefoot{The columns are as follows: (1) CO transition, (2)--(4) number of sources fitted using (2) a single Gaussian component, (3) two and (4) three components. In each column, the values in parenthesis indicate the corresponding number of \fiq, \irq, \irb and \hii regions.}
\end{table}
\setlength{\tabcolsep}{6pt}

    Each component was classified as narrow (N) or broad (B), adopting the scheme from \citet{SanJose13}.
    According to their definition, the narrow component has a full-width at half maximum (FWHM) narrower than 7.5\,km\,s$^{-1}$, otherwise it is classified as a broad component. Results of the Gaussian fit are presented in Tables\,\ref{tbl_gauss_fitting_co43_fixbeam}--\ref{tbl_gauss_fitting_co76_fixbeam}.
    In several cases, two broad components are needed to fit the spectrum.
    For the CO\,(6--5) data, 29 of the profiles required 3 components and, thus, two or three components have received the same classification. In these cases, they were named as, for example, B1, B2; ordered by their width. The P2 features mark secondary velocity components not associated with \atgtop clumps (see Sect.\,\ref{sec_selfabs}).

    As a consequence of high opacity and self-absorption, the Gaussian decomposition of the line profile can be somewhat dubious.
    In some cases, and in particular for the CO\,(4--3) transition, the fit is unreliable (e.g. 
AGAL305.192$-$00.006 and AGAL333.134$-$00.431 in Fig.\,\ref{fig_fixbeam_gaussian_fit}).
    The sources associated with unreliable Gaussian decomposed CO profiles (32, 8 and 4 for CO\,(4--3), CO\,(6--5) and CO\,(7--6), respectively) are not shown in Tables\,\ref{tbl_gauss_fitting_co43_fixbeam} to  \ref{tbl_gauss_fitting_co76_fixbeam} and their data are not included in the analysis presented in Sect.\,\ref{sec_COprofile}, as well as in that of the integrated properties of their line profiles (e.g. their integrated intensities and corresponding line luminosities, see Sect.\,\ref{sec_CO_correlations}).
	
    The general overview of the fits are given in Table\,\ref{table_gauss_ncomp} and the statistics of FWHM of the narrow and broad Gaussian  components are listed in Table\,\ref{table:co_components}. 
    The spectrum of each source with its corresponding decomposition into Gaussian components is presented in Fig.\,\ref{fig_fixbeam_gaussian_fit}.
    
\begin{table}
\centering
\setlength{\tabcolsep}{4pt}	
  \caption{\label{table:co_components}Statistics of the FWHM of the Gaussian components fitted on the CO line profiles convolved to a common angular resolution of 13\farcs4.}
  \begin{tabular}{clc|cccc}
\hline\hline
			&		&		&	\multicolumn{4}{c}{FWHM (km\,s$^{-1}$)}	 \\ 
Transition	&	Class.	&	$N$	&	Range			&	Mean	&	Median	&	$\sigma$ \\ 
\hline
\multirow{2}{*}{CO\,(4--3)} 	&	Narrow	&	28	&	3.23-7.47	&	6.27	&	6.57	&	1.14 \\ 
			&	Broad	&	83	&	7.5-86.0	&	21.2	&	14.9	&	16.5 \\ 
\hline
\multirow{2}{*}{CO\,(6--5)}	&	Narrow	&	48	&	2.55-7.48	&	5.69	&	6.91	&	1.36 \\ 
			&	Broad	&	148	&	7.5-97.1	&	24.5	&	17.8	&	19.0 \\ 
\hline
\multirow{2}{*}{CO\,(7--6)} &	Narrow	&	32	&	2.00-7.38	&	5.90	&	6.32	&	1.25 \\ 
			&	Broad	&	133	&	7.5-120.2	&	27.8	&	16.4	&	26.6 \\ 
\hline
 \end{tabular}
\tablefoot{The table presents the statistics on the FWHM of the narrow and broad Gaussian velocity component, classified according to Sect.\,\ref{sec_gaussfit}. The columns are as follows: (1) referred CO transition; (2) classification of the Gaussian component; (3) number of fitted components per class ($N$); (4) the minimum and maximum value per class; (5) the mean, (6) the median, and (7) the standard deviation of the distribution.}
\end{table}
\setlength{\tabcolsep}{6pt}

\section{Observational results}
\label{sec_COprofile}

The whole sample is detected above a 3-$\sigma$ threshold in the single-pointing CO\,(4--3) data (source AGAL301.136$-$00.226 was not observed with \flash) and in the 13\farcs4 {CO\,(6--5)} and {CO\,(7--6)} spectra, with three \fiq sources (AGAL030.893+00.139, AGAL351.571+00.762 and AGAL353.417$-$00.079) only  marginally detected above the 3-$\sigma$ limit in CO\,(7--6).

    In the rest of this section we characterise the CO emission towards the \atgtop sample through the maps of CO\,(6--5) (Sect.\,\ref{sec_maps}) and the analysis of the CO line profiles for the spectra convolved to 13\farcs4 (Sect.\,\ref{sec_intpropCO}). 
  In Sect.\,\ref{sec_CO_correlations}, we compute the CO line luminosities and compare them with the clump properties. 
 Finally, in Sect.\,\ref{sec_texc} we compute the excitation temperature of the gas.

\subsection{Extent of the CO emission}
\label{sec_maps}

\begin{figure}[ht!]
    \centering
 \includegraphics[width=\linewidth]{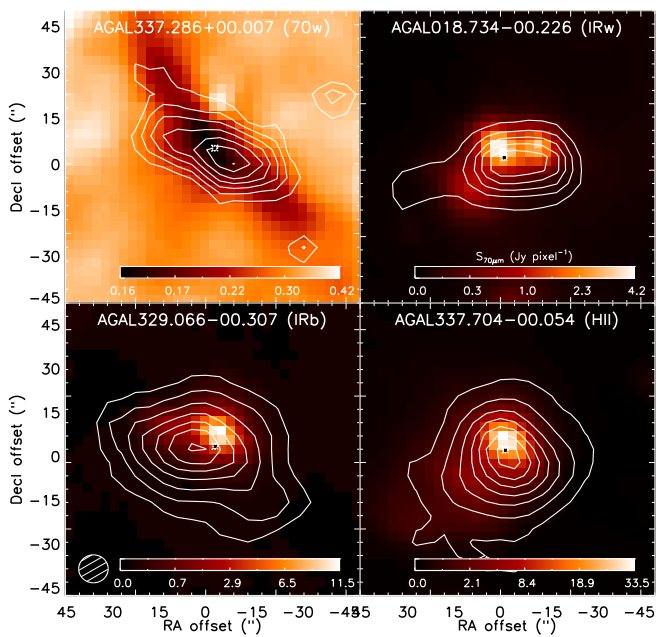} \\ [-2ex]
 \caption{Distribution of the CO\,(6--5) emission of four representative clumps of each evolutionary classes of the \atgtop clumps.
The CO contours are presented on top of the {\it Herschel}/PACS maps at 70\,$\mu$m.
Each 70\,\um map is scaled in according to the colour bar shown in the corresponding panel.
The CO contours correspond to the emission integrated over the full-width at zero power (FWZP) of the CO\,(6--5) line, and the contour levels are shown from 20\% to 90\% of the peak emission of each map, in steps of 10\%.
    The position of the CSC source from \citet{Contreras13} is shown as a $\times$ symbol.
    The beam size of the CO\,(6--5) observations is indicated in the left bottom region.
}
 \label{fig_co_maps}
\end{figure}

    In Fig.\,\ref{fig_co_maps} we present examples of the integrated intensity maps of the CO\,(6--5) emission as a function of the evolutionary class of the \atgtop clumps.
  The CO\,(6--5) maps of the full \atgtop sample are presented in  Appendix\,\ref{appendix_co_fixbeam} (see Fig.\,\ref{fig_fixbeam_gaussian_fit}).

\begin{figure}
 \centering
 \includegraphics[width=0.95\linewidth]{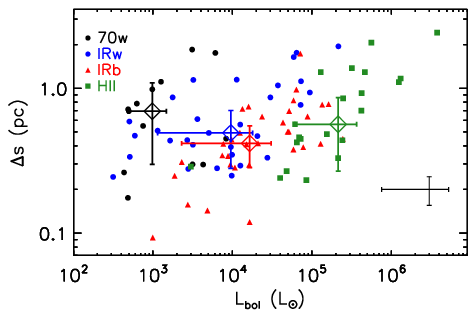} \\[-0.9ex]
 \includegraphics[width=0.95\linewidth]{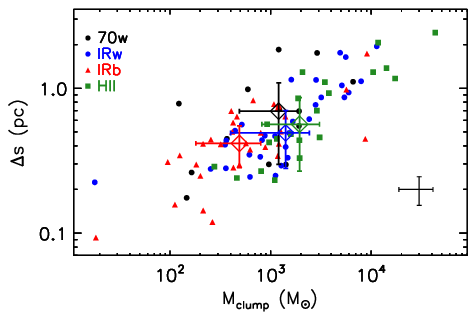} \\[-1.5ex]
 \caption{Size of the CO\,(6--5) emission towards the \atgtop sample versus the bolometric luminosity (top) and the mass (bottom) of the sources. 
 The median values for each class are shown as open diamonds and their error bars correspond to the absolute deviation of the data from their median value.
 The typical uncertainty is shown by the error bars on the bottom right of each plot.}
 \label{fig_extension_co}
\end{figure}

    We estimated the linear size of the CO emission, $\Delta s$, defined as the  average between the maximum and minimum elongation of the half-power peak intensity (50\%) contour level of the CO\,(6--5) integrated intensity (see Table\,\ref{table_co_extension}). The uncertainty on $\Delta s$ was estimated as the dispersion between the major and minor axis of the CO extent.
    The linear sizes of the CO emission ranges between 0.1 and 2.4\,pc, with a median value of 0.5\,pc.
    In order to investigate if $\Delta s$ varies with evolution, we performed a non-parametric two-sided Kolmogorov-Smirnov (KS) test between pairs of classes (i.e. \fiq vs. \irb; \irq vs. \hii). 
    The sub-samples were considered statistically different if their KS rank factor is close to 1 and associated with a low probability value, $p$, ($p$\,$\leq$\,0.05 for a significance $\geq$\,2\,$\sigma$).
    Our analysis indicates that there is no significant change in  the  extension of CO with evolution  (KS\,$\leq$\,0.37, $p$\,$\geq$\,0.05 for all comparisons).
    The CO extent was further compared with the bolometric luminosity, \lbol, and the mass of the clumps, \mclump, reported by \citet{Koenig15}. The results are presented in Fig.\,\ref{fig_extension_co}. 
    $\Delta s$ shows a large scatter as a function of \lbol while it increases with \mclump ($\rho$\,=\,0.72, $p$\,$<$\,0.001 for the correlation with \mclump\,  $\rho$\,=\,0.42, $p$\,$<$\,0.001 for \lbol, where the $\rho$ is the Spearman rank correlation factor and $p$ its associated probability).
    This confirms that the extent of the CO emission is likely dependent of the amount of gas within the clumps, but not on their bolometric luminosity.
    
\begin{figure}
 \centering
 \includegraphics[width=0.95\linewidth]{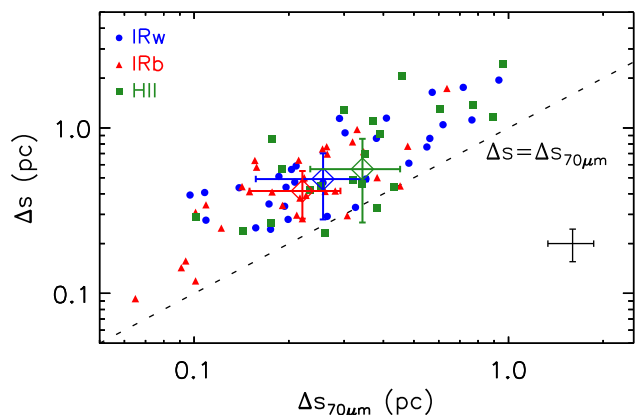} \\[-1.5ex]
 \caption{Size of the CO\,(6--5) emission versus the size of the 70\,\um emission from the {\it Herschel}-PACS images towards the 70\,\um-bright clumps. The dashed line indicates $y$\,=\,$x$. 
 The median values for the 70\,\um-bright classes are shown as open diamonds and their error bars correspond to the absolute deviation of the data from their median value.
 The typical uncertainty, computed as the dispersion between the major and minor axis of the emission, is shown by the error bars on the bottom right of each plot.
 }
 \label{fig_extension_co_pacs}
\end{figure}

We derived the extent of the 70\,\um emission ($\Delta s_{\rm{70\,\mu m}}$) towards the 70\,\um-bright clumps by cross-matching the position of the \atgtop clumps with the sources from \citet{Molinari16}. Then, $\Delta s_{\rm{70\,\mu m}}$ was obtained by computing the average between the maximum and minimum FWHM reported on their work and the corresponding error was obtained as the standard deviation of the FWHM values.
%%%%%%%%%%%%%%%%%%%%%
    The values are also reported in Table\,\ref{table_co_extension}. 
 Figure\,\ref{fig_extension_co_pacs}  compares the extent of the CO\,(6--5) emission with that of the 70\,\um emission towards the 70\,\um bright clumps. The extent of the emission of CO\,(6--5) and of the 70\,\um continuum emission are correlated (Fig.\,\ref{fig_extension_co_pacs}, $\rho$\,=\,0.67, $p$\,$<$\,0.001), and in the majority of cases, the points are located above the equality line, suggesting that the gas probed by the CO\,(6--5) transition tends to be more extended than the dust emission probed by the PACS data towards the 70\,\um-bright clumps.

\subsection{Line profiles}
\label{sec_intpropCO}

    In the majority of the cases, the CO profiles are well fit with two Gaussian components, one for the envelope, one for high-velocity emission (see Table\,\ref{table_gauss_ncomp}). A third component is  required in some cases, in particular for the CO\,(6--5) data, which have the highest signal-to-noise ratio.
    The majority of sources fitted with a single Gaussian component are in the earliest stages of evolution (\fiq and \irq clumps), suggesting that the CO emission is less complex in earlier stages of high-mass star formation.
    We also detect non-Gaussian high-velocity wings likely associated with outflows in most of the  CO\,(6--5) profiles.
    A detailed discussion of the outflow content in the \atgtop sample and of their properties will be presented in a forthcoming paper (Navarete et al., in prep.).

 \begin{figure*}[!ht]
 \centering
\includegraphics[width=0.4\linewidth]{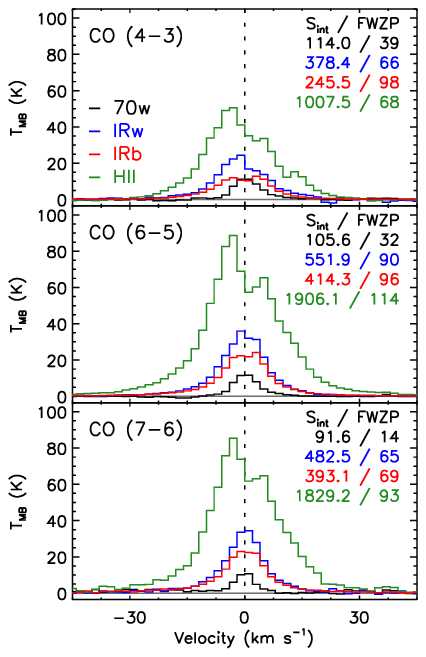}
\includegraphics[width=0.4\linewidth]{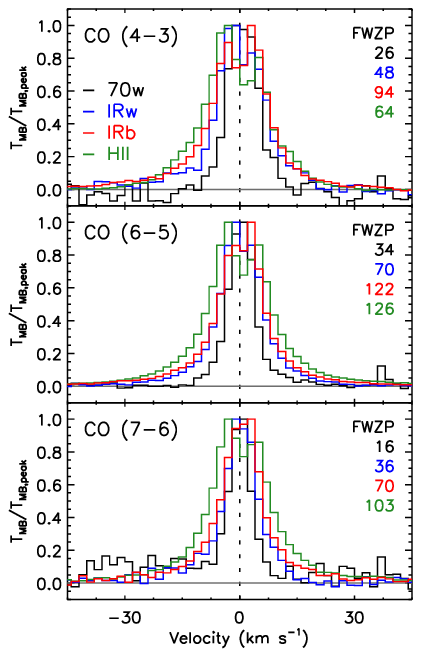} \\[-1.5ex]
 \caption{Left: Average CO\,(4--3), CO\,(6--5) and CO\,(7--6) spectra convolved to {13\farcs4} beam of each evolutionary class scaled to the median distance of the whole sample ($d$\,=\,3.80\,kpc). Right: Same plot, but the average CO spectra are normalised by their peak intensity.
    The baseline level is indicated by the solid grey line and the black dashed line is placed at 0\,km\,s$^{-1}$.
 The FWZP of the profiles are shown in the upper right side of the panels (in \kms units), together with the integrated intensities (\sint, in K\,\kms units) of the CO profiles shown in the left panels.
}
 \label{fig_avgspc_fixbeam}
\end{figure*}

    To minimise biases due to different sensitivities in the analysis of single spectra, we computed the average CO spectrum of each evolutionary class and normalised it by its peak intensity.
    The spectra were shifted to 0\,\kms using the correspondent \vlsr given in Table\,\ref{tbl_observations_short}. 
    Then, the averaging was performed by scaling the intensity of each spectrum to the median distance of the sub-sample ($d$\,=\,3.26\,kpc for the distance-limited sample, $d$\,=\,3.80\,kpc for the full sample).
    The resulting spectra of the 13\farcs4 dataset are shown in Fig.\,\ref{fig_avgspc_fixbeam} while those of the distance-limited sub-sample are presented in Appendix\,\ref{fig_avgspc_fixscale} (Fig.\,\ref{fig_avgspc_fixscale}).
%%%%%%%%%%%%
    While the 13\farcs4 data show no significant difference between the average profiles of \irq and \irb classes, in the distance-limited sub-sample the width (expressed through the full width at zero power, FWZP, to avoid any assumption on the profile) and the intensity of the CO lines progressively increase with the evolution of the sources (from \fiq clumps towards \hii regions) especially when the normalised profiles are considered.
%%%%%%%%%%%%%%%%%%%    
    The difference between the two datasets is due to sources at large distances (d\,>\,12\,kpc; AGAL018.606$-$00.074, AGAL018.734$-$00.226 and AGAL342.484+00.182) for which the observations sample a much larger volume of gas. The increase of line width with evolution is confirmed by the analysis of the individual FWZP values of the three CO lines, presented in Table\,\ref{tbl_intpropCO_fixbeam} (see Table\,\ref{table_limits_intpropCO} for the statistics on the full \atgtop sample).

\begin{table}
 \centering
 \caption{\label{table_limits_intpropCO}Statistics on the CO line profiles, convolved to a common angular resolution of 13\farcs4.}
\begin{tabular}{l|cccc}
 \hline
 \hline
 		& \multicolumn{4}{c}{FWZP (\kms)} \\
	&		&	 &		&	Standard 	\\
Transition	&	Range	&	Mean &	Median	&	 deviation	\\
 \hline
CO\,(4--3)	&	10-134	&	47	&	42	&	25	\\
CO\,(6--5)	&	14-162	&	62	&	54	&	34	\\
CO\,(7--6)	&	 4-142	&	39	&	30	&	27	\\
 \hline
 \end{tabular}
\tablefoot{The table presents the FWZP of the CO lines. 
The mid-$J$ CO lines were convolved to a common angular resolution of 13\farcs4. The columns are as follows: (1) referred CO line; (2) the minimum and maximum values, (3) the mean, (4) the median, and (5) the standard deviation of the distribution.}
\end{table}

    Despite the possible biases in the analysis of the line profiles (e.g. different sensitivities, different excitation conditions, complexity of the profiles), our data indicate that the CO emission is brighter in late evolutionary phases.
    The average spectra per class show also that the CO lines becomes broader towards more evolved phases likely due to the presence of outflows.
Our study extends the work of \citet{Leurini13} on one source of our sample, AGAL327.293$-$00.579. They mapped in CO\,(3--2), CO\,(6--5), CO\,(7--6) and in $^{13}$CO\,(6--5), $^{13}$CO\,(8--7) and $^{13}$CO\,(10--9) a larger area of the source than that presented here  and found that, for all transitions, the spectra are dominated in intensity by the \hii region rather than by younger sources (a hot core and an infrared dark cloud are also present in the area). They interpreted this result as an evidence that the bulk of the Galactic CO line emission comes from PDRs around massive stars, as suggested by \citet{Cubick08} for FIR line emission. Based on this, we suggest that the increase in mid-$J$ CO brightness in the later stages of the \atgtop is due to a major contribution of PDR to the line emission.  We notice however that the increase of width and of intensity of the CO lines with evolution can also be due to an increase with time of multiplicity of sources in the beam.
%%%%%%%%%%%%%%%%%%%%%%%%%%%%%%%%%%%%%%%

\subsection{The CO line luminosities}
\label{sec_CO_correlations}

    The intensity of the CO profiles (\sint, in K\,km\,s$^{-1}$) was computed by integrating the CO emission over the velocity channels within the corresponding FWZP range.
    Then, the line luminosity  (\lco, in K\,km\,s$^{-1}$\,pc$^2$) of each CO line was calculated using Eq.\,2 from \citet{Wu05}, assuming a source of size equal to the beam size of the data (see Sect.\,\ref{sec_CO_convolution}).
    The derived \lco values are reported in Table\,\ref{tbl_intpropCO_fixbeam}. The errors in the \lco values are estimated by error propagation on the integrated flux (see Sect.\,\ref{sec_obs_champ}) and considering an uncertainty of 20\% in the distance.
%%%%%%%%%%%%%%%
    The median values of \lco, \lbol, \mclump and   \lmratio, the luminosity-to-mass ratio, per evolutionary class are summarised in Table\,\ref{table_median_class}.
    We also performed the same analysis on the data convolved to a common linear scale of 0.24\,pc (assuming the corresponding angular source size of 0.24\,pc to derive the line luminosity) and no significant differences in the slope of the trends were found. Therefore, the distance-limited sample will not be discussed any further in this section.
 
 \begin{table*}[!]
\caption{\label{table_median_class}Median values per class of the clump and CO profile properties.}
\centering
\setlength{\tabcolsep}{4pt}
\begin{tabular}{l|cccc}	
	\hline\hline
Property						&	\fiq	&	\irq	&	\irb	&	\hii	\\
\hline
\lbol ($10^3$\,L$_\odot$)		&	1.26$\pm$0.83	&	9.6$\pm$8.4	&	16.5$\pm$1.4	&	21.4$\pm$1.5	\\
\mclump ($10^3$\,M$_\odot$)		&	1.22$\pm$0.70	&	1.4$\pm$1.1	&	0.49$\pm$0.31	&	1.9$\pm$1.1	\\
\lmratio (L$_\odot$/M$_\odot$)	&	2.58$\pm$0.93	&	9.0$\pm$6.8	&	40$\pm$23		&	76$\pm$28	\\
\hline
$L_{\rm{CO\,(4-3)}}$ (K\,km\,s$^{-1}$\,pc$^2$)	&	9.8$\pm$8.5	&	30$\pm$23	&	21$\pm$15	&	119$\pm$58	\\
$L_{\rm{CO\,(6-5)}}$	&	5.1$\pm$4.0	&	16$\pm$12	&	19$\pm$12	&	51$\pm$44	\\
$L_{\rm{CO\,(7-6)}}$	&	4.7$\pm$3.6	&	11.8$\pm$8.6	&	14.8$\pm$9.7	&	48$\pm$45	\\
\hline
FWZP$_{\rm{CO\,(4-3)}}$ (km\,s$^{-1}$)	&	24.0$\pm$6.0	&	34$\pm$12	&	52$\pm$16	&	62$\pm$18	\\
FWZP$_{\rm{CO\,(6-5)}}$ &	26.0$\pm$6.0	&	42$\pm$14		&	72$\pm$22	&	102$\pm$28	\\
FWZP$_{\rm{CO\,(7-6)}}$ &	12.0$\pm$2.0	&	24.0$\pm$6.0	&	38$\pm$14	&	66$\pm$20	\\
\hline
\tex(K) &	22.4$\pm$5.0	&	29.9$\pm$8.1	&	45$\pm$15	&	95$\pm$21	\\
\hline
\end{tabular}
\tablefoot{The median and the absolute deviation of the data from their median value are shown for the clump properties (bolometric luminosity, mass and luminosity-to-mass ratio), for the line luminosity and full width at zero power of the low-$J$ and mid-$J$ CO profiles convolved to the same angular size of 13\farcs4, and for the excitation temperature of the CO\,(6--5) emission.}
\end{table*}
\setlength{\tabcolsep}{6pt}

\begin{figure*}[!]
 \centering
  \includegraphics[width=0.485\linewidth]{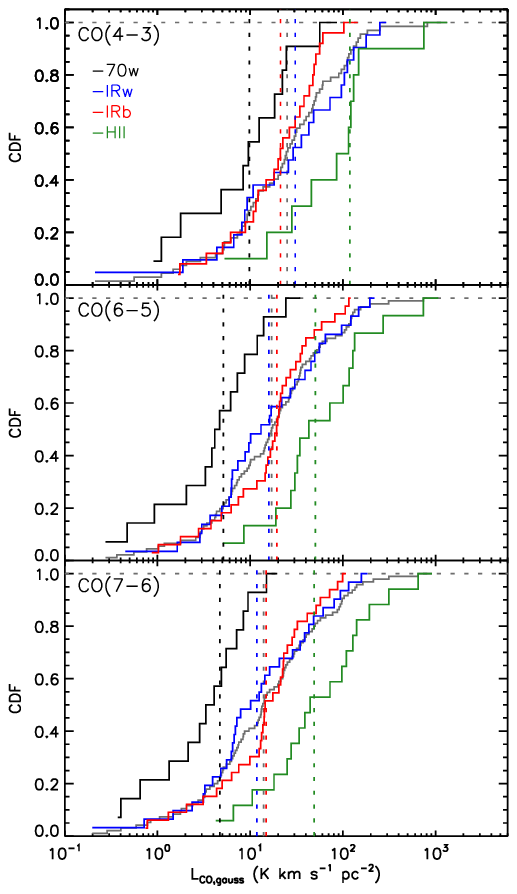}
  \includegraphics[width=0.485\linewidth]{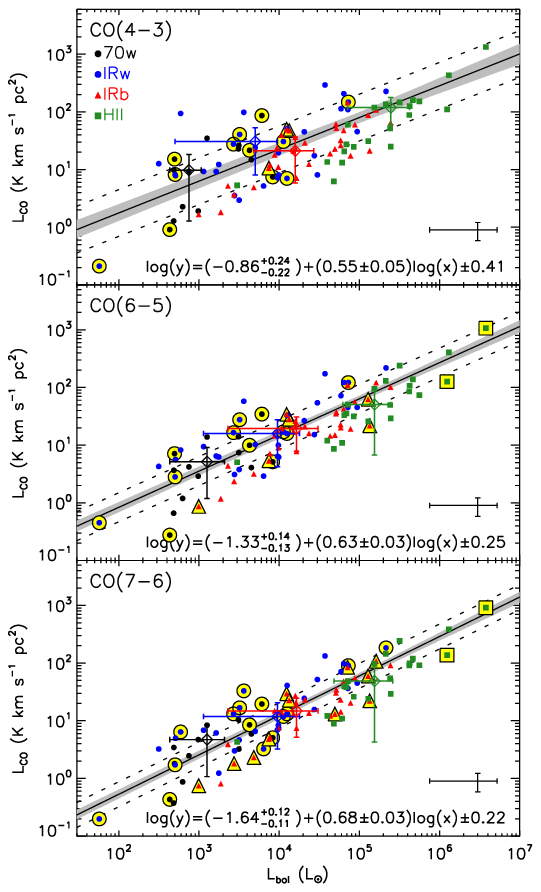}  \\[-1.5ex]
 \caption{Left panels: Cumulative distribution function of the line luminosity of the CO\,(4--3) (upper panel), CO\,(6--5) (middle) and CO\,(7--6) emission (bottom) towards the \atgtop sample. The median values per class are shown as vertical dashed lines in their corresponding colours. Right panels: Line luminosity of  the same CO $J$-transitions versus the bolometric luminosity of the \atgtop sources.
 The median values for each class are shown as open diamonds and their error bars correspond to the absolute deviation of the data from their median value.
Data points highlighted in yellow indicate those sources from which no signs of self-absorption features where identified in the spectrum convolved to 13\farcs4.
 The typical error bars are shown at the bottom right side of the plots. The black solid line is the best fit, the light grey shaded area indicates the 68\% uncertainty, and the dashed lines show the intrinsic scatter ($\epsilon$) of the relation.}
 \label{fig_lco_correlation_fixbeam_lbol}
\end{figure*}

In Fig.\,\ref{fig_lco_correlation_fixbeam_lbol} we show the cumulative distribution function (CDF) of the line luminosities for the three CO transitions:  \lco increases from \fiq sources towards \hii regions. Each evolutionary class was tested against the others by computing their two-sided KS coefficient (see Table\,\ref{table_lco_classes}).
    The most significant differences are found when comparing the earlier and later evolutionary classes (\fiq and \irb, \fiq and \hii, $\rho$\,$\geq$\,0.66 for the CO\,(6--5) line), while no strong differences are found among the other classes (KS\,$\leq$\,0.5 and $p$\,$\geq$0.003 for the CO\,(6--5) transition).
    These results indicate that, although we observe an increase on the CO line luminosity from \fiq clumps towards \hii regions, no clear separation is found in the intermediate classes (\irq and \irb, see also Table\,\ref{table_median_class}).

\begin{table}
 \centering
\setlength{\tabcolsep}{3pt}	
 \caption{\label{table_lco_classes}Kolmogorov-Smirnov statistics of the CO line luminosity as a function of the evolutionary class of the clumps.}
\begin{tabular}{c|lll}
 \hline
 \hline
Classes	& \multicolumn{1}{c}{CO\,(4--3)} & \multicolumn{1}{c}{CO\,(6--5)} & \multicolumn{1}{c}{CO\,(7--6)}	\\
 \hline
\fiq-\irq &	0.48, $p$\,=\,0.05 	&	0.45, $p$\,=\,0.03 		&	0.46, $p$\,=\,0.02 		\\
\fiq-\irb &	0.35, $p$\,=\,0.25 	&	0.66, $p$\,$<$\,0.001 	&	0.66, $p$\,$<$\,0.001 	\\
\fiq-\hii &	0.72, $p$\,=\,0.004 &	0.80, $p$\,$<$\,0.001 	&	0.82, $p$\,$<$\,0.001 	\\
\irq-\irb &	0.29, $p$\,=\,0.23 	&	0.21, $p$\,=\,0.46 		&	0.24, $p$\,=\,0.26 		\\
\irq-\hii &	0.41, $p$\,=\,0.15 	&	0.46, $p$\,=\,0.02 		&	0.47, $p$\,=\,0.01 		\\
\irb-\hii &	0.62, $p$\,=\,0.004 &	0.53, $p$\,=\,0.003 	&	0.52, $p$\,=\,0.003 	\\
\hline
 \end{tabular}
\tablefoot{The rank KS and its corresponding probability ($p$) are shown for each comparison. A $p$-value of $<$\,0.001 indicate a correlation at 0.001 significance level. $p$-values of 0.05, 0.002 and $<$\,0.001 represent the $\sim$\,2, 3 and $>$\,3\,$\sigma$ confidence levels.}
\end{table}
\setlength{\tabcolsep}{6pt}	

\begin{table}[h!]
\caption{\label{table_lco_fit}Parameters of the fits of \lco as a function of the clump properties.}
\centering
\begin{tabular}{cl|ccc}
	\hline\hline
Transition	&	Property	&	$\alpha$	&	$\beta$	&	$\epsilon$	\\
\hline
			&	\lbol	&	$-$0.86$^{+0.24}_{-0.22}$	&	0.55$\pm$0.05 	&	0.41 	\\
CO\,(4--3)	&	\mclump	&	$-$1.37$^{+0.23}_{-0.19}$	&	0.92$\pm$0.06 	&	0.34 	\\
			&	\lmratio&	$+$1.08$^{+0.12}_{-0.14}$	&	0.28$\pm$0.12 	&	0.63 	\\
\hline																
			&	\lbol	&	$-$1.33$^{+0.14}_{-0.13}$	&	0.63$\pm$0.03 	&	0.25 	\\
CO\,(6--5)	&	\mclump	&	$-$1.58$^{+0.23}_{-0.22}$	&	0.92$\pm$0.07 	&	0.37 	\\
			&	\lmratio&	$+$0.74$^{+0.09}_{-0.08}$	&	0.46$\pm$0.09 	&	0.55 	\\
\hline																
			&	\lbol	&	$-$1.64$^{+0.12}_{-0.11}$	&	0.68$\pm$0.03 	&	0.22 	\\
CO\,(7--6)	&	\mclump	&	$-$1.64$^{+0.22}_{-0.24}$	&	0.92$\pm$0.08 	&	0.43 	\\
			&	\lmratio&	$+$0.55$^{+0.10}_{-0.08}$	&	0.55$\pm$0.10 	&	0.54 	\\
\hline 
\end{tabular}
\tablefoot{The fits were performed by adjusting a model with three free parameters in the form of $\log(y) = \alpha + \beta \log(x) \pm \epsilon$, where $\alpha$, $\beta$ and $\epsilon$ correspond to the intercept, the slope and the intrinsic scatter, respectively.}
\end{table}

\begin{figure}
 \centering
 \includegraphics[width=0.95\linewidth]{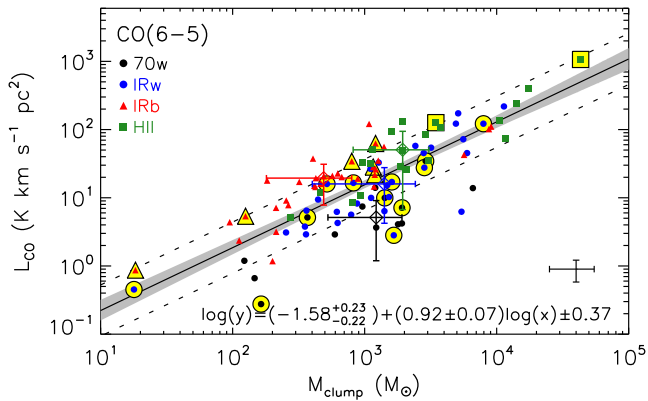}  \\[-1.5ex]
 \caption{Same as the right panel of Fig.\,\ref{fig_lco_correlation_fixbeam_lbol}, but displaying the line luminosity of the CO\,(6--5) emission as a function of the mass of the clumps. Data points highlighted in yellow indicate those sources from which no signs of self-absorption features where identified in the spectrum convolved to 13\farcs4.}
 \label{fig_lco_correlation_fixbeam_mclump}
\end{figure}

\begin{figure}
 \centering
 \includegraphics[width=0.95\linewidth]{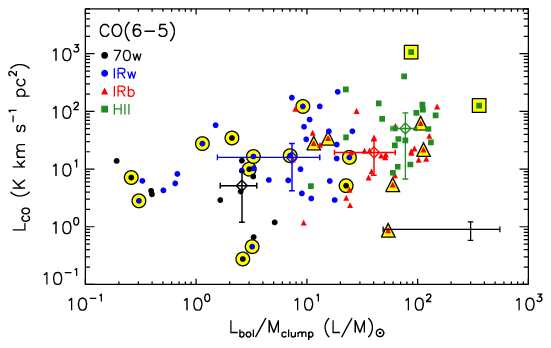} \\[-1.5ex]
 \caption{Same as the right panel of  Fig.\,\ref{fig_lco_correlation_fixbeam_lbol}, but displaying \lco of the CO\,(6--5) line as a function of the \lmratio ratio of the \atgtop sources. Data points highlighted in yellow indicate those sources from which no signs of self-absorption features where identified in the spectrum convolved to 13\farcs4.
}
 \label{fig_lco_correlation_fixbeam_lmratio}
\end{figure}

    We also plot \lco against the bolometric luminosity of the clumps (Fig.\,\ref{fig_lco_correlation_fixbeam_lbol}), their mass and their luminosity-to-mass ratio (Figs.\,\ref{fig_lco_correlation_fixbeam_mclump} and \ref{fig_lco_correlation_fixbeam_lmratio} for the CO\,(6--5) line). The \lmratio ratio is believed to be a rough estimator of evolution in the star formation process for both low- \citep{Saraceno96} and high-mass regimes \citep[e.g.][]{Molinari08}, with small \lmratio values corresponding to embedded  regions where (proto-)stellar activity is just starting, and high \lmratio values in sources with stronger radiative flux and that have accreted most of the mass \citep{Molinari16,Giannetti17,2018MNRAS.473.1059U}. In addition, the \lmratio ratio also reflects the properties of the most massive young stellar object embedded in the clump \citep{Faundez04,Urquhart13}.
    The fits  were performed using a Bayesian approach, by adjusting a model with three free parameters (the intercept, $\alpha$, the slope, $\beta$, and the intrinsic scatter, $\epsilon$). In order to obtain a statistically reliable solution, we computed a total of 100\,000 iterations per fit. The parameters of the fits are summarised in Table\,\ref{table_lco_fit}.
  The correlation between \lco and the clump properties was checked by computing their Spearman rank correlation factor and its associated probability ($\rho$ and $p$, respectively, see Table\,\ref{table_co_correlation}).
  Since \lco  with \lbol and \mclump have the same dependence on the distance of the source, a partial Spearman correlation test was computed and the partial coefficient, $\rho_{\rm p}$, was obtained (see Table\,\ref{table_co_correlation}).

\begin{table}[h!]
\caption{\label{table_co_correlation}Spearman rank correlation statistics for the CO line luminosity as a function of the clump properties towards the \atgtop sample.}
\centering
\setlength{\tabcolsep}{4pt}
\begin{tabular}{l|ccc}			
	\hline\hline
Property				&	CO\,(4--3)				&	CO\,(6--5)				&	CO\,(7--6)					\\
\hline
\multirow{2}{*}{\lbol}		&	0.70, $p$\,$<$\,0.001; 	&	0.85, $p$\,$<$\,0.001; 	&	0.89, $p$\,$<$\,0.001;	\\
							&	$\rho_p$\,=\,0.81		&	$\rho_p$\,=\,0.91		&	 $\rho_p$\,=\,0.92		\\
\multirow{2}{*}{\mclump}	&	0.75, $p$\,$<$\,0.001;	&	0.70, $p$\,$<$\,0.001;	&	0.67, $p$\,$<$\,0.001; 	\\
							&	$\rho_p$\,=\,0.48		&	$\rho_p$\,=\,0.55		&	 $\rho_p$\,=\,0.57		\\
\lmratio					&   0.24, $p$\,=\,0.05	&   0.45, $p$\,$<$\,0.001	&   0.50, $p$\,$<$\,0.001		\\
\hline
\end{tabular}
\tablefoot{The rank $\rho$ and its corresponding probability ($p$) are shown for each comparison. A $p$-value of $<$\,0.001 indicate a correlation at 0.001 significance level. $p$-values of 0.05, 0.002 and $<$\,0.001 represent the $\sim$\,2, 3 and $>$\,3\,$\sigma$ confidence levels. For \lbol and \mclump, the partial correlation coefficient, $\rho_p$, is also shown.}
\end{table}
\setlength{\tabcolsep}{6pt}

    In the right panel of Fig.\,\ref{fig_lco_correlation_fixbeam_lbol}, we show the CO line luminosity versus the bolometric luminosity of the \atgtop clumps. The plot indicates that \lco increases with \lbol over the entire \lbol range covered by the \atgtop clumps ($\sim$10$^2$-10$^6$\,\lsun).
    The Spearman rank test confirms that both quantities are well correlated for all CO lines ($\rho$\,$\geq$\,0.7, with $p$\,$<$\,0.001), even when excluding the mutual dependence on distance  ($\rho_p$\,$\geq$\,0.81).
    The results of the fits  indicate a systematic increase in the slope of \lco versus \lbol for higher-$J$ transitions: 0.55$\pm$0.05, 0.63$\pm$0.03 and 0.68$\pm$0.03 for the CO\,(4--3), CO\,(6--5) and CO\,(7--6), respectively. For the CO\,(6--5) and CO\,(7--6) lines, however, the slopes are consistent in within 2-$\sigma$.
    Concerning the dependence of the CO luminosity on  \mclump (see Fig.\,\ref{fig_lco_correlation_fixbeam_mclump}), the partial correlation tests indicates that the distance of the clumps plays a more substantial role in the correlation found between \lco and \mclump (0.48$\leq$\,$\rho_p$\,$\leq$\,0.57) than in the correlations found for \lco vs. \lbol.
    Finally, we do not find any strong correlation between the CO line luminosity and \lmratio ($\rho$\,$\leq$\,0.5 for all transitions) although the median \lco values per class do increase with \lmratio (Fig.\,\ref{fig_lco_correlation_fixbeam_lmratio}).
    These findings are discussed in more detail in Sect.\,\ref{sec_discussion}.

We further tested whether the steepness of the relations between \lco and the clump properties is not affected by self-absorption by selecting only those clumps which do not show clear signs of self-absorption (see Sect.\,\ref{sec_gaussfit}). This defines a sub-sample of 15 sources in the CO\,(4--3) line,  18 in the CO\,(6--5) line and 26 objects in the CO\,(7--6) transition. These sources are highlighted in Figs.\,\ref{fig_lco_correlation_fixbeam_lbol}, \ref{fig_lco_correlation_fixbeam_mclump} and \ref{fig_lco_correlation_fixbeam_lmratio}.
    Then, we repeated the fit of the relations between \lco and the clump properties using these sub-samples, finding no significant differences in the slopes of the relations presented in Table\,\ref{table_lco_fit}. 
    The result of the fits for the sub-sample of sources with no signs of self-absorption in their 13\farcs4 spectra are summarised in Table\,\ref{table_lco_fit_nosabs} and the correlations between \lco and the clump properties are listed in Table\,\ref{table_co_correlation_nosabs}.
    The correlations are systematically weaker due to the smaller number of points than those obtained for the whole \atgtop sample (see Table\,\ref{table_co_correlation}).
% Lbol
     We found that the derived slopes for the relations between \lco and \lbol increases from 0.58$\pm$0.09 to 0.71$\pm$0.07, from the CO\,(4--3) to the CO\,(7--6) transition.
     Despite the larger errors in these relations, the slopes of the fits performed on these sources are not significantly different from those found for the whole \atgtop sample, confirming that at least the slopes of the relations found for the whole \atgtop sample are robust in terms of self-absorption effects.
    In addition, similar results were also found for the relations between \lco and the mass of the clumps, while no strong correlation between \lco and \lmratio was found for this sub-sample.
    
\begin{table}[h!]
\caption{\label{table_lco_fit_nosabs}Parameters of the fits of \lco as a function of the clump properties for the \atgtop clumps that are not affected by self-absorption features.}
\centering
\begin{tabular}{cl|ccc}
	\hline\hline
Transition	&	Property	&	$\alpha$	&	$\beta$	&	$\epsilon$	\\
\hline
			&	\lbol	&	$-$0.95$^{+0.39}_{-0.39}$	&	0.58$\pm$0.09 	&	0.56 	\\
CO\,(4--3)	&	\mclump	&	$-$1.60$^{+0.28}_{-0.27}$	&	1.00$\pm$0.09 	&	0.31 	\\
			&	\lmratio&	$+$1.19$^{+0.23}_{-0.29}$	&	0.19$\pm$0.24 	&	0.86 	\\
\hline																
			&	\lbol	&	$-$1.54$^{+0.30}_{-0.29}$	&	0.68$\pm$0.08 	&	0.43 	\\
CO\,(6--5)	&	\mclump	&	$-$1.92$^{+0.30}_{-0.28}$	&	1.05$\pm$0.09 	&	0.38 	\\
			&	\lmratio&	$+$0.76$^{+0.24}_{-0.23}$	&	0.43$\pm$0.21 	&	0.83 	\\
\hline																
			&	\lbol	&	$-$1.71$^{+0.28}_{-0.29}$	&	0.71$\pm$0.07 	&	0.33 	\\
CO\,(7--6)	&	\mclump	&	$-$1.95$^{+0.34}_{-0.32}$	&	1.02$\pm$0.11 	&	0.41 	\\
			&	\lmratio&	$+$0.57$^{+0.22}_{-0.23}$	&	0.53$\pm$0.20 	&	0.79 	\\
\hline 
\end{tabular}
\tablefoot{The fits were performed by adjusting a model with three free parameters in the form of $\log(y) = \alpha + \beta \log(x) \pm \epsilon$, where $\alpha$, $\beta$ and $\epsilon$ correspond to the intercept, the slope and the intrinsic scatter, respectively.}
\end{table}

\begin{table}[h!]
\caption{\label{table_co_correlation_nosabs}Spearman rank correlation statistics for the CO line luminosity as a function of the clump properties for the \atgtop clumps that are not affected by self-absorption features.}
\centering
\setlength{\tabcolsep}{4pt}
\begin{tabular}{l|ccc}			
	\hline\hline
Property				&	CO\,(4--3)				&	CO\,(6--5)				&	CO\,(7--6)					\\
\hline
\multirow{2}{*}{\lbol}		&	0.56, $p$\,=\,0.03; 	&	0.81, $p$\,$<$\,0.001; 	&	0.83, $p$\,$<$\,0.001;	\\
							&	$\rho_p$\,=\,0.54		&	$\rho_p$\,=\,0.83		&	 $\rho_p$\,=\,0.89		\\
\multirow{2}{*}{\mclump}	&	0.72, $p$\,=\,0.002;	&	0.73, $p$\,$<$\,0.001;	&	0.79, $p$\,$<$\,0.001; 	\\
							&	$\rho_p$\,=\,0.24		&	$\rho_p$\,=\,0.50		&	 $\rho_p$\,=\,0.45		\\
\lmratio					&   $-$0.02, $p$\,=\,0.95	&   0.39, $p$\,=\,0.09		&   0.30, $p$\,=\,0.11		\\
\hline
\end{tabular}
\tablefoot{The rank $\rho$ and its corresponding probability ($p$) are shown for each comparison. A $p$-value of $<$\,0.001 indicate a correlation at 0.001 significance level. $p$-values of 0.05, 0.002 and $<$\,0.001 represent the $\sim$\,2, 3 and $>$\,3\,$\sigma$ confidence levels. For \lbol and \mclump, the partial correlation coefficient, $\rho_p$, is also shown.}
\end{table}
\setlength{\tabcolsep}{6pt}

\subsection{The excitation temperature of the CO gas}
\label{sec_texc}

    The increase of \lco with the bolometric luminosity of the source (see Fig.\,\ref{fig_lco_correlation_fixbeam_lbol}) suggests that the intensity of the CO transitions may depend on an average warmer temperature of the gas in the clumps due to an increase of the radiation field from the central source \citep[see e.g.][]{vanKempen09}.
    To confirm this scenario, we computed the excitation temperature of the gas, \tex, and compared it with the properties of the clumps.
    
    Ideally, the intensity ratio of different CO transitions well separated in energy (e.g. CO\,(4--3) and CO\,(7--6)) allows a determination of the excitation temperature of the gas.
    However, most of the CO profiles in the \atgtop clumps are affected by self-absorption (see Sect.\,\ref{sec_selfabs}), causing a considerable underestimate of the flux  especially in CO\,(4--3)  and leading to unreliable ratios. Moreover, the CO\,(6--5) and CO\,(7--6) lines are too close in energy to allow a reliable estimate of the temperature.
    Alternatively, the excitation temperature can be estimated using the peak intensity of optically thick lines.
     From the equation of radiative transport, the observed main beam temperature ($T_{\rm{mb}}$) can be written in terms of \tex as:
\begin{equation}
T_{\rm{mb}} = \frac{h \nu}{k} \left[ J_\nu(T_{\rm{ex}})-J_\nu(T_{\rm{bg}})\right] \left[ 1 - exp\left(-\tau_\nu \right) \right]
\label{eq.line}
\end{equation}
\noindent where $J_\nu(T) = \left[{\rm exp}(h \nu / k T) -1\right]^{-1}$, $T_{\rm{bg}}$ is the background temperature and $\tau_\nu$ is the opacity of the source at the frequency $\nu$. In the following, we include only the cosmic background as background radiation.
    Assuming optically thick emission ($\tau_\nu$\,$\gg$\,1), $T_{\rm{ex}}$ is given by:
\begin{equation}
T_{\rm{ex}} = \frac{h \nu / k}{{\rm ln}\left[ 1 + \frac{h \nu / k}{T_{\rm{mb}} + (h \nu / k) J_\nu(T_{\rm{bg}})} \right]}
\label{eq.tex}
\end{equation}

    We computed \tex using  the peak intensity of the CO\,(6--5) line from  the Gaussian fit (Sect.\,\ref{sec_gaussfit}) and  also from its maximum observed value. Since CO\,(6--5) may be  affected by self-absorption, the maximum observed  intensity likely results in  a lower limit of  the excitation temperature.
    The values derived using both methods are reported in Table\,\ref{tbl_excitation_temperature}.
    \tex derived from the peak intensity of the Gaussian fit ranges between 14 and 143\,K, with a median value of 35\,K. The analysis based on the observed intensity delivers similar results (\tex values range between 14 and 147\,K, with a median value of 34\,K).

\begin{figure*}
 \centering
 \subfigure[]{
  \includegraphics[width=0.485\linewidth]{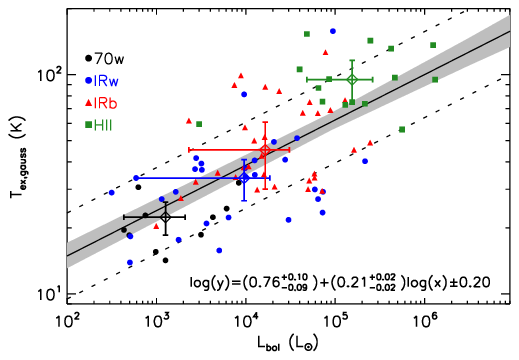}
  \label{fig_tex_lbol}}
 \subfigure[]{
  \includegraphics[width=0.485\linewidth]{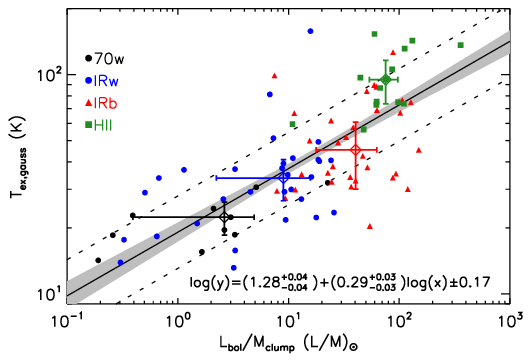}
  \label{fig_tex_lmratio}} \\[-2.0ex]
 \subfigure[]{
  \includegraphics[width=0.485\linewidth]{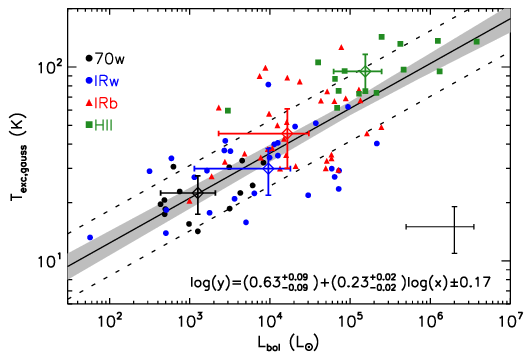}
  \label{fig_gtex_lbol}}
 \subfigure[]{
  \includegraphics[width=0.485\linewidth]{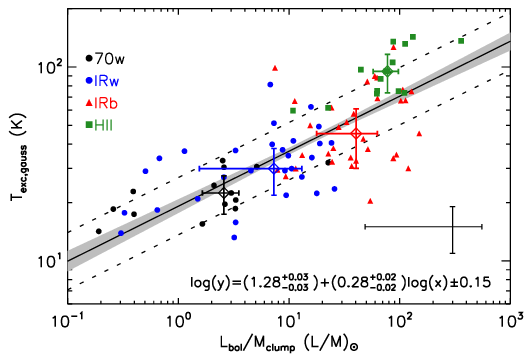}
  \label{fig_gtex_lmratio}} \\[-2.0ex]
 \caption{Excitation temperature of the CO\,(6--5) line versus the bolometric luminosity of the \atgtop clumps (Left) and the luminosity-to-mass ratio (Right).
 The excitation temperature was derived using the peak of the Gaussian fit of the CO profiles. 
 The median values for each class are shown as open diamonds and their error bars correspond to the median absolute deviation of the data from their median value.
The black solid line is the best fit, the light grey shaded area indicates the 68\% uncertainty, and the dashed lines show the intrinsic scatter ($\epsilon$) of the relation.
The best fit to the data is indicated by the filled black line.}
 \label{fig_tex}
\end{figure*}

    The temperature of the gas increases with the evolutionary stage of the clumps  and is well correlated with \lbol ($\rho$\,=\,0.69, $p$\,$<$\,0.001, see Fig.\,\ref{fig_tex_lbol}). No significant correlation is found with  \mclump  ($\rho$\,=\,0.09, $p$\,=\,0.37). On the other hand, the excitation temperature is strongly correlated with \lmratio ($\rho$\,=\,0.72, $p$\,$<$\,0.001), suggesting a progressive warm-up of the gas in more evolved clumps.
We further compared the \tex values obtained from CO with temperature estimates based on other tracers (C$^{17}$O\,(3--2), methyl acetylene, CH$_3$CCH, ammonia, and the dust, \citealt{Giannetti14,Giannetti17,Koenig15,Wienen12}).
    All temperatures are well correlated ($\rho$\,$\geq$\,0.44, $p$,$<$\,0.001), however, the warm-up of the gas is more evident in the other molecular species than in CO \citep[cf. ][]{Giannetti17}.

\begin{table}
 \centering
\setlength{\tabcolsep}{3pt}	
 \caption{\label{table_tex_classes}Kolmogorov-Smirnov statistics of the excitation temperature of the CO\,(6--5) line as a function of the evolutionary class of the clumps.}
\begin{tabular}{c|ll}
 \hline
 \hline
Classes	& \multicolumn{1}{c}{Observed} & \multicolumn{1}{c}{Gaussian}	\\
 \hline
\fiq-\irq & 0.43, $p$\,=\,0.1		& 0.48, $p$\,=\,0.02	\\
\fiq-\irb & 0.83, $p$\,$<$\,0.001	& 0.76, $p$\,$<$\,0.001	\\
\fiq-\hii & 1.00, $p$\,$<$\,0.001	& 1.00, $p$\,$<$\,0.001	\\
\irq-\irb & 0.56, $p$\,$<$\,0.001	& 0.42, $p$\,=\,0.005	\\
\irq-\hii & 1.00, $p$\,$<$\,0.001	& 0.93, $p$\,$<$\,0.001	\\
\irb-\hii & 0.68, $p$\,=\,0.001		& 0.63, $p$\,=\,0.001	\\    
 \hline
 \end{tabular}
 \tablefoot{The rank KS and its corresponding probability ($p$) are shown for each comparison. A $p$-value of $<$\,0.001 indicate a correlation at 0.001 significance level. $p$-values of 0.05, 0.002 and $<$\,0.001 represent the $\sim$\,2, 3 and $>$\,3\,$\sigma$ confidence levels.}
\end{table}
\setlength{\tabcolsep}{6pt}	

\section{Discussion}
\label{sec_discussion}

\subsection{Opacity effects}
\label{sec_opacity}

In Sect.\,\ref{sec_selfabs} we found that self-absorption features are present in most of the CO spectra analysed in this work.
To address this, we investigated the effects of self-absorption on our analysis and concluded that they are negligible since more than 80\% of the CO integrated intensities are recovered in the majority of the sources (Sect.\,\ref{sec_gaussfit}).
We also verified that  the steepness of the relations between \lco and the clump properties is not affected by self-absorption (Sect.\,\ref{sec_CO_correlations}).
%%%%%%%%%%%%%%%%%

In addition, the CO lines under examination are certainly optically thick, and their opacity is likely to decrease with $J$.
    Indeed, the comparison between \lco for different CO transitions and the bolometric luminosity of the clumps (see Sect.\,\ref{sec_CO_correlations}) suggests a systematic increase in the slope of the relations as a function of $J$ (see Table\,\ref{table_lco_fit} for the derived power-law indices for \lco versus \lbol).
    Such a steepening of the slopes with $J$ is even more evident when including the relation found by \citet{SanJose13} for CO\,(10--9) line luminosity in a complementary sample of the \atgtop. For the CO\,(10--9) transition, they derived, $\log(L_{\rm CO})$\,=\,($-$2.9$\pm$0.2)\,+\,(0.84$\pm$0.06)\,$\log(L_{\rm bol})$, which is steeper than the relations found towards lower-$J$ transitions reported in this work.
    In Sect.\,\ref{sec_CO_highJ} we further discuss \sanjose results by analysing their low- and high-mass YSO sub-samples.
    Our findings suggest that there is a significant offset between the sub-samples, leading to a much steeper relation between \lco and \lbol when considering their whole sample. However, the individual sub-samples follow similar power-law distributions, with power-law indices of (0.70$\pm$0.08) and (0.69$\pm$0.21) for the high- and low-mass YSOs, respectively. 

\begin{figure}
 \centering
 \includegraphics[width=0.85\linewidth]{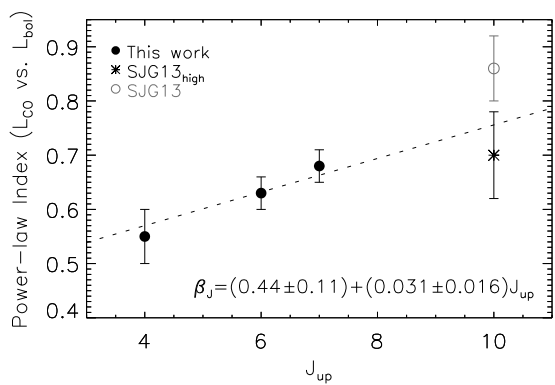}  \\[-1.5ex]
 \caption{Power-law indices of the \lco versus \lbol relations for different $J$ transitions as a function of the upper-level $J$ number.
 The $\beta$ indices from Table\,\ref{table_lco_fit} (filled black circles) are plotted together with data from \sanjose (open grey circle, excluded from the fit) and the exponent derived for their high-luminosity sub-sample ($\ast$ symbol, see Fig.\,\ref{fig_lco_low_to_high_champres}).
 The best fit is indicated by the dashed black line.}
 \label{fig_lco_powerlaw}
\end{figure}

    In Fig.\,\ref{fig_lco_powerlaw}, we present the distribution of the power-law indices of the \lco versus \lbol relations, $\beta_{\rm J}$, as a function of their corresponding upper-level $J$ number, $J_{\rm up}$. We include also the datapoint from the $J_{\rm up}$\,=\,10 line  for the high-mass sources of  \sanjose (see also discussion in Sect.\,\ref{sec_CO_highJ}.
    The best fit to the data, $\beta_{\rm J}$\,=\,(0.44$\pm$0.11)+(0.03$\pm$0.02)\,$J_{\rm up}$,
confirms that the power-law index $\beta_{\rm J}$ gets steeper with $J$.

    The fact that the opacity decreases with $J$ could result in different behaviours of the line luminosities with \lbol for different transitions. This effect was recently discussed by \citet{Benz16} who found that the value of the power-law exponents of the line luminosity of particular molecules and transitions depends mostly on the radius where the line gets optically thick.
    In the case of CO lines, the systematic increase on the steepness of the \lco versus \lbol relation with $J$ (see Table\,\ref{table_lco_fit}) suggests that higher $J$ lines trace more compact gas closer to the source and, thus, a smaller volume of gas is responsible for their emission. 
    Therefore, observations of distinct $J$ transitions of the CO molecule, from CO\,(4--3) to CO\,(7--6) (and even higher $J$ transitions, considering the CO\,(10--9) data from \sanjose), suggest that the line emission arises and gets optically thick at different radii from the central sources, in agreement with \citet{Benz16}.

\subsection{Evolution of CO properties with time}

In Sect.\,\ref{sec_CO_correlations}, we showed that \lco does not correlate with the evolutionary indicator \lmratio. This result is unexpected if we consider that  in \atgtop the  evolutionary classes are quite well separated in \lmratio \citep[with median values of 2.6, 9.0, 40 and 76 for \fiq, \irq, \irb and \hii regions, respectively,][]{Koenig15}.

Previous work on SiO in sources with similar  values of \lmratio as those of the \atgtop (e.g. \citealt{Leurini14} and \citealt{Csengeri16})  found that the line luminosity of  low-excitation SiO transitions does not increase with \lmratio, while the line luminosity of higher excitation SiO lines (i.e. $J_{\rm{up}}>3$) seems to increase with time. Those authors interpreted these findings in terms of a change of excitation conditions with time which is not reflected in low excitation transitions. This effect likely applies  also to low- and mid-$J$ CO lines with relatively low energies ($\le$\,155\,K); higher-$J$ CO transitions could be more sensitive to changes in excitation  since they have upper level energies in excess of 300\,K (e.g. CO\,(10--9) or higher $J$ CO transitions).
    This hypothesis is strengthened by our finding that the excitation temperature of the gas increases with \lmratio (see Fig.\,\ref{fig_tex_lmratio}).
    This scenario can be tested  with observations of high-$J$ CO transitions which are now made possible by the SOFIA telescope in the range $J_{\rm up}$\,=\,11--16. Also {\it Herschel}-PACS archive data could  be used despite their coarse spectral resolution.

\subsection{Do embedded H\,{\sc{ii}} regions still actively power molecular outflows?}
\label{sec_hii}

    In Sect.\,\ref{sec_intpropCO} we showed that \hii regions have broad CO lines (see Fig.\,\ref{fig_avgspc_fixbeam}) likely associated with high-velocity outflowing gas. The \hii sources in our sample are either compact or unresolved objects in the continuum emission at 5\,GHz.
    Their envelopes still largely consist of molecular gas and have not yet been significantly dispersed by the energetic feedback of the YSOs. Our observations suggest that high-mass YSOs in this phase of evolution still power molecular outflows and are therefore accreting.
   This result is in agreement with the recent study of \citet{Urquhart13b} and \citet{Cesaroni15} who suggested that accretion might still be present during the early stages of evolution of \hii regions based on the finding that the Lyman continuum luminosity of several \hii regions appears in excess of that expected for a zero-age main-sequence star with the same bolometric luminosity.
   Such excess could be due to the so-called flashlight effect \citep[e.g.][]{Yorke99}, where most of the photons escape along the axis of a bipolar outflow.
     Indeed \citet{Cesaroni16} further investigated the origin of the Lyman excess looking for infall and outflow signatures in the same sources. They found evidence for both phenomena although with low-angular resolution data. Alternatively, the high-velocity emission seen in CO in the \atgtop in this work and in SiO in the ultra-compact \hii regions of \citet{Cesaroni16}  could be associated with other younger unresolved sources in the clump and not directly associated with the most evolved object in the cluster. Clearly, high angular resolution observations (e.g. with ALMA) are needed to shed light on the origin of the high-velocity emission and confirm whether indeed the high-mass YSO ionising the surrounding gas is still actively accreting. 
 
\subsection{CO line luminosities from low- to high-luminosity sources}
\label{sec_CO_highJ}

	In this section, we further study the correlation between \lco and the bolometric luminosity of the clumps for different CO transitions to investigate the possible biases  that can arise when comparing  data with very different linear resolutions. This is important in particular when comparing  galactic observations to the increasing number of extragalactic studies of mid- and high-$J$ CO lines \citep[e.g.][]{Weiss07,Decarli16}.
    We use results from  \sanjose for the high energy CO\,(10--9) line (with a resolution of $\sim 20$\arcsec) and from the CO\,(6--5) and CO\,(7--6) transitions  observed with APEX by \vankempen. The sources presented by \sanjose cover a broad range of luminosities (from $<$\,1\,L$_\odot$ to $\sim$\,10$^5$\,L$_\odot$) and are in different evolutionary phases. On the other hand, the sample studied by \vankempen consists of eight low-mass YSOs with bolometric luminosities $\lesssim$\,30\,\lsun.

 To investigate the dependence of the line luminosity in different CO transitions on \lbol from low- to high-mass star-forming clumps, we first divided the sources from \sanjose into low- (\lbol$<$\,50\,\lsun) and  high-luminosity (\lbol$>$\,50\,\lsun) objects.
 In this way and assuming the limit \lbol=\,50\,\lsun adopted by \sanjose as a separation between low- and high-mass YSOs, we defined a sub-sample of low-mass sources (the targets of \vankempen for CO\,(6--5) and CO\,(7--6), and those of \sanjose with \lbol$<$\,50\,\lsun for CO $J$\,=\,10--9) and one of intermediate- to high-mass clumps (the \atgtop for the mid-$J$ CO lines and  the sources from \sanjose with \lbol$>$\,50\,\lsun for CO\,(10--9)).
% The limit \lbol=\,50\,\lsun was chosen  as separation between low- and intermediate/high-mass YSOs following \sanjose.

\begin{figure}[!ht]
 \centering
    \includegraphics[width=\linewidth]{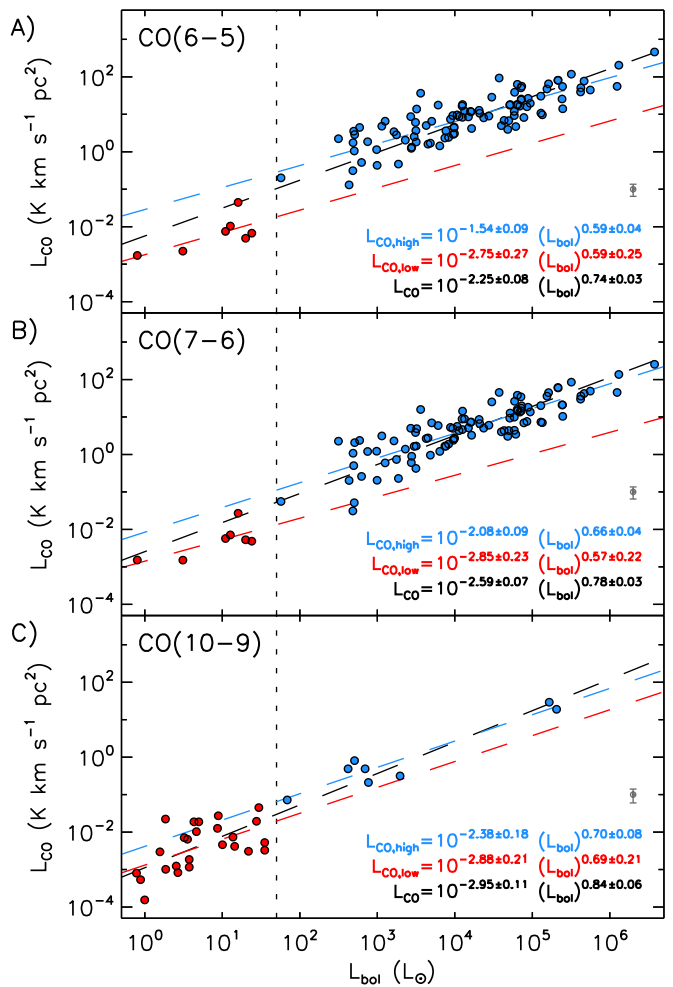} \\[-2.0ex]
 \caption{CO line luminosity as a function of the bolometric luminosity for the CO\,(6--5) (upper panel), CO\,(7--6) (middle) and CO\,(10--9) (bottom) transitions.
 The fits were performed on the whole dataset (all points, shown with black contours), on the low- and high-luminosity sub-samples (points filled in red and blue, respectively). 
 The CO\,(6--5) and CO\,(7--6) data towards low-luminosity sources are from \citet{vanKempen09}; the CO\,(10--9) data are from \citet{SanJose13}. The dashed vertical line  at \lbol=\,50\,\lsun marks the transition from low- to high-mass YSOs. The typical error bars are shown in the bottom right side of the plots.} 
 \label{fig_lco_low_to_high_champres}
\end{figure}

     In the upper panels of Fig.\,\ref{fig_lco_low_to_high_champres} we compare our data with those of \vankempen for the CO\,(6--5) and CO\,(7--6) transitions. We could not include the sources of \sanjose in this analysis because observations in the CO\,(6--5) or CO\,(7--6) lines are not available.
	We calculated the CO line luminosity of their eight low-mass YSOs using the integrated intensities centred on the YSO on scales of $\sim$0.01\,pc (see  their Table\,3). 
    In order to limit  biases due to different beam sizes, we recomputed the  CO luminosities from the central position of our map at the original resolution of the \champ data (see Table\,\ref{table:champ_setup}), probing  linear scales ranging from $\sim$0.04 to 0.6\,pc.
    We performed three fits on the data: we first considered only the original sources of \vankempen and the \atgtop separately, and then combined both samples.
     The derived power-law indices of the CO\,(6--5) data are 0.59\,$\pm$\,0.25 and 0.59\,$\pm$\,0.04 for the low- and high-luminosity sub-samples, respectively.
   Although the power-law indices derived for the two sub-samples are consistent within 1-$\sigma$, the fits are offset by roughly one order of magnitude (from $-2.75$ to $-1.54$\,dex), indicating that \lco values are systematically larger towards high-luminosity sources.
      Indeed, the change on the offsets explains reasonably well the steeper power-law index found when combining both sub-samples (0.74\,$\pm$0.03).
      Similar results are found for  CO\,(7--6), although the difference between the offsets are slight smaller ($\sim$0.8\,dex).
The bottom panel of Fig.\,\ref{fig_lco_low_to_high_champres} presents the CO\,(10--9) luminosity for the \sanjose sample with the best fit of their low- and high-luminosity sources separately. The derived power-law indices are 0.69\,$\pm$\,0.21 and 0.70\,$\pm$\,0.08 for the low- and high-luminosity sub-samples, respectively. The fits are offset by roughly 0.3\,dex, which also explains the steeper slope of 0.84\,$\pm$\,0.06 found by \sanjose when fitting the two sub-samples simultaneously.
    
We interpret the shift in CO line luminosities between low- and high-luminosity sources as a consequence of the varying linear resolution and sampled volume of gas of the data across the \lbol axis.
    In high-mass sources, mid-$J$ CO lines trace extended gas (see the maps presented in Figs.\,\ref{fig_fixbeam_gaussian_fit} for the \atgtop) probably due to the effect of clustered star formation. Since the data presented in Fig.\,\ref{fig_lco_low_to_high_champres} are taken with comparable angular resolutions, the volume of gas sampled by the data is increasing with \lbol because sources with high luminosities are on average more distant.
    For the CO\,(10--9) data, the two sub-samples are likely differently affected by beam dilution. In close-by low-mass YSOs, the CO\,(10--9) line is dominated by emission from UV heated outflow cavities \citep{vanKempen10} and therefore is extended.
    In high-mass YSOs, the CO\,(10--9) line is probably emitted in the inner part of passively heated envelope \citep{Karska14} and therefore could suffer from beam dilution.
    This could explain the smaller offset in the CO\,(10--9) line luminosity between the low- and high-luminosity sub-samples.
    
\sanjose and \citet{Sanjose16} found a similar increase in the slope of the line luminosity of the CO\,(10--9) and of H$_2$O transitions versus \lbol when including extragalactic sources (see Fig.\,14 of \sanjose and Fig.\,9 of \citealt{Sanjose16}).
    These findings clearly outline the difficulties of comparing observations of such different scales and the problems to extend Galactic relations to extragalactic objects.

\section{Summary}
\label{sec_summary}

A sample of 99 sources, selected from the ATLASGAL 870\,\um survey and representative of the Galactic population of star-forming clumps in different evolutionary stages (from 70\,\um-weak clumps to \hii regions), was characterised in terms of their CO\,(4--3), CO\,(6--5) and CO\,(7--6) emission.

We first investigated the effects of different linear resolutions on our data.
By taking advantage of our relatively high angular resolution maps in the CO\,(6--5) and CO\,(7--6) lines, we could study the influence of different beam sizes on the observed line profiles and on the integrated emission.
We first convolved the CO\,(6--5) and CO\,(7--6) data to a common linear size of $\sim$0.24\,pc using a distance limited sub-sample of clumps and then to a common angular resolution of 13\farcs4, including the single-pointing CO\,(4--3) data.
We verified that the results typically do not depend on the spatial resolution of the data, at least in the range of distances sampled by our sources.
The only difference between the two methods is found when comparing the average spectra for each evolutionary class: indeed, only when using spectra that sample the same volume of gas (i.e. same linear resolution) it is possible to detect an increase in line width from \fiq clumps to \hii regions, while the line widths of each evolutionary class are less distinct in the spectra smoothed to the same angular size due to sources at large distances ($>$12\,kpc). 
This result is encouraging for studies of large samples of SF regions across the Galaxy based on single-pointing observations. 

The analysis of the CO emission led to the following results:
\begin{enumerate}
  \item All the sources were detected in the CO\,(4--3), CO\,(6--5) and CO\,(7--6) transitions.
  \item The spatial distribution of the CO\,(6--5) emission ranges between 0.1 and 2.4\,pc. The sizes of mid-$J$ CO emission display a moderate correlation with the sub-mm dust mass of the clumps, suggesting that the extension of the gas probed by the CO is linked to the available amount of the total gas in the region. In addition, the CO\,(6--5) extension is also correlated with the infrared emission probed by the {\it Herschel}-PACS 70\,\um maps towards the 70\,\um-bright clumps.
  \item  The CO profiles can be decomposed using up to three velocity components. The majority of the spectra are well fitted by two components, one narrow (FWHM\,$<$\,7.5\,km\,s$^{-1}$) and one broad;  30\% of the sources need a third and broader component for the CO\,(6--5) line profile.
  \item The FWZP of the CO lines increases with the evolution of the clumps (with median values of 26, 42, 72 and 94 for \fiq, \irq, \irb clumps and \hii regions, respectively, for the CO\,(6--5) transition).
  \hii regions are often associated with broad velocity components, with FWHM values up to $\sim$100\,\kms. This suggests that accretion, resulting in outflows, is still undergoing in the more evolved clumps of the \atgtop.
    \item The CO line luminosity increases with the bolometric luminosity of the sources, although it does not seem to increase neither with the mass nor with the \lmratio ratio of the clumps.
  \item The dependence of the CO luminosity as a function of the bolometric luminosity of the source seems to get steeper with $J$. This likely reflects the fact that higher $J$ CO transitions are more sensitive to the temperature of the gas and likely arise from an inner part of the envelope.
  These findings are quite robust in terms of self-absorption present in most of the $^{12}$CO emission.
  \item The excitation temperature of the clumps was evaluated based on the peak intensity of the Gaussian fit of the CO\,(6--5) spectra.
  We found that \tex increases as a function of the bolometric luminosity and the luminosity-to-mass ratio of the clumps, as expected for a warming up of the gas from \fiq clumps towards \hii regions.
     The observed CO emission towards more luminous and distant objects likely originates from multiple sources within the linear scale probed by the size of beam (up to 0.84\,pc), thus, are systematically larger than the emission from resolved and nearby less luminous objects, from which the CO emission is integrated over smaller linear scales ($\sim$0.01\,pc).
     We found that the line luminosity of the CO lines shows similar slopes as a function of the bolometric luminosity for low-mass and high-mass star-forming sources.
     However, as a consequence, the distribution of the CO line luminosity versus the bolometric luminosity follows steeper power-laws when combining low- and high-luminosity sources.
\end{enumerate}

\begin{acknowledgements}
  F.N. thanks to {\it Funda\c{c}\~{a}o de Amparo \`a Pesquisa do Estado de S\~{a}o Paulo} (FAPESP) for support through processes 2013/11680-2, 2014/20522-4 and 2017/18191-8.
  T.Cs. acknowledges support from the \emph{Deut\-sche For\-schungs\-ge\-mein\-schaft, DFG\/}  via the SPP (priority programme) 1573 'Physics of the ISM'. 
  We thank the useful comments and suggestions made by an anonymous referee that led to a much improved version of this work.
\end{acknowledgements}

\bibliographystyle{aa} % style aa.bst

\Online

\begin{onecolumn}
  \begin{appendix}
  
\section{Full tables}
\label{appendix_co_tables}

Here, we present the full version of tables shown in the paper.
Table\,\ref{tbl_observations_short} presents the properties of the observed clumps.
Tables\,\ref{tbl_gauss_fitting_co43_fixbeam}, \ref{tbl_gauss_fitting_co65_fixbeam} and \ref{tbl_gauss_fitting_co76_fixbeam} present the Gaussian components of each source for the CO\,(4--3), CO\,(6--5) and CO\,(7--6) transitions, respectively.
The extension of the CO\,(6--5) emission is listed in Table\,\ref{table_co_extension}.
Table\,\ref{tbl_intpropCO_fixbeam} displays the integrated properties of the CO lines studied in this work.
Finally, the excitation temperature derived from the CO\,(6--5) spectra is presented in Table\,\ref{tbl_excitation_temperature}.

\begin{landscape}	
\begin{longtab}	
\setcounter{table}{0}	
\captionsetup{width=0.9\textwidth}	
\begin{longtable}{rcrrcccrrrrc}	
\caption{\label{tbl_observations_short} Properties of the \atgtop sources.}\\	
\hline\hline                                                            	
\multirow{2}{*}{ID} & \multirow{2}{*}{CSC Name}   & RA(J2000)       & DEC(J2000)     & Offset CSC      & \multirow{2}{*}{GCSC Name}     & Offset GCSC  &	\vlsr	&	$d$	&	\multirow{2}{*}{$\log\left(\frac{L_{\rm{bol}}}{L_\odot}\right)$}	&	\multirow{2}{*}{$\log\left(\frac{M_{\rm{clump}}}{M_\odot }\right)$}	&	\multirow{2}{*}{Class}	\\	
                    &                        	& (HH:MM:SS)      & (DD:MM:SS)      &  (\arcsec,\arcsec)	&	&  (\arcsec,\arcsec)	&	(km\,s$^{-1}$)	&	(kpc)	&	&	&	\\	
\hline	
\endfirsthead	
\caption{continued.} \\	
\hline\hline                                                                                                    	
\multirow{2}{*}{ID} & \multirow{2}{*}{CSC Name}   & RA(J2000)       & DEC(J2000)     & Offset CSC      & \multirow{2}{*}{GCSC Name}     & Offset GCSC  &	\vlsr	&	$d$	&	\multirow{2}{*}{$\log\left(\frac{L_{\rm{bol}}}{L_\odot}\right)$}	&	\multirow{2}{*}{$\log\left(\frac{M_{\rm{clump}}}{M_\odot }\right)$}	&	\multirow{2}{*}{Class}	\\	
                    &                        	& (HH:MM:SS)      & (DD:MM:SS)      &  (\arcsec,\arcsec)	&	&  (\arcsec,\arcsec)	&	(km\,s$^{-1}$)	&	(kpc)	&	&	&	\\	
\hline	
\endhead	
\hline	
\endfoot	
1	&	AGAL008.684$-$00.367	&	18:06:23.27	&	$-$21:37:12.7	&	(+2.5,+6.8)	&	G008.6834$-$0.3675	&	(+1.9,+4.7)	&	37.3	&	4.8	&	4.44	&	3.17	&	\irq	\\
2	&	AGAL008.706$-$00.414	&	18:06:36.81	&	$-$21:37:18.1	&	($-$1.9,$-$1.7)	&	G008.7064$-$0.4136	&	($-$1.3,+1.5)	&	37.6	&	4.8	&	2.70	&	3.22	&	\irq	\\
3	&	AGAL010.444$-$00.017	&	18:08:44.94	&	$-$19:54:32.0	&	($-$3.2,$-$4.7)	&	G010.4446$-$0.0178	&	($-$0.8,$-$4.0)	&	74.8	&	8.6	&	4.05	&	3.21	&	\irq	\\
4	&	AGAL010.472+00.027	&	18:08:38.24	&	$-$19:51:49.6	&	($-$1.0,+0.6)	&	G010.4722+0.0277	&	($-$3.1,+0.3)	&	66.3	&	8.6	&	5.67	&	4.02	&	\hii	\\
5	&	AGAL010.624$-$00.384	&	18:10:28.87	&	$-$19:55:47.4	&	(+0.0,$-$0.7)	&	G010.6237$-$0.3833	&	($-$4.6,$-$0.6)	&	$-$2.8	&	5.0	&	5.63	&	3.58	&	\hii	\\
6	&	AGAL012.804$-$00.199	&	18:14:13.75	&	$-$17:55:31.2	&	($-$5.4,$-$13.6)	&	G012.8057$-$0.1994	&	($-$2.9,$-$9.2)	&	35.3	&	2.4	&	5.39	&	3.27	&	\hii	\\
7	&	AGAL013.178+00.059	&	18:14:00.77	&	$-$17:28:37.8	&	(+6.8,$-$2.1)	&	G013.1768+0.0599	&	(+2.9,$-$3.0)	&	49.3	&	2.4	&	3.92	&	2.57	&	\fiq	\\
8	&	AGAL013.658$-$00.599	&	18:17:24.25	&	$-$17:22:11.9	&	(+1.2,+2.1)	&	G013.6570$-$0.5992	&	(+1.6,+0.7)	&	47.4	&	4.5	&	4.32	&	2.76	&	\irb	\\
9	&	AGAL014.114$-$00.574	&	18:18:13.21	&	$-$16:57:17.4	&	($-$0.5,$-$2.3)	&	G014.1145$-$0.5745	&	(+1.6,$-$2.0)	&	19.5	&	2.6	&	3.50	&	2.55	&	\irq	\\
10	&	AGAL014.194$-$00.194	&	18:16:58.81	&	$-$16:42:15.6	&	(+0.1,$-$1.7)	&	G014.1944$-$0.1939	&	($-$0.2,$-$0.6)	&	38.9	&	3.9	&	3.43	&	2.91	&	\irq	\\
11	&	AGAL014.492$-$00.139	&	18:17:22.19	&	$-$16:25:00.3	&	($-$0.4,+2.0)	&	G014.4918$-$0.1389	&	(+0.2,+0.6)	&	39.5	&	3.9	&	2.88	&	3.28	&	\fiq	\\
12	&	AGAL014.632$-$00.577	&	18:19:14.82	&	$-$16:30:02.0	&	(+7.7,+0.2)	&	G014.6323$-$0.5763	&	(+6.4,+1.7)	&	17.9	&	1.8	&	3.44	&	2.40	&	\irq	\\
13	&	AGAL015.029$-$00.669	&	18:20:22.64	&	$-$16:11:42.7	&	($-$3.6,+5.7)	&	G015.0292$-$0.6706	&	(+4.2,+1.8)	&	18.5	&	2.0	&	5.13	&	3.08	&	\irb	\\
14	&	AGAL018.606$-$00.074	&	18:25:08.35	&	$-$12:45:22.8	&	($-$0.8,$-$0.5)	&	G018.6057$-$0.0747	&	(+0.8,$-$2.0)	&	44.9	&	4.3	&	2.77	&	2.94	&	\irq	\\
15	&	AGAL018.734$-$00.226	&	18:25:56.21	&	$-$12:42:49.3	&	($-$2.6,$-$0.4)	&	G018.7344$-$0.2261	&	(+1.1,$-$0.3)	&	40.8	&	12.5	&	4.86	&	3.90	&	\irq	\\
16	&	AGAL018.888$-$00.474	&	18:27:07.58	&	$-$12:41:39.5	&	(+1.2,+1.5)	&	G018.8870$-$0.4741	&	($-$0.3,$-$0.3)	&	65.4	&	4.7	&	3.51	&	3.45	&	\irq	\\
17	&	AGAL019.882$-$00.534	&	18:29:14.71	&	$-$11:50:25.4	&	($-$6.5,$-$1.6)	&	G019.8832$-$0.5347	&	($-$1.1,$-$0.4)	&	43.7	&	3.7	&	4.09	&	2.90	&	\irb	\\
18	&	AGAL022.376+00.447	&	18:30:24.22	&	$-$09:10:38.9	&	($-$1.4,+3.2)	&	G022.3752+0.4472	&	($-$1.6,+1.0)	&	52.9	&	4.0	&	2.50	&	2.80	&	\irq	\\
19	&	AGAL023.206$-$00.377	&	18:34:55.09	&	$-$08:49:18.1	&	($-$0.4,+1.2)	&	G023.2056$-$0.3772	&	($-$3.7,+1.1)	&	76.8	&	4.6	&	4.10	&	3.11	&	\irq	\\
20	&	AGAL024.629+00.172	&	18:35:35.71	&	$-$07:18:08.7	&	($-$2.7,$-$7.6)	&	G024.6294+0.1731	&	($-$2.9,$-$6.0)	&	114.5	&	7.7	&	3.70	&	3.18	&	\irq	\\
21	&	AGAL028.564$-$00.236	&	18:44:17.89	&	$-$03:59:44.3	&	(+1.8,+5.1)	&	G028.5637$-$0.2358	&	($-$0.2,+3.8)	&	86.5	&	5.5	&	3.25	&	3.73	&	\irq	\\
22	&	AGAL028.861+00.066	&	18:43:46.20	&	$-$03:35:29.2	&	($-$2.0,$-$3.8)	&	G028.8614+0.0664	&	($-$0.0,$-$0.9)	&	103.0	&	7.4	&	5.21	&	3.03	&	\irb	\\
23	&	AGAL030.818$-$00.056	&	18:47:46.60	&	$-$01:54:30.1	&	(+2.0,+4.4)	&	G030.8166$-$0.0561	&	(+1.2,+1.0)	&	97.8	&	4.9	&	4.80	&	3.75	&	\irb	\\
24	&	AGAL030.848$-$00.081	&	18:47:55.43	&	$-$01:53:37.7	&	($-$1.1,+7.0)	&	G030.8472$-$0.0817	&	(+3.7,+4.4)	&	93.8	&	4.9	&	3.49	&	3.08	&	\fiq	\\
25	&	AGAL030.893+00.139	&	18:47:13.69	&	$-$01:45:07.6	&	($-$6.1,+2.4)	&	G030.8930+0.1383	&	($-$1.9,+2.6)	&	106.7	&	4.9	&	2.70	&	3.28	&	\fiq	\\
26	&	AGAL031.412+00.307	&	18:47:34.40	&	$-$01:12:46.5	&	($-$1.9,+3.5)	&	G031.4120+0.3076	&	($-$4.8,+2.3)	&	97.1	&	4.9	&	4.84	&	3.49	&	\hii	\\
27	&	AGAL034.258+00.154	&	18:53:18.68	&	+01:14:58.5	&	($-$3.9,+1.6)	&	G034.2572+0.1535	&	($-$1.2,$-$0.3)	&	58.1	&	1.6	&	4.68	&	2.91	&	\hii	\\
28	&	AGAL034.401+00.226	&	18:53:18.78	&	+01:24:38.7	&	(+0.4,$-$1.7)	&	G034.4005+0.2262	&	($-$2.6,$-$2.4)	&	56.9	&	1.6	&	3.48	&	2.44	&	\hii	\\
29	&	AGAL034.411+00.234	&	18:53:18.31	&	+01:25:24.6	&	($-$2.8,$-$1.9)	&	G034.4112+0.2344	&	($-$1.0,$-$0.0)	&	57.6	&	1.6	&	3.68	&	2.33	&	\irb	\\
30	&	AGAL034.821+00.351	&	18:53:38.29	&	+01:50:28.2	&	($-$3.2,$-$0.4)	&	G034.8206+0.3504	&	($-$3.8,$-$2.1)	&	56.5	&	1.6	&	3.44	&	2.05	&	\irb	\\
31	&	AGAL035.197$-$00.742	&	18:58:13.09	&	+02:40:38.9	&	($-$2.6,$-$0.5)	&	G035.1976$-$0.7427	&	($-$0.2,$-$0.7)	&	33.4	&	2.2	&	4.37	&	2.67	&	\irb	\\
32	&	AGAL037.554+00.201	&	18:59:10.06	&	+04:12:18.5	&	($-$0.9,$-$2.0)	&	G037.5537+0.2006	&	($-$1.5,$-$3.8)	&	85.2	&	6.7	&	4.71	&	3.10	&	\irb	\\
33	&	AGAL043.166+00.011	&	19:10:13.64	&	+09:06:16.7	&	($-$3.5,$-$6.3)	&	G043.1668+0.0115	&	($-$3.0,$-$2.3)	&	4.9	&	11.1	&	6.58	&	4.64	&	\hii	\\
34	&	AGAL049.489$-$00.389	&	19:23:43.30	&	+14:30:26.5	&	(+13.5,+3.2)	&	G049.4888$-$0.3882	&	(+12.9,+3.7)	&	56.6	&	5.4	&	5.75	&	4.07	&	\hii	\\
35	&	AGAL053.141+00.069	&	19:29:17.52	&	+17:56:22.3	&	(+3.4,$-$4.8)	&	G053.1415+0.0701	&	(+0.1,$-$1.2)	&	21.5	&	1.6	&	3.36	&	1.98	&	\irb	\\
36	&	AGAL059.782+00.066	&	19:43:11.06	&	+23:44:05.4	&	($-$2.4,$-$1.6)	&	G059.7830+0.0657	&	($-$2.1,+0.6)	&	22.2	&	2.2	&	3.99	&	2.41	&	\irb	\\
37	&	AGAL301.136$-$00.226	&	12:35:35.51	&	$-$63:02:30.5	&	($-$10.5,$-$1.6)	&	G301.1365$-$0.2256	&	($-$5.2,$-$1.4)	&	$-$39.3	&	4.4	&	5.33	&	3.29	&	\hii	\\
38	&	AGAL305.192$-$00.006	&	13:11:14.92	&	$-$62:47:26.6	&	($-$14.3,+0.2)	&	G305.1935$-$0.0059	&	($-$4.3,+0.1)	&	$-$34.2	&	3.8	&	4.10	&	2.71	&	\irq	\\
39	&	AGAL305.209+00.206	&	13:11:13.72	&	$-$62:34:38.5	&	(+6.7,$-$3.6)	&	G305.2083+0.2063	&	($-$2.3,$-$2.3)	&	$-$42.2	&	3.8	&	4.95	&	3.15	&	\irb	\\
40	&	AGAL305.562+00.014	&	13:14:26.54	&	$-$62:44:27.3	&	($-$1.7,+3.2)	&	G305.5628+0.0137	&	($-$0.3,+1.5)	&	$-$39.8	&	3.8	&	4.71	&	2.61	&	\irb	\\
41	&	AGAL305.794$-$00.096	&	13:16:34.48	&	$-$62:49:45.7	&	($-$25.3,+5.1)	&	G305.7949$-$0.0965	&	($-$18.6,+2.8)	&	$-$40.9	&	3.8	&	2.99	&	2.77	&	\fiq	\\
42	&	AGAL309.384$-$00.134	&	13:47:22.79	&	$-$62:18:09.8	&	(+21.0,+0.1)	&	G309.3826$-$0.1332	&	(+7.4,+1.9)	&	$-$51.3	&	5.3	&	4.20	&	3.08	&	\irb	\\
43	&	AGAL310.014+00.387	&	13:51:38.19	&	$-$61:39:17.3	&	(+3.0,+4.1)	&	G310.0135+0.3877	&	($-$0.4,+4.0)	&	$-$41.3	&	3.6	&	4.70	&	2.62	&	\irb	\\
44	&	AGAL313.576+00.324	&	14:20:08.44	&	$-$60:42:07.0	&	($-$3.2,+2.0)	&	G313.5763+0.3243	&	(+1.1,+2.7)	&	$-$46.9	&	3.8	&	3.97	&	2.26	&	\irb	\\
45	&	AGAL316.641$-$00.087	&	14:44:18.46	&	$-$59:55:17.8	&	(+4.9,+4.5)	&	G316.6403$-$0.0877	&	(+2.1,+3.0)	&	$-$17.7	&	1.2	&	3.00	&	1.26	&	\irb	\\
46	&	AGAL317.867$-$00.151	&	14:53:16.76	&	$-$59:26:36.9	&	($-$1.7,+4.1)	&	G317.8680$-$0.1514	&	(+6.0,+3.2)	&	$-$40.6	&	3.0	&	3.22	&	2.56	&	\irq	\\
47	&	AGAL318.779$-$00.137	&	14:59:33.29	&	$-$59:00:36.6	&	($-$0.8,+1.9)	&	G318.7790$-$0.1376	&	(+1.0,+1.0)	&	$-$39.4	&	2.8	&	3.81	&	2.56	&	\irq	\\
48	&	AGAL320.881$-$00.397	&	15:14:33.61	&	$-$58:11:31.9	&	($-$7.8,+1.5)	&	G320.8803$-$0.3970	&	($-$9.1,+2.2)	&	$-$46.0	&	10.0	&	3.78	&	3.46	&	\fiq	\\
49	&	AGAL326.661+00.519	&	15:45:03.20	&	$-$54:09:13.9	&	($-$5.2,+2.7)	&	G326.6607+0.5190	&	($-$4.7,+2.4)	&	$-$39.8	&	1.8	&	3.87	&	2.10	&	\irb	\\
50	&	AGAL326.987$-$00.032	&	15:49:08.36	&	$-$54:23:04.7	&	($-$0.4,$-$3.1)	&	G326.9871$-$0.0317	&	($-$4.8,$-$1.3)	&	$-$58.6	&	4.0	&	3.06	&	2.65	&	\irq	\\
51	&	AGAL327.119+00.509	&	15:47:33.56	&	$-$53:52:41.9	&	($-$6.7,$-$2.0)	&	G327.1197+0.5099	&	($-$8.4,+0.8)	&	$-$83.7	&	5.5	&	4.77	&	2.83	&	\irb	\\
52	&	AGAL327.393+00.199	&	15:50:19.55	&	$-$53:57:04.6	&	($-$10.6,$-$0.3)	&	G327.3928+0.1984	&	($-$8.8,$-$2.2)	&	$-$89.2	&	5.9	&	4.13	&	3.07	&	\irb	\\
53	&	AGAL329.029$-$00.206	&	16:00:31.50	&	$-$53:12:38.8	&	(+6.7,$-$14.3)	&	G329.0303$-$0.2022	&	($-$2.7,$-$1.8)	&	$-$43.2	&	11.5	&	5.33	&	4.06	&	\irq	\\
54	&	AGAL329.066$-$00.307	&	16:01:09.93	&	$-$53:16:05.4	&	($-$2.6,+2.1)	&	G329.0656$-$0.3076	&	($-$1.3,+1.0)	&	$-$41.9	&	11.6	&	4.85	&	3.96	&	\irb	\\
55	&	AGAL330.879$-$00.367	&	16:10:20.62	&	$-$52:06:11.0	&	($-$4.9,+3.9)	&	G330.8788$-$0.3681	&	($-$5.2,+1.3)	&	$-$62.6	&	4.2	&	5.19	&	3.20	&	\hii	\\
56	&	AGAL330.954$-$00.182	&	16:09:53.25	&	$-$51:54:54.8	&	($-$6.0,+0.1)	&	G330.9545$-$0.1828	&	($-$5.0,+0.1)	&	$-$92.0	&	9.3	&	6.12	&	4.24	&	\hii	\\
57	&	AGAL331.709+00.582	&	16:10:06.19	&	$-$50:50:28.6	&	(+7.8,+0.9)	&	G331.7084+0.5834	&	(+0.7,+1.4)	&	$-$67.3	&	10.5	&	4.57	&	3.71	&	\irq	\\
58	&	AGAL332.094$-$00.421	&	16:16:16.62	&	$-$51:18:26.0	&	(+0.2,+1.7)	&	G332.0946$-$0.4210	&	(+0.8,+2.3)	&	$-$57.5	&	3.6	&	4.77	&	2.80	&	\irb	\\
59	&	AGAL332.826$-$00.549	&	16:20:10.69	&	$-$50:53:19.6	&	(+3.3,+4.1)	&	G332.8262$-$0.5493	&	(+5.2,+4.6)	&	$-$57.4	&	3.6	&	5.38	&	3.29	&	\hii	\\
60	&	AGAL333.134$-$00.431	&	16:21:02.20	&	$-$50:35:12.6	&	(+4.0,+2.3)	&	G333.1341$-$0.4314	&	(+4.7,$-$0.6)	&	$-$53.5	&	3.6	&	5.62	&	3.46	&	\hii	\\
61	&	AGAL333.284$-$00.387	&	16:21:30.64	&	$-$50:26:54.3	&	(+5.8,$-$4.7)	&	G333.2841$-$0.3868	&	(+4.7,$-$3.9)	&	$-$52.4	&	3.6	&	5.11	&	3.32	&	\hii	\\
62	&	AGAL333.314+00.106	&	16:19:28.79	&	$-$50:04:42.9	&	($-$1.5,+1.2)	&	G333.3139+0.1057	&	($-$3.4,$-$0.4)	&	$-$46.5	&	3.6	&	4.03	&	2.63	&	\irb	\\
63	&	AGAL333.604$-$00.212	&	16:22:09.58	&	$-$50:06:01.1	&	($-$0.6,+1.7)	&	G333.6036$-$0.2130	&	($-$1.5,$-$1.4)	&	$-$47.1	&	3.6	&	6.09	&	3.54	&	\hii	\\
64	&	AGAL333.656+00.059	&	16:21:11.83	&	$-$49:52:16.7	&	($-$4.5,+0.6)	&	G333.6563+0.0587	&	(+0.8,+0.4)	&	$-$85.2	&	5.3	&	3.63	&	3.15	&	\fiq	\\
65	&	AGAL335.789+00.174	&	16:29:47.62	&	$-$48:15:51.4	&	($-$5.0,$-$0.4)	&	G335.7896+0.1737	&	(+2.1,+0.0)	&	$-$50.6	&	3.7	&	4.31	&	3.04	&	\irq	\\
66	&	AGAL336.958$-$00.224	&	16:36:17.29	&	$-$47:40:49.1	&	($-$3.7,+3.2)	&	G336.9574$-$0.2247	&	($-$0.1,+1.5)	&	$-$71.3	&	10.9	&	3.56	&	3.38	&	\irq	\\
67	&	AGAL337.176$-$00.032	&	16:36:18.70	&	$-$47:23:24.5	&	(+2.5,+4.3)	&	G337.1751$-$0.0324	&	(+0.3,+2.3)	&	$-$68.2	&	11.0	&	4.77	&	3.75	&	\irq	\\
68	&	AGAL337.258$-$00.101	&	16:36:56.58	&	$-$47:22:29.1	&	($-$4.8,+1.2)	&	G337.2580$-$0.1012	&	($-$0.9,+1.3)	&	$-$68.3	&	11.0	&	4.48	&	3.50	&	\irq	\\
69	&	AGAL337.286+00.007	&	16:36:34.63	&	$-$47:16:50.9	&	($-$0.7,+0.6)	&	G337.2860+0.0083	&	($-$1.0,+2.9)	&	$-$107.5	&	9.4	&	3.10	&	3.82	&	\fiq	\\
70	&	AGAL337.406$-$00.402	&	16:38:51.00	&	$-$47:27:58.8	&	($-$0.2,+1.0)	&	G337.4052$-$0.4024	&	(+0.1,$-$0.6)	&	$-$40.9	&	3.3	&	4.93	&	3.04	&	\hii	\\
71	&	AGAL337.704$-$00.054	&	16:38:29.69	&	$-$47:00:38.2	&	($-$0.8,$-$2.9)	&	G337.7045$-$0.0535	&	(+1.0,$-$0.3)	&	$-$47.4	&	12.3	&	5.50	&	4.15	&	\hii	\\
72	&	AGAL337.916$-$00.477	&	16:41:10.51	&	$-$47:08:06.7	&	(+2.5,+2.3)	&	G337.9154$-$0.4773	&	(+2.5,+1.5)	&	$-$39.5	&	3.2	&	5.11	&	3.08	&	\irb	\\
73	&	AGAL338.066+00.044	&	16:39:28.79	&	$-$46:40:30.4	&	($-$4.2,$-$4.2)	&	G338.0663+0.0445	&	($-$5.2,$-$2.1)	&	$-$70.1	&	4.7	&	3.50	&	2.98	&	\fiq	\\
74	&	AGAL338.786+00.476	&	16:40:22.30	&	$-$45:51:05.3	&	(+3.6,$-$0.4)	&	G338.7851+0.4767	&	($-$3.3,+0.3)	&	$-$64.0	&	4.5	&	2.69	&	3.09	&	\fiq	\\
75	&	AGAL338.926+00.554	&	16:40:34.50	&	$-$45:41:46.7	&	($-$3.2,+4.9)	&	G338.9249+0.5539	&	($-$2.7,+2.9)	&	$-$61.6	&	4.4	&	4.97	&	3.78	&	\irq	\\
76	&	AGAL339.623$-$00.122	&	16:46:06.21	&	$-$45:36:49.5	&	(+6.0,+2.9)	&	G339.6225$-$0.1220	&	(+5.8,+3.8)	&	$-$34.6	&	3.0	&	4.18	&	2.50	&	\irb	\\
77	&	AGAL340.374$-$00.391	&	16:50:02.85	&	$-$45:12:45.2	&	($-$5.8,+3.1)	&	G340.3736$-$0.3904	&	($-$6.7,+2.4)	&	$-$43.4	&	3.6	&	2.71	&	2.90	&	\irq	\\
78	&	AGAL340.746$-$01.001	&	16:54:04.02	&	$-$45:18:46.7	&	($-$7.1,+1.8)	&	G340.7456$-$1.0014	&	($-$6.3,$-$0.4)	&	$-$29.4	&	2.8	&	3.89	&	2.33	&	\irb	\\
79	&	AGAL340.784$-$00.097	&	16:50:15.36	&	$-$44:42:30.1	&	($-$4.7,$-$2.1)	&	G340.7848$-$0.0968	&	($-$5.5,+1.3)	&	$-$101.7	&	10.0	&	4.86	&	3.45	&	\irq	\\
80	&	AGAL341.217$-$00.212	&	16:52:18.19	&	$-$44:26:53.1	&	($-$3.2,$-$1.2)	&	G341.2179$-$0.2122	&	($-$1.0,+0.7)	&	$-$43.6	&	3.7	&	4.21	&	2.69	&	\irb	\\
81	&	AGAL342.484+00.182	&	16:55:02.31	&	$-$43:12:59.2	&	(+2.3,$-$2.0)	&	G342.4836+0.1831	&	($-$0.4,$-$2.7)	&	$-$41.6	&	12.6	&	4.81	&	3.69	&	\irq	\\
82	&	AGAL343.128$-$00.062	&	16:58:17.47	&	$-$42:52:09.3	&	($-$3.1,+4.8)	&	G343.1271$-$0.0632	&	($-$1.0,+2.0)	&	$-$30.3	&	3.0	&	4.86	&	3.06	&	\hii	\\
83	&	AGAL343.756$-$00.164	&	17:00:50.14	&	$-$42:26:14.7	&	($-$0.8,+2.0)	&	G343.7559$-$0.1640	&	($-$1.5,+2.3)	&	$-$28.2	&	2.9	&	4.00	&	2.79	&	\irq	\\
84	&	AGAL344.227$-$00.569	&	17:04:07.71	&	$-$42:18:41.3	&	(+2.3,+0.8)	&	G344.2275$-$0.5688	&	($-$0.4,+1.7)	&	$-$22.3	&	2.5	&	3.99	&	3.05	&	\irq	\\
85	&	AGAL345.003$-$00.224	&	17:05:11.26	&	$-$41:29:06.6	&	($-$3.6,$-$1.2)	&	G345.0029$-$0.2241	&	($-$3.3,$-$1.2)	&	$-$26.9	&	3.0	&	4.81	&	2.99	&	\hii	\\
86	&	AGAL345.488+00.314	&	17:04:28.26	&	$-$40:46:26.1	&	($-$0.6,+0.7)	&	G345.4871+0.3142	&	($-$0.9,$-$0.6)	&	$-$17.7	&	2.2	&	4.79	&	2.97	&	\hii	\\
87	&	AGAL345.504+00.347	&	17:04:23.18	&	$-$40:44:23.3	&	($-$2.3,$-$1.6)	&	G345.5045+0.3481	&	($-$5.8,+0.4)	&	$-$17.8	&	2.3	&	4.64	&	2.63	&	\irb	\\
88	&	AGAL345.718+00.817	&	17:03:06.25	&	$-$40:17:04.2	&	(+0.5,$-$2.1)	&	G345.7172+0.8176	&	($-$1.9,$-$2.5)	&	$-$11.2	&	1.6	&	3.27	&	2.30	&	\irb	\\
89	&	AGAL351.131+00.771	&	17:19:34.58	&	$-$35:56:47.7	&	($-$0.6,+0.3)	&	G351.1314+0.7709	&	(+1.5,+1.8)	&	$-$5.3	&	1.8	&	2.80	&	2.09	&	\fiq	\\
90	&	AGAL351.161+00.697	&	17:19:56.97	&	$-$35:57:52.2	&	(+7.0,+2.2)	&	G351.1598+0.6982	&	(+0.4,+0.4)	&	$-$6.5	&	1.8	&	3.94	&	3.07	&	\irb	\\
91	&	AGAL351.244+00.669	&	17:20:19.14	&	$-$35:54:42.0	&	($-$10.5,$-$0.3)	&	G351.2437+0.6687	&	($-$8.1,$-$3.0)	&	$-$3.4	&	1.8	&	4.89	&	2.95	&	\irb	\\
92	&	AGAL351.416+00.646	&	17:20:53.65	&	$-$35:47:00.8	&	($-$8.3,$-$2.2)	&	G351.4161+0.6464	&	($-$12.8,$-$0.1)	&	$-$7.4	&	1.3	&	4.60	&	2.67	&	\hii	\\
93	&	AGAL351.444+00.659	&	17:20:55.49	&	$-$35:45:07.8	&	($-$12.7,$-$4.0)	&	G351.4441+0.6579	&	($-$7.9,$-$6.9)	&	$-$4.3	&	1.3	&	3.98	&	3.15	&	\irq	\\
94%	&	AGAL351.496+00.691	&	17:20:56.61	&	$-$35:40:33.8	&	($-$14.4,$-$60.3)	&	G351.5113+0.6984	&	($-$3.2,+1.0)	&	0.0	&	1.3	&	3.18	&	1.79	&	\irq	\\
94	&	AGAL351.571+00.762	&	17:20:51.05	&	$-$35:35:22.4	&	($-$2.4,$-$2.9)	&	G351.5719+0.7631	&	($-$0.7,+1.7)	&	$-$3.4	&	1.3	&	2.64	&	2.22	&	\fiq	\\
95	&	AGAL351.581$-$00.352	&	17:25:25.30	&	$-$36:12:47.2	&	($-$4.7,+2.1)	&	G351.5815$-$0.3528	&	(+0.3,+3.0)	&	$-$95.4	&	6.8	&	5.39	&	3.94	&	\irb	\\
96	&	AGAL351.774$-$00.537	&	17:26:42.55	&	$-$36:09:21.5	&	(+2.7,+0.7)	&	G351.7747$-$0.5369	&	(+2.5,+3.2)	&	$-$2.8	&	1.0	&	4.22	&	2.42	&	\irb	\\
97	&	AGAL353.066+00.452	&	17:26:13.58	&	$-$34:31:54.8	&	($-$3.5,+1.3)	&	G353.0670+0.4519	&	($-$0.4,+3.6)	&	1.5	&	0.9	&	1.76	&	1.25	&	\irq	\\
98	&	AGAL353.417$-$00.079	&	17:29:19.13	&	$-$34:32:14.6	&	($-$1.3,+9.7)	&	G353.4173$-$0.0803	&	(+2.9,+6.8)	&	$-$54.9	&	6.1	&	3.65	&	3.25	&	\fiq	\\
99	&	AGAL354.944$-$00.537	&	17:35:12.04	&	$-$33:30:28.0	&	($-$3.3,+3.9)	&	G354.9437$-$0.5381	&	($-$4.2,+1.0)	&	$-$5.6	&	1.9	&	2.68	&	2.17	&	\fiq	\\
\end{longtable}	
\tablefoot{The columns are as follows: (1) ID of each source; (2) name from the ATLASGAL-CSC catalogue from \citet{Contreras13}; (3)--(4) Equatorial coordinates (J2000) of the central position of the CHAMP$^+$ maps; (5) offset between the CSC and the central position of the CHAMP$^+$ maps; (6) name from the ATLASGAL-GCSC catalogue from \citet{Csengeri14}; (7) offset between GCSC and the central position of the \champ maps; (8) the local standard rest velocity (\vlsr) from the C$^{17}$O\,(3--2) data from \citet{Giannetti14}; (9)--(11) distances, bolometric luminosities and clump masses from \citet{Koenig15}; (12) classification of the clump: 70\,\um weak (\fiq), infrared weak (\irq), infrared bright (\irb) or \hii regions (\hii) from \citet{Koenig15}.}	
\end{longtab}	
\end{landscape}	
 			% A.1

\begin{longtab}
\setcounter{table}{1}
\setlength{\tabcolsep}{4pt}
\captionsetup{width=0.9\textwidth}
\begin{longtable}{cl|lrrr|lrrr|lrrr} \caption{\label{tbl_gauss_fitting_co43_fixbeam}Gaussian decomposition of the CO\,(4--3) profiles towards the \atgtop sample.}\\
\hline\hline
ID	&	\multicolumn{1}{c|}{CSC Name}	&	C$_1$	&	\vpeak	&	FWHM	&	\tpeak	&	C$_2$	&	\vpeak	&	FWHM	&	\tpeak	&	C$_3$	&	\vpeak	&	FWHM	&	\tpeak	 \\ 
	&	&	&	(km\,s$^{-1}$)	&	(km\,s$^{-1}$)	&	(K)	&		&	(km\,s$^{-1}$)	&	(km\,s$^{-1}$)	&	(K)	&		&	(km\,s$^{-1}$)	&	(km\,s$^{-1}$)	&	(K)	 \\ 
\hline
\endfirsthead
\caption{continued.} \\
\hline\hline
ID	&	\multicolumn{1}{c|}{CSC Name}	&	C$_1$	&	\vpeak	&	FWHM	&	\tpeak	&	C$_2$	&	\vpeak	&	FWHM	&	\tpeak	&	C$_3$	&	\vpeak	&	FWHM	&	\tpeak	 \\ 
	&	&	&	(km\,s$^{-1}$)	&	(km\,s$^{-1}$)	&	(K)	&		&	(km\,s$^{-1}$)	&	(km\,s$^{-1}$)	&	(K)	&		&	(km\,s$^{-1}$)	&	(km\,s$^{-1}$)	&	(K)	 \\ 
\hline
\endhead
\hline
\endfoot
%1	&	AGAL008.684$-$00.367$^\ast$	&	N	&	36.7	&	6.4	&	16.4	&	B	&	40.6	&	14.3	&	3.2	&	--	&	--	&	--	&	--	\\
%1	&	AGAL008.684$-$00.367$^\ast$	&	--	&	--	&	--	&	--	&	--	&	--	&	--	&	--	&	--	&	--	&	--	&	--	\\
2	&	AGAL008.706$-$00.414	&	N	&	38.1	&	7.1	&	4.9	&	--	&	--	&	--	&	--	&	--	&	--	&	--	&	--	\\
3	&	AGAL010.444$-$00.017	&	B	&	73.8	&	9.3	&	3.1	&	P2	&	65.2	&	4.9	&	2.1	&	--	&	--	&	--	&	--	\\
4	&	AGAL010.472$+$00.027	&	N	&	64.6	&	6.2	&	22.8	&	B	&	68.0	&	12.7	&	11.1	&	--	&	--	&	--	&	--	\\
5	&	AGAL010.624$-$00.384	&	B	&	$-$2.8	&	12.7	&	51.2	&	--	&	--	&	--	&	--	&	--	&	--	&	--	&	--	\\
6	&	AGAL012.804$-$00.199	&	B	&	36.6	&	14.9	&	43.6	&	--	&	--	&	--	&	--	&	--	&	--	&	--	&	--	\\
7	&	AGAL013.178$+$00.059	&	B	&	48.3	&	11.8	&	9.2	&	--	&	--	&	--	&	--	&	--	&	--	&	--	&	--	\\
8	&	AGAL013.658$-$00.599	&	N	&	48.6	&	7.1	&	8.7	&	B	&	46.4	&	49.5	&	1.6	&	--	&	--	&	--	&	--	\\
%9	&	AGAL014.114$-$00.574$^\ast$	&	N	&	19.6	&	5.1	&	15.1	&	B	&	21.1	&	16.7	&	1.0	&	--	&	--	&	--	&	--	\\
%9	&	AGAL014.114$-$00.574$^\ast$	&	--	&	--	&	--	&	--	&	--	&	--	&	--	&	--	&	--	&	--	&	--	&	--	\\
10	&	AGAL014.194$-$00.194	&	B1	&	40.1	&	9.1	&	10.4	&	B2	&	39.7	&	23.4	&	3.4	&	--	&	--	&	--	&	--	\\
11	&	AGAL014.492$-$00.139	&	N	&	40.1	&	7.5	&	7.9	&	B	&	27.2	&	70.9	&	0.7	&	--	&	--	&	--	&	--	\\
12	&	AGAL014.632$-$00.577	&	B1	&	17.4	&	7.6	&	14.0	&	B2	&	22.1	&	13.8	&	1.2	&	--	&	--	&	--	&	--	\\
13	&	AGAL015.029$-$00.669	&	B	&	19.6	&	9.3	&	55.0	&	--	&	--	&	--	&	--	&	--	&	--	&	--	&	--	\\
14	&	AGAL018.606$-$00.074	&	B	&	45.2	&	7.5	&	9.9	&	--	&	--	&	--	&	--	&	--	&	--	&	--	&	--	\\
15	&	AGAL018.734$-$00.226	&	N	&	39.4	&	4.9	&	11.0	&	B	&	39.6	&	27.8	&	1.4	&	--	&	--	&	--	&	--	\\
16	&	AGAL018.888$-$00.474	&	B	&	65.7	&	11.9	&	14.3	&	--	&	--	&	--	&	--	&	--	&	--	&	--	&	--	\\
17	&	AGAL019.882$-$00.534	&	B1	&	44.3	&	11.0	&	18.2	&	B2	&	43.1	&	27.8	&	6.5	&	B3	&	86.6	&	29.2	&	0.3	\\
%18	&	AGAL022.376$+$00.447$^\ast$	&	N	&	53.3	&	3.4	&	20.3	&	B	&	56.9	&	16.9	&	3.8	&	--	&	--	&	--	&	--	\\
%18	&	AGAL022.376$+$00.447$^\ast$	&	--	&	--	&	--	&	--	&	--	&	--	&	--	&	--	&	--	&	--	&	--	&	--	\\
19	&	AGAL023.206$-$00.377	&	B1	&	78.1	&	16.7	&	13.0	&	B2	&	87.5	&	64.1	&	1.4	&	--	&	--	&	--	&	--	\\
20	&	AGAL024.629$+$00.172	&	N	&	114.0	&	3.2	&	3.3	&	B	&	116.7	&	13.9	&	2.5	&	--	&	--	&	--	&	--	\\
21	&	AGAL028.564$-$00.236	&	B	&	87.2	&	8.4	&	5.9	&	--	&	--	&	--	&	--	&	--	&	--	&	--	&	--	\\
%22	&	AGAL028.861$+$00.066$^\ast$	&	N	&	107.4	&	6.9	&	13.8	&	B	&	104.0	&	25.5	&	5.0	&	--	&	--	&	--	&	--	\\
%22	&	AGAL028.861$+$00.066$^\ast$	&	--	&	--	&	--	&	--	&	--	&	--	&	--	&	--	&	--	&	--	&	--	&	--	\\
23	&	AGAL030.818$-$00.056	&	B1	&	97.3	&	8.0	&	25.3	&	B2	&	101.7	&	19.2	&	5.6	&	--	&	--	&	--	&	--	\\
24	&	AGAL030.848$-$00.081	&	N1	&	102.9	&	7.0	&	3.6	&	N2	&	94.4	&	7.5	&	9.2	&	--	&	--	&	--	&	--	\\
25	&	AGAL030.893$+$00.139	&	N	&	106.7	&	6.2	&	8.3	&	P2	&	94.3	&	2.0	&	4.2	&	--	&	--	&	--	&	--	\\
%26	&	AGAL031.412$+$00.307$^\ast$	&	N	&	98.2	&	7.3	&	24.7	&	B	&	96.5	&	20.5	&	3.0	&	--	&	--	&	--	&	--	\\
%26	&	AGAL031.412$+$00.307$^\ast$	&	--	&	--	&	--	&	--	&	--	&	--	&	--	&	--	&	--	&	--	&	--	&	--	\\
%27	&	AGAL034.258$+$00.154$^\ast$	&	B1	&	57.7	&	8.8	&	25.9	&	B2	&	58.4	&	23.4	&	6.7	&	--	&	--	&	--	&	--	\\
%27	&	AGAL034.258$+$00.154$^\ast$	&	--	&	--	&	--	&	--	&	--	&	--	&	--	&	--	&	--	&	--	&	--	&	--	\\
28	&	AGAL034.401$+$00.226	&	B1	&	57.5	&	11.6	&	19.1	&	B2	&	56.8	&	51.0	&	1.1	&	--	&	--	&	--	&	--	\\
29	&	AGAL034.411$+$00.234	&	B1	&	57.9	&	10.7	&	14.3	&	B2	&	62.9	&	42.9	&	1.6	&	--	&	--	&	--	&	--	\\
%30	&	AGAL034.821$+$00.351$^\ast$	&	B1	&	57.2	&	12.4	&	9.9	&	B2	&	61.5	&	56.2	&	1.5	&	--	&	--	&	--	&	--	\\
%30	&	AGAL034.821$+$00.351$^\ast$	&	--	&	--	&	--	&	--	&	--	&	--	&	--	&	--	&	--	&	--	&	--	&	--	\\
31	&	AGAL035.197$-$00.742	&	B1	&	34.5	&	9.0	&	27.1	&	B2	&	31.2	&	15.7	&	11.9	&	--	&	--	&	--	&	--	\\
32	&	AGAL037.554$+$00.201	&	N	&	85.7	&	7.1	&	11.3	&	B	&	82.7	&	16.9	&	7.2	&	--	&	--	&	--	&	--	\\
33	&	AGAL043.166$+$00.011	&	B1	&	$-$1.1	&	11.8	&	39.0	&	B2	&	13.2	&	17.3	&	24.0	&	B3	&	2.8	&	33.4	&	6.6	\\
%34	&	AGAL049.489$-$00.389$^\ast$	&	N	&	52.2	&	6.8	&	27.7	&	B	&	56.6	&	20.1	&	16.1	&	--	&	--	&	--	&	--	\\
%34	&	AGAL049.489$-$00.389$^\ast$	&	--	&	--	&	--	&	--	&	--	&	--	&	--	&	--	&	--	&	--	&	--	&	--	\\
35	&	AGAL053.141$+$00.069	&	B1	&	23.3	&	9.2	&	20.6	&	B2	&	23.4	&	42.1	&	1.4	&	--	&	--	&	--	&	--	\\
36	&	AGAL059.782$+$00.066	&	B1	&	22.2	&	8.1	&	24.1	&	B2	&	21.0	&	26.0	&	3.0	&	--	&	--	&	--	&	--	\\
%37	&	AGAL301.136$-$00.226	&	--	&	--	&	--	&	--	&	--	&	--	&	--	&	--	&	--	&	--	&	--	&	--	\\
%38	&	AGAL305.192$-$00.006$^\ast$	&	B	&	$-$30.5	&	7.8	&	6.3	&	--	&	--	&	--	&	--	&	--	&	--	&	--	&	--	\\
%38	&	AGAL305.192$-$00.006$^\ast$	&	--	&	--	&	--	&	--	&	--	&	--	&	--	&	--	&	--	&	--	&	--	&	--	\\
39	&	AGAL305.209$+$00.206	&	N	&	$-$44.6	&	6.1	&	15.7	&	B	&	$-$40.3	&	17.3	&	18.2	&	--	&	--	&	--	&	--	\\
40	&	AGAL305.562$+$00.014	&	N	&	$-$38.8	&	6.7	&	25.1	&	B	&	$-$41.0	&	14.2	&	13.6	&	--	&	--	&	--	&	--	\\
%41	&	AGAL305.794$-$00.096$^\ast$	&	B	&	$-$42.4	&	8.5	&	2.2	&	--	&	--	&	--	&	--	&	--	&	--	&	--	&	--	\\
%41	&	AGAL305.794$-$00.096$^\ast$	&	--	&	--	&	--	&	--	&	--	&	--	&	--	&	--	&	--	&	--	&	--	&	--	\\
42	&	AGAL309.384$-$00.134	&	B	&	$-$50.6	&	11.8	&	13.2	&	--	&	--	&	--	&	--	&	--	&	--	&	--	&	--	\\
%43	&	AGAL310.014$+$00.387$^\ast$	&	B1	&	$-$43.8	&	12.3	&	9.0	&	B2	&	$-$53.6	&	29.1	&	3.0	&	--	&	--	&	--	&	--	\\
%43	&	AGAL310.014$+$00.387$^\ast$	&	--	&	--	&	--	&	--	&	--	&	--	&	--	&	--	&	--	&	--	&	--	&	--	\\
44	&	AGAL313.576$+$00.324	&	B1	&	$-$47.9	&	9.3	&	10.0	&	B2	&	$-$40.4	&	17.0	&	4.5	&	--	&	--	&	--	&	--	\\
45	&	AGAL316.641$-$00.087	&	N	&	$-$16.9	&	7.2	&	9.9	&	B	&	$-$22.2	&	26.9	&	2.7	&	--	&	--	&	--	&	--	\\
46	&	AGAL317.867$-$00.151	&	B	&	$-$38.4	&	10.3	&	11.9	&	--	&	--	&	--	&	--	&	--	&	--	&	--	&	--	\\
47	&	AGAL318.779$-$00.137	&	B1	&	$-$38.4	&	8.8	&	6.0	&	B2	&	$-$47.4	&	72.7	&	1.0	&	--	&	--	&	--	&	--	\\
48	&	AGAL320.881$-$00.397	&	N	&	$-$45.5	&	6.4	&	11.7	&	--	&	--	&	--	&	--	&	--	&	--	&	--	&	--	\\
49	&	AGAL326.661$+$00.519	&	N1	&	$-$39.7	&	3.8	&	31.6	&	N2	&	$-$38.0	&	7.2	&	13.7	&	--	&	--	&	--	&	--	\\
50	&	AGAL326.987$-$00.032	&	B1	&	$-$57.9	&	9.1	&	7.0	&	B2	&	$-$57.6	&	14.4	&	2.5	&	--	&	--	&	--	&	--	\\
%51	&	AGAL327.119$+$00.509$^\ast$	&	B1	&	$-$84.2	&	10.5	&	8.9	&	B2	&	$-$85.6	&	33.8	&	1.6	&	--	&	--	&	--	&	--	\\
%51	&	AGAL327.119$+$00.509$^\ast$	&	--	&	--	&	--	&	--	&	--	&	--	&	--	&	--	&	--	&	--	&	--	&	--	\\
52	&	AGAL327.393$+$00.199	&	B1	&	$-$89.3	&	9.4	&	10.5	&	B2	&	$-$86.3	&	86.0	&	1.0	&	--	&	--	&	--	&	--	\\
%53	&	AGAL329.029$-$00.206$^\ast$	&	B1	&	$-$46.6	&	12.3	&	15.5	&	B2	&	$-$52.9	&	26.4	&	1.6	&	--	&	--	&	--	&	--	\\
%53	&	AGAL329.029$-$00.206$^\ast$	&	--	&	--	&	--	&	--	&	--	&	--	&	--	&	--	&	--	&	--	&	--	&	--	\\
54	&	AGAL329.066$-$00.307	&	B1	&	$-$41.9	&	11.1	&	7.2	&	B2	&	$-$47.6	&	15.0	&	3.1	&	--	&	--	&	--	&	--	\\
55	&	AGAL330.879$-$00.367	&	B1	&	$-$64.7	&	17.5	&	26.7	&	B2	&	$-$78.6	&	32.2	&	7.0	&	--	&	--	&	--	&	--	\\
%56	&	AGAL330.954$-$00.182$^\ast$	&	B1	&	$-$91.4	&	13.4	&	16.0	&	B2	&	$-$96.8	&	24.3	&	16.0	&	--	&	--	&	--	&	--	\\
%56	&	AGAL330.954$-$00.182$^\ast$	&	--	&	--	&	--	&	--	&	--	&	--	&	--	&	--	&	--	&	--	&	--	&	--	\\
57	&	AGAL331.709$+$00.582	&	B1	&	$-$67.2	&	9.9	&	15.3	&	B2	&	$-$64.7	&	24.2	&	9.2	&	--	&	--	&	--	&	--	\\
58	&	AGAL332.094$-$00.421	&	B1	&	$-$57.6	&	12.2	&	16.9	&	B2	&	$-$56.4	&	29.9	&	3.2	&	--	&	--	&	--	&	--	\\
%59	&	AGAL332.826$-$00.549$^\ast$	&	N	&	$-$56.1	&	6.9	&	77.9	&	B	&	$-$59.0	&	19.3	&	22.0	&	--	&	--	&	--	&	--	\\
%59	&	AGAL332.826$-$00.549$^\ast$	&	--	&	--	&	--	&	--	&	--	&	--	&	--	&	--	&	--	&	--	&	--	&	--	\\
%60	&	AGAL333.134$-$00.431$^\ast$	&	B1	&	$-$47.4	&	12.3	&	18.4	&	B2	&	$-$58.8	&	20.6	&	26.8	&	--	&	--	&	--	&	--	\\
%60	&	AGAL333.134$-$00.431$^\ast$	&	--	&	--	&	--	&	--	&	--	&	--	&	--	&	--	&	--	&	--	&	--	&	--	\\
61	&	AGAL333.284$-$00.387	&	B	&	$-$52.3	&	8.0	&	41.8	&	--	&	--	&	--	&	--	&	--	&	--	&	--	&	--	\\
62	&	AGAL333.314$+$00.106	&	B1	&	$-$45.7	&	11.3	&	16.5	&	B2	&	$-$49.9	&	51.2	&	2.9	&	--	&	--	&	--	&	--	\\
63	&	AGAL333.604$-$00.212	&	N	&	$-$46.8	&	7.2	&	25.2	&	B	&	$-$46.4	&	24.0	&	28.8	&	--	&	--	&	--	&	--	\\
64	&	AGAL333.656$+$00.059	&	N	&	$-$83.9	&	6.5	&	10.7	&	--	&	--	&	--	&	--	&	--	&	--	&	--	&	--	\\
65	&	AGAL335.789$+$00.174	&	B1	&	$-$50.0	&	10.2	&	13.6	&	B2	&	$-$49.7	&	24.2	&	9.4	&	--	&	--	&	--	&	--	\\
66	&	AGAL336.958$-$00.224	&	B	&	$-$71.9	&	8.8	&	7.1	&	--	&	--	&	--	&	--	&	--	&	--	&	--	&	--	\\
67	&	AGAL337.176$-$00.032	&	N	&	$-$70.4	&	4.5	&	5.9	&	B	&	$-$70.5	&	16.9	&	3.1	&	P2	&	$-$79.3	&	2.9	&	4.1	\\
68	&	AGAL337.258$-$00.101	&	N	&	$-$69.1	&	6.2	&	6.1	&	--	&	--	&	--	&	--	&	--	&	--	&	--	&	--	\\
%69	&	AGAL337.286$+$00.007$^\ast$	&	B	&	$-$102.6	&	11.0	&	2.7	&	P2	&	-73.3	&	3.9	&	3.8	&	--	&	--	&	--	&	--	\\
%69	&	AGAL337.286$+$00.007$^\ast$	&	--	&	--	&	--	&	--	&	--	&	--	&	--	&	--	&	--	&	--	&	--	&	--	\\
%70	&	AGAL337.406$-$00.402$^\ast$	&	B1	&	$-$41.3	&	7.6	&	33.7	&	B2	&	$-$35.6	&	30.6	&	4.5	&	--	&	--	&	--	&	--	\\
%70	&	AGAL337.406$-$00.402$^\ast$	&	--	&	--	&	--	&	--	&	--	&	--	&	--	&	--	&	--	&	--	&	--	&	--	\\
%71	&	AGAL337.704$-$00.054$^\ast$	&	N	&	$-$48.8	&	4.6	&	12.9	&	B	&	$-$52.8	&	20.2	&	2.9	&	--	&	--	&	--	&	--	\\
%71	&	AGAL337.704$-$00.054$^\ast$	&	--	&	--	&	--	&	--	&	--	&	--	&	--	&	--	&	--	&	--	&	--	&	--	\\
%72	&	AGAL337.916$-$00.477$^\ast$	&	B1	&	$-$41.0	&	12.7	&	21.7	&	B2	&	$-$47.4	&	38.4	&	12.2	&	B3	&	2.9	&	91.2	&	2.3	\\
%72	&	AGAL337.916$-$00.477$^\ast$	&	--	&	--	&	--	&	--	&	--	&	--	&	--	&	--	&	--	&	--	&	--	&	--	\\
73	&	AGAL338.066$+$00.044	&	N	&	$-$68.9	&	7.3	&	4.2	&	B	&	$-$63.9	&	34.6	&	1.8	&	P2	&	$-$39.0	&	9.7	&	3.8	\\
74	&	AGAL338.786$+$00.476	&	N	&	$-$62.4	&	6.8	&	7.2	&	--	&	--	&	--	&	--	&	--	&	--	&	--	&	--	\\
%75	&	AGAL338.926$+$00.554$^\ast$	&	B1	&	$-$62.3	&	8.6	&	56.7	&	B2	&	$-$62.4	&	34.5	&	4.2	&	--	&	--	&	--	&	--	\\
%75	&	AGAL338.926$+$00.554$^\ast$	&	--	&	--	&	--	&	--	&	--	&	--	&	--	&	--	&	--	&	--	&	--	&	--	\\
76	&	AGAL339.623$-$00.122	&	B1	&	$-$32.3	&	11.2	&	15.6	&	B2	&	$-$29.1	&	30.9	&	3.2	&	--	&	--	&	--	&	--	\\
%77	&	AGAL340.374$-$00.391$^\ast$	&	B1	&	$-$42.7	&	8.0	&	2.2	&	B2	&	$-$43.4	&	37.0	&	1.7	&	--	&	--	&	--	&	--	\\
%77	&	AGAL340.374$-$00.391$^\ast$	&	--	&	--	&	--	&	--	&	--	&	--	&	--	&	--	&	--	&	--	&	--	&	--	\\
78	&	AGAL340.746$-$01.001	&	N	&	$-$29.4	&	7.4	&	15.5	&	B2	&	$-$23.8	&	13.7	&	3.5	&	--	&	--	&	--	&	--	\\
79	&	AGAL340.784$-$00.097	&	B1	&	$-$102.4	&	9.7	&	10.0	&	B2	&	$-$102.6	&	35.5	&	1.0	&	--	&	--	&	--	&	--	\\
%80	&	AGAL341.217$-$00.212$^\ast$	&	N	&	$-$42.0	&	7.1	&	22.1	&	B	&	$-$48.5	&	32.5	&	3.8	&	--	&	--	&	--	&	--	\\
%80	&	AGAL341.217$-$00.212$^\ast$	&	--	&	--	&	--	&	--	&	--	&	--	&	--	&	--	&	--	&	--	&	--	&	--	\\
81	&	AGAL342.484$+$00.182	&	B1	&	$-$42.4	&	8.2	&	7.5	&	B2	&	$-$47.5	&	15.3	&	4.0	&	--	&	--	&	--	&	--	\\
82	&	AGAL343.128$-$00.062	&	B1	&	$-$29.2	&	15.7	&	25.1	&	B2	&	$-$29.3	&	35.4	&	10.5	&	--	&	--	&	--	&	--	\\
%83	&	AGAL343.756$-$00.164$^\ast$	&	B	&	$-$27.3	&	10.5	&	13.4	&	--	&	--	&	--	&	--	&	--	&	--	&	--	&	--	\\
%83	&	AGAL343.756$-$00.164$^\ast$	&	--	&	--	&	--	&	--	&	--	&	--	&	--	&	--	&	--	&	--	&	--	&	--	\\
%84	&	AGAL344.227$-$00.569$^\ast$	&	B1	&	$-$23.8	&	16.4	&	10.0	&	B2	&	$-$23.9	&	87.3	&	2.6	&	--	&	--	&	--	&	--	\\
%84	&	AGAL344.227$-$00.569$^\ast$	&	--	&	--	&	--	&	--	&	--	&	--	&	--	&	--	&	--	&	--	&	--	&	--	\\
%85	&	AGAL345.003$-$00.224$^\ast$	&	B1	&	$-$27.4	&	10.1	&	26.6	&	B2	&	$-$26.9	&	31.0	&	6.6	&	--	&	--	&	--	&	--	\\
%85	&	AGAL345.003$-$00.224$^\ast$	&	--	&	--	&	--	&	--	&	--	&	--	&	--	&	--	&	--	&	--	&	--	&	--	\\
%86	&	AGAL345.488$+$00.314$^\ast$	&	B1	&	$-$18.7	&	9.6	&	28.8	&	B2	&	$-$15.1	&	11.3	&	7.0	&	--	&	--	&	--	&	--	\\
%86	&	AGAL345.488$+$00.314$^\ast$	&	--	&	--	&	--	&	--	&	--	&	--	&	--	&	--	&	--	&	--	&	--	&	--	\\
87	&	AGAL345.504$+$00.347	&	N	&	$-$16.9	&	6.7	&	28.6	&	B	&	$-$16.6	&	20.7	&	15.1	&	--	&	--	&	--	&	--	\\
88	&	AGAL345.718$+$00.817	&	B	&	$-$13.0	&	9.0	&	11.4	&	--	&	--	&	--	&	--	&	--	&	--	&	--	&	--	\\
89	&	AGAL351.131$+$00.771	&	N	&	$-$5.6	&	5.0	&	15.3	&	--	&	--	&	--	&	--	&	--	&	--	&	--	&	--	\\
%90	&	AGAL351.161$+$00.697$^\ast$	&	B1	&	$-$4.7	&	10.9	&	28.5	&	B2	&	$-$11.4	&	19.9	&	13.3	&	--	&	--	&	--	&	--	\\
%90	&	AGAL351.161$+$00.697$^\ast$	&	--	&	--	&	--	&	--	&	--	&	--	&	--	&	--	&	--	&	--	&	--	&	--	\\
91	&	AGAL351.244$+$00.669	&	B1	&	$-$3.3	&	9.3	&	47.8	&	B2	&	3.6	&	26.0	&	5.5	&	--	&	--	&	--	&	--	\\
%92	&	AGAL351.416$+$00.646$^\ast$	&	B1	&	$-$6.7	&	9.4	&	46.8	&	B2	&	$-$18.9	&	28.6	&	14.8	&	--	&	--	&	--	&	--	\\
%92	&	AGAL351.416$+$00.646$^\ast$	&	--	&	--	&	--	&	--	&	--	&	--	&	--	&	--	&	--	&	--	&	--	&	--	\\
%93	&	AGAL351.444$+$00.659$^\ast$	&	B1	&	$-$4.2	&	11.6	&	27.0	&	B2	&	$-$9.7	&	32.4	&	5.4	&	--	&	--	&	--	&	--	\\
%93	&	AGAL351.444$+$00.659$^\ast$	&	--	&	--	&	--	&	--	&	--	&	--	&	--	&	--	&	--	&	--	&	--	&	--	\\
94	&	AGAL351.571$+$00.762	&	N	&	$-$3.6	&	6.0	&	8.1	&	--	&	--	&	--	&	--	&	--	&	--	&	--	&	--	\\
%95	&	AGAL351.581$-$00.352$^\ast$	&	B1	&	$-$95.6	&	13.5	&	9.0	&	B2	&	$-$108.6	&	54.1	&	1.6	&	--	&	--	&	--	&	--	\\
%95	&	AGAL351.581$-$00.352$^\ast$	&	--	&	--	&	--	&	--	&	--	&	--	&	--	&	--	&	--	&	--	&	--	&	--	\\
96	&	AGAL351.774$-$00.537	&	B1	&	$-$1.7	&	17.4	&	28.4	&	B2	&	$-$10.7	&	53.3	&	7.2	&	--	&	--	&	--	&	--	\\
97	&	AGAL353.066$+$00.452	&	N	&	$-$1.6	&	5.2	&	5.4	&	--	&	--	&	--	&	--	&	--	&	--	&	--	&	--	\\
%98	&	AGAL353.417$-$00.079$^\ast$	&	B	&	$-$49.7	&	10.0	&	3.3	&	P2	&	$-$37.3	&	3.4	&	2.4	&	--	&	--	&	--	&	--	\\
%98	&	AGAL353.417$-$00.079$^\ast$	&	--	&	--	&	--	&	--	&	--	&	--	&	--	&	--	&	--	&	--	&	--	&	--	\\
99	&	AGAL354.944$-$00.537	&	N	&	$-$5.9	&	5.4	&	5.9	&	--	&	--	&	--	&	--	&	--	&	--	&	--	&	--	\\
\end{longtable}
\tablefoot{The columns are as follows: (1)-(2) ID and CSC name of the source (given in Table\,\ref{tbl_observations_short}); %a $\ast$ symbol indicates the cases where the Gaussian fit is not reliable, as discussed in Sect.\,\ref{sec_gaussfit};
(3), (7), (11) the classification of each fitted Gaussian component (C$_1$, C$_2$ and C$_3$), into narrow (N), broad (B) or secondary peaks (P2) as discussed in the main text; (4)-(6), (8)-(10), (12)-(14) the central velocity (\vpeak), full width at half maximum (FWHM) and peak temperature (\tpeak) is presented for each component.}
\end{longtab}
\setlength{\tabcolsep}{6pt} % A.2
	
\begin{longtab}
\setcounter{table}{2}
\setlength{\tabcolsep}{4pt}
\captionsetup{width=0.9\textwidth}
\begin{longtable}{cl|lrrr|lrrr|lrrr}
    \caption{\label{tbl_gauss_fitting_co65_fixbeam}Gaussian decomposition of the CO\,(6--5) profiles towards the \atgtop sample.}\\
\hline\hline
ID	&	\multicolumn{1}{c|}{CSC Name}	&	C$_1$	&	\vpeak	&	FWHM	&	\tpeak	&	C$_2$	&	\vpeak	&	FWHM	&	\tpeak	&	C$_3$	&	\vpeak	&	FWHM	&	\tpeak	 \\ 
	&	&	&	(km\,s$^{-1}$)	&	(km\,s$^{-1}$)	&	(K)	&		&	(km\,s$^{-1}$)	&	(km\,s$^{-1}$)	&	(K)	&		&	(km\,s$^{-1}$)	&	(km\,s$^{-1}$)	&	(K)	 \\ 
\hline
\endfirsthead
\caption{continued.} \\
\hline\hline
ID	&	\multicolumn{1}{c|}{CSC Name}	&	C$_1$	&	\vpeak	&	FWHM	&	\tpeak	&	C$_2$	&	\vpeak	&	FWHM	&	\tpeak	&	C$_3$	&	\vpeak	&	FWHM	&	\tpeak	 \\ 
	&	&	&	(km\,s$^{-1}$)	&	(km\,s$^{-1}$)	&	(K)	&		&	(km\,s$^{-1}$)	&	(km\,s$^{-1}$)	&	(K)	&		&	(km\,s$^{-1}$)	&	(km\,s$^{-1}$)	&	(K)	 \\ 
\hline
\endhead
\hline
\endfoot
%1	&	AGAL008.684$-$00.367$^\ast$	&	N	&	38.0	&	7.2	&	24.9	&	B	&	42.4	&	17.8	&	3.3	&	--	&	--	&	--	&	--	\\
%1	&	AGAL008.684$-$00.367$^\ast$	&	--	&	--	&	--	&	--	&	--	&	--	&	--	&	--	&	--	&	--	&	--	&	--	\\
2	&	AGAL008.706$-$00.414	&	N	&	38.3	&	6.3	&	3.4	&	--	&	--	&	--	&	--	&	--	&	--	&	--	&	--	\\
3	&	AGAL010.444$-$00.017	&	N	&	74.3	&	3.3	&	3.1	&	B	&	74.7	&	12.0	&	2.6	&	P2	&	65.4	&	2.0	&	1.7	\\
4	&	AGAL010.472$+$00.027	&	N	&	66.0	&	7.1	&	72.7	&	B	&	69.0	&	14.9	&	13.3	&	--	&	--	&	--	&	--	\\
5	&	AGAL010.624$-$00.384	&	B1	&	$-$2.4	&	11.4	&	108.1	&	B2	&	$-$2.5	&	24.1	&	7.8	&	--	&	--	&	--	&	--	\\
6	&	AGAL012.804$-$00.199	&	B1	&	35.1	&	10.3	&	105.9	&	B2	&	39.8	&	18.5	&	28.4	&	--	&	--	&	--	&	--	\\
7	&	AGAL013.178$+$00.059	&	B1	&	49.0	&	9.7	&	17.8	&	B2	&	59.6	&	26.0	&	0.8	&	--	&	--	&	--	&	--	\\
8	&	AGAL013.658$-$00.599	&	N	&	48.2	&	5.5	&	12.2	&	B1	&	46.6	&	15.5	&	4.4	&	B2	&	42.1	&	62.3	&	1.4	\\
%9	&	AGAL014.114$-$00.574$^\ast$	&	N	&	19.8	&	6.3	&	22.2	&	B	&	19.6	&	20.0	&	3.0	&	--	&	--	&	--	&	--	\\
%9	&	AGAL014.114$-$00.574$^\ast$	&	--	&	--	&	--	&	--	&	--	&	--	&	--	&	--	&	--	&	--	&	--	&	--	\\
10	&	AGAL014.194$-$00.194	&	N	&	39.8	&	5.3	&	13.6	&	B1	&	40.5	&	14.4	&	8.0	&	B2	&	35.8	&	29.1	&	1.8	\\
11	&	AGAL014.492$-$00.139	&	N	&	40.1	&	5.6	&	8.7	&	B	&	40.1	&	23.5	&	1.5	&	--	&	--	&	--	&	--	\\
12	&	AGAL014.632$-$00.577	&	B1	&	18.2	&	7.9	&	26.4	&	B2	&	18.4	&	32.6	&	1.3	&	--	&	--	&	--	&	--	\\
13	&	AGAL015.029$-$00.669	&	N	&	20.9	&	6.9	&	75.3	&	B	&	18.6	&	10.9	&	60.3	&	--	&	--	&	--	&	--	\\
14	&	AGAL018.606$-$00.074	&	N	&	45.8	&	4.4	&	15.0	&	B	&	45.6	&	10.4	&	4.8	&	--	&	--	&	--	&	--	\\
15	&	AGAL018.734$-$00.226	&	N	&	40.1	&	6.4	&	12.9	&	B	&	40.1	&	26.6	&	3.2	&	--	&	--	&	--	&	--	\\
16	&	AGAL018.888$-$00.474	&	B1	&	65.6	&	9.1	&	19.9	&	B2	&	69.4	&	20.7	&	3.1	&	--	&	--	&	--	&	--	\\
17	&	AGAL019.882$-$00.534	&	N	&	45.3	&	6.5	&	18.5	&	B1	&	45.5	&	17.8	&	15.5	&	B3	&	64.1	&	70.9	&	1.8	\\
18	&	AGAL022.376$+$00.447	&	N	&	52.5	&	3.1	&	11.6	&	B	&	53.8	&	10.4	&	4.1	&	--	&	--	&	--	&	--	\\
19	&	AGAL023.206$-$00.377	&	B1	&	78.3	&	12.7	&	17.6	&	B2	&	77.8	&	37.3	&	3.5	&	--	&	--	&	--	&	--	\\
20	&	AGAL024.629$+$00.172	&	B1	&	114.8	&	7.8	&	4.3	&	B2	&	120.5	&	40.1	&	0.5	&	--	&	--	&	--	&	--	\\
21	&	AGAL028.564$-$00.236	&	N	&	86.1	&	5.6	&	4.6	&	B	&	87.6	&	16.1	&	1.6	&	--	&	--	&	--	&	--	\\
22	&	AGAL028.861$+$00.066	&	B1	&	105.0	&	10.5	&	24.8	&	B2	&	102.0	&	32.0	&	6.4	&	--	&	--	&	--	&	--	\\
23	&	AGAL030.818$-$00.056	&	N	&	97.5	&	6.6	&	35.9	&	B1	&	98.2	&	14.2	&	13.7	&	B2	&	100.3	&	52.5	&	2.4	\\
24	&	AGAL030.848$-$00.081	&	N	&	95.5	&	5.5	&	5.1	&	B	&	96.5	&	12.5	&	7.5	&	--	&	--	&	--	&	--	\\
25	&	AGAL030.893$+$00.139	&	N	&	106.8	&	5.8	&	7.0	&	P2	&	97.0	&	5.2	&	4.5	&	--	&	--	&	--	&	--	\\
26	&	AGAL031.412$+$00.307	&	N	&	97.5	&	7.3	&	42.8	&	B	&	100.2	&	29.3	&	4.2	&	--	&	--	&	--	&	--	\\
%27	&	AGAL034.258$+$00.154$^\ast$	&	N	&	58.4	&	7.1	&	110.1	&	B1	&	57.9	&	15.1	&	23.2	&	B2	&	67.5	&	42.9	&	5.5	\\
%27	&	AGAL034.258$+$00.154$^\ast$	&	--	&	--	&	--	&	--	&	--	&	--	&	--	&	--	&	--	&	--	&	--	&	--	\\
28	&	AGAL034.401$+$00.226	&	N	&	58.5	&	7.0	&	17.0	&	B1	&	58.0	&	12.1	&	26.1	&	B2	&	59.4	&	42.8	&	1.6	\\
29	&	AGAL034.411$+$00.234	&	B1	&	57.9	&	10.1	&	19.6	&	B2	&	57.4	&	41.3	&	2.0	&	--	&	--	&	--	&	--	\\
30	&	AGAL034.821$+$00.351	&	N	&	57.9	&	5.9	&	11.3	&	B	&	56.4	&	18.9	&	7.4	&	--	&	--	&	--	&	--	\\
31	&	AGAL035.197$-$00.742	&	B1	&	34.3	&	9.5	&	46.2	&	B2	&	32.5	&	20.0	&	22.9	&	--	&	--	&	--	&	--	\\
32	&	AGAL037.554$+$00.201	&	N	&	85.7	&	6.4	&	12.2	&	B	&	84.3	&	15.6	&	7.1	&	--	&	--	&	--	&	--	\\
33	&	AGAL043.166$+$00.011	&	B1	&	$-$0.5	&	12.9	&	56.0	&	B2	&	15.2	&	13.5	&	32.3	&	B3	&	2.1	&	42.3	&	13.1	\\
%34	&	AGAL049.489$-$00.389$^\ast$	&	N	&	50.8	&	6.4	&	34.1	&	B	&	58.6	&	29.3	&	10.9	&	--	&	--	&	--	&	--	\\
%34	&	AGAL049.489$-$00.389$^\ast$	&	--	&	--	&	--	&	--	&	--	&	--	&	--	&	--	&	--	&	--	&	--	&	--	\\
35	&	AGAL053.141$+$00.069	&	N	&	22.3	&	7.2	&	37.2	&	B	&	21.0	&	17.0	&	10.8	&	--	&	--	&	--	&	--	\\
36	&	AGAL059.782$+$00.066	&	B1	&	22.8	&	8.4	&	37.2	&	B2	&	22.0	&	24.6	&	5.3	&	--	&	--	&	--	&	--	\\
37	&	AGAL301.136$-$00.226	&	B1	&	$-$38.3	&	9.9	&	34.4	&	B2	&	$-$37.3	&	31.1	&	14.0	&	B3	&	$-$28.8	&	65.4	&	10.4	\\
38	&	AGAL305.192$-$00.006	&	N	&	$-$33.7	&	7.5	&	22.4	&	B	&	$-$34.4	&	19.9	&	4.6	&	--	&	--	&	--	&	--	\\
39	&	AGAL305.209$+$00.206	&	B1	&	$-$42.4	&	10.9	&	43.7	&	B2	&	$-$40.1	&	33.2	&	10.1	&	--	&	--	&	--	&	--	\\
40	&	AGAL305.562$+$00.014	&	N	&	$-$38.9	&	5.7	&	38.7	&	B	&	$-$40.0	&	14.9	&	21.2	&	--	&	--	&	--	&	--	\\
41	&	AGAL305.794$-$00.096	&	B	&	$-$41.7	&	9.2	&	4.6	&	--	&	--	&	--	&	--	&	--	&	--	&	--	&	--	\\
42	&	AGAL309.384$-$00.134	&	B1	&	$-$51.0	&	8.6	&	18.3	&	B2	&	$-$51.0	&	22.3	&	3.0	&	--	&	--	&	--	&	--	\\
43	&	AGAL310.014$+$00.387	&	B1	&	$-$42.6	&	10.1	&	13.9	&	B2	&	$-$46.8	&	35.0	&	3.0	&	--	&	--	&	--	&	--	\\
44	&	AGAL313.576$+$00.324	&	N	&	$-$46.7	&	7.4	&	19.4	&	B	&	$-$43.5	&	27.7	&	4.2	&	--	&	--	&	--	&	--	\\
45	&	AGAL316.641$-$00.087	&	B1	&	$-$16.7	&	9.1	&	5.4	&	B2	&	$-$21.5	&	28.3	&	3.1	&	--	&	--	&	--	&	--	\\
46	&	AGAL317.867$-$00.151	&	B	&	$-$39.9	&	11.2	&	15.7	&	--	&	--	&	--	&	--	&	--	&	--	&	--	&	--	\\
47	&	AGAL318.779$-$00.137	&	N	&	$-$40.0	&	5.4	&	7.7	&	B	&	$-$38.7	&	20.7	&	2.2	&	--	&	--	&	--	&	--	\\
48	&	AGAL320.881$-$00.397	&	N	&	$-$45.7	&	4.6	&	8.9	&	B	&	$-$44.6	&	10.5	&	2.9	&	--	&	--	&	--	&	--	\\
49	&	AGAL326.661$+$00.519	&	N	&	$-$39.4	&	3.3	&	57.6	&	B	&	$-$38.4	&	7.6	&	18.1	&	--	&	--	&	--	&	--	\\
50	&	AGAL326.987$-$00.032	&	N	&	$-$58.2	&	5.9	&	9.9	&	B1	&	$-$55.2	&	16.3	&	3.6	&	B2	&	$-$54.1	&	56.4	&	0.6	\\
51	&	AGAL327.119$+$00.509	&	N	&	$-$83.9	&	6.3	&	18.1	&	B	&	$-$83.4	&	20.5	&	2.9	&	--	&	--	&	--	&	--	\\
52	&	AGAL327.393$+$00.199	&	B1	&	$-$88.9	&	8.2	&	14.8	&	B2	&	$-$90.7	&	27.7	&	1.5	&	--	&	--	&	--	&	--	\\
53	&	AGAL329.029$-$00.206	&	B1	&	$-$44.2	&	10.5	&	20.9	&	B2	&	$-$49.7	&	17.0	&	6.5	&	B3	&	$-$64.1	&	62.7	&	0.6	\\
54	&	AGAL329.066$-$00.307	&	B1	&	$-$42.2	&	8.0	&	12.6	&	B2	&	$-$43.7	&	17.7	&	3.3	&	--	&	--	&	--	&	--	\\
55	&	AGAL330.879$-$00.367	&	B1	&	$-$60.3	&	10.0	&	42.7	&	B2	&	$-$68.8	&	13.0	&	23.3	&	B3	&	$-$73.2	&	38.6	&	11.3	\\
56	&	AGAL330.954$-$00.182	&	B1	&	$-$92.5	&	9.8	&	50.3	&	B2	&	$-$93.6	&	26.7	&	29.1	&	--	&	--	&	--	&	--	\\
57	&	AGAL331.709$+$00.582	&	N	&	$-$66.1	&	5.7	&	8.6	&	B1	&	$-$66.3	&	13.4	&	14.9	&	B3	&	$-$63.9	&	36.3	&	2.9	\\
58	&	AGAL332.094$-$00.421	&	B1	&	$-$57.6	&	12.4	&	12.5	&	B2	&	$-$54.5	&	23.7	&	9.2	&	--	&	--	&	--	&	--	\\
%59	&	AGAL332.826$-$00.549$^\ast$	&	B1	&	$-$58.0	&	8.8	&	170.7	&	B2	&	$-$58.9	&	19.1	&	25.7	&	B3	&	$-$60.7	&	51.3	&	3.0	\\
%59	&	AGAL332.826$-$00.549$^\ast$	&	--	&	--	&	--	&	--	&	--	&	--	&	--	&	--	&	--	&	--	&	--	&	--	\\
%60	&	AGAL333.134$-$00.431$^\ast$	&	B1	&	$-$50.0	&	9.6	&	34.7	&	B2	&	$-$55.9	&	22.3	&	47.5	&	--	&	--	&	--	&	--	\\
%60	&	AGAL333.134$-$00.431$^\ast$	&	--	&	--	&	--	&	--	&	--	&	--	&	--	&	--	&	--	&	--	&	--	&	--	\\
61	&	AGAL333.284$-$00.387	&	B	&	$-$51.1	&	8.0	&	57.9	&	--	&	--	&	--	&	--	&	--	&	--	&	--	&	--	\\
62	&	AGAL333.314$+$00.106	&	N	&	$-$45.1	&	7.1	&	11.1	&	B1	&	$-$47.1	&	15.2	&	12.2	&	B2	&	$-$45.9	&	64.9	&	1.2	\\
63	&	AGAL333.604$-$00.212	&	B1	&	$-$44.1	&	11.9	&	85.0	&	B2	&	$-$48.1	&	25.3	&	38.4	&	--	&	--	&	--	&	--	\\
64	&	AGAL333.656$+$00.059	&	N	&	$-$84.6	&	6.4	&	9.0	&	B	&	$-$84.5	&	23.4	&	0.8	&	--	&	--	&	--	&	--	\\
65	&	AGAL335.789$+$00.174	&	N	&	$-$50.4	&	6.1	&	20.1	&	B1	&	$-$49.9	&	19.6	&	14.9	&	B2	&	$-$55.9	&	68.3	&	0.8	\\
66	&	AGAL336.958$-$00.224	&	N	&	$-$72.4	&	4.8	&	6.7	&	B	&	$-$73.1	&	29.7	&	2.3	&	--	&	--	&	--	&	--	\\
67	&	AGAL337.176$-$00.032	&	N	&	$-$69.8	&	3.4	&	12.2	&	B	&	$-$72.0	&	13.8	&	4.5	&	P2	&	$-$79.4	&	3.1	&	4.5	\\
68	&	AGAL337.258$-$00.101	&	B	&	$-$67.8	&	9.6	&	9.2	&	--	&	--	&	--	&	--	&	--	&	--	&	--	&	--	\\
69	&	AGAL337.286$+$00.007	&	N	&	$-$106.0	&	7.4	&	3.6	&	P2	&	$-$73.9	&	3.8	&	1.5	&	--	&	--	&	--	&	--	\\
70	&	AGAL337.406$-$00.402	&	N	&	$-$40.7	&	6.8	&	69.8	&	B1	&	$-$40.3	&	23.5	&	8.7	&	B2	&	$-$28.2	&	83.2	&	3.1	\\
%71	&	AGAL337.704$-$00.054$^\ast$	&	N	&	$-$50.8	&	3.9	&	15.0	&	B1	&	$-$45.5	&	13.0	&	15.2	&	B2	&	$-$56.2	&	24.0	&	2.5	\\
%71	&	AGAL337.704$-$00.054$^\ast$	&	--	&	--	&	--	&	--	&	--	&	--	&	--	&	--	&	--	&	--	&	--	&	--	\\
72	&	AGAL337.916$-$00.477	&	B1	&	$-$40.1	&	10.7	&	38.1	&	B1	&	$-$43.7	&	28.1	&	18.5	&	B2	&	$-$52.0	&	56.9	&	6.5	\\
73	&	AGAL338.066$+$00.044	&	N	&	$-$69.3	&	4.9	&	4.5	&	B	&	$-$68.7	&	21.7	&	2.3	&	P2	&	$-$39.9	&	8.2	&	3.7	\\
74	&	AGAL338.786$+$00.476	&	N	&	$-$63.8	&	7.1	&	6.0	&	--	&	--	&	--	&	--	&	--	&	--	&	--	&	--	\\
75	&	AGAL338.926$+$00.554	&	N	&	$-$62.5	&	6.4	&	131.4	&	B1	&	$-$60.4	&	16.3	&	11.3	&	B2	&	$-$62.5	&	49.1	&	2.5	\\
76	&	AGAL339.623$-$00.122	&	N	&	$-$32.9	&	6.2	&	14.1	&	B1	&	$-$31.9	&	18.0	&	12.7	&	B2	&	$-$25.6	&	58.6	&	1.8	\\
77	&	AGAL340.374$-$00.391	&	B1	&	$-$44.5	&	10.7	&	5.3	&	B	&	$-$38.7	&	50.8	&	1.3	&	--	&	--	&	--	&	--	\\
78	&	AGAL340.746$-$01.001	&	N	&	$-$29.3	&	6.9	&	18.3	&	B1	&	$-$23.8	&	13.4	&	2.7	&	B2	&	$-$15.3	&	49.0	&	1.0	\\
79	&	AGAL340.784$-$00.097	&	N	&	$-$101.4	&	6.9	&	8.1	&	B	&	$-$101.0	&	18.3	&	2.6	&	--	&	--	&	--	&	--	\\
80	&	AGAL341.217$-$00.212	&	N	&	$-$43.5	&	4.1	&	26.6	&	B1	&	$-$41.5	&	12.3	&	7.5	&	B2	&	$-$44.9	&	37.9	&	3.6	\\
81	&	AGAL342.484$+$00.182	&	N	&	$-$42.2	&	5.9	&	8.4	&	B	&	$-$45.1	&	17.2	&	6.4	&	--	&	--	&	--	&	--	\\
82	&	AGAL343.128$-$00.062	&	B1	&	$-$29.2	&	16.5	&	46.2	&	B2	&	$-$26.2	&	39.8	&	14.1	&	--	&	--	&	--	&	--	\\
83	&	AGAL343.756$-$00.164	&	B1	&	$-$27.2	&	8.4	&	19.1	&	B2	&	$-$25.5	&	23.2	&	1.2	&	--	&	--	&	--	&	--	\\
84	&	AGAL344.227$-$00.569	&	B1	&	$-$22.3	&	10.1	&	17.8	&	B2	&	$-$27.8	&	29.3	&	6.1	&	--	&	--	&	--	&	--	\\
85	&	AGAL345.003$-$00.224	&	B1	&	$-$26.9	&	7.5	&	51.2	&	B2	&	$-$27.3	&	20.6	&	17.8	&	B3	&	$-$13.4	&	97.1	&	2.6	\\
%86	&	AGAL345.488$+$00.314$^\ast$	&	N	&	$-$18.7	&	7.1	&	87.0	&	B	&	$-$15.6	&	13.5	&	15.3	&	--	&	--	&	--	&	--	\\
%86	&	AGAL345.488$+$00.314$^\ast$	&	--	&	--	&	--	&	--	&	--	&	--	&	--	&	--	&	--	&	--	&	--	&	--	\\
87	&	AGAL345.504$+$00.347	&	B1	&	$-$16.8	&	8.2	&	31.3	&	B2	&	$-$16.0	&	19.4	&	19.7	&	B3	&	$-$1.6	&	43.3	&	1.5	\\
88	&	AGAL345.718$+$00.817	&	B	&	$-$12.2	&	7.8	&	14.4	&	--	&	--	&	--	&	--	&	--	&	--	&	--	&	--	\\
89	&	AGAL351.131$+$00.771	&	N	&	$-$5.1	&	4.7	&	17.5	&	--	&	--	&	--	&	--	&	--	&	--	&	--	&	--	\\
90	&	AGAL351.161$+$00.697	&	B1	&	$-$5.3	&	9.2	&	60.2	&	B2	&	$-$8.5	&	22.2	&	26.6	&	--	&	--	&	--	&	--	\\
91	&	AGAL351.244$+$00.669	&	B1	&	$-$2.9	&	8.6	&	96.6	&	B2	&	0.6	&	25.4	&	13.4	&	B3	&	8.7	&	72.3	&	1.4	\\
92	&	AGAL351.416$+$00.646	&	B1	&	$-$6.5	&	9.5	&	70.5	&	B2	&	$-$12.4	&	34.2	&	18.8	&	B3	&	$-$25.1	&	96.5	&	3.2	\\
93	&	AGAL351.444$+$00.659	&	B1	&	$-$4.1	&	8.9	&	37.7	&	B2	&	$-$4.1	&	15.0	&	19.9	&	B3	&	$-$8.3	&	40.0	&	8.4	\\
94	&	AGAL351.571$+$00.762	&	N	&	$-$4.1	&	3.5	&	8.2	&	--	&	--	&	--	&	--	&	--	&	--	&	--	&	--	\\
95	&	AGAL351.581$-$00.352	&	B1	&	$-$95.4	&	10.7	&	23.9	&	B2	&	$-$97.2	&	29.2	&	10.2	&	B3	&	$-$131.9	&	70.0	&	0.6	\\
96	&	AGAL351.774$-$00.537	&	B1	&	$-$1.3	&	10.6	&	36.0	&	B2	&	$-$0.8	&	26.3	&	22.9	&	B3	&	$-$10.2	&	61.3	&	14.9	\\
97	&	AGAL353.066$+$00.452	&	B	&	$-$1.0	&	7.6	&	16.5	&	--	&	--	&	--	&	--	&	--	&	--	&	--	&	--	\\
98	&	AGAL353.417$-$00.079	&	N	&	$-$55.6	&	2.7	&	1.9	&	B	&	$-$53.8	&	9.0	&	1.8	&	--	&	--	&	--	&	--	\\
99	&	AGAL354.944$-$00.537	&	N	&	$-$5.7	&	2.6	&	13.1	&	B	&	$-$5.2	&	8.8	&	2.0	&	--	&	--	&	--	&	--	\\
\end{longtable}
\tablefoot{The columns are as follows: (1)-(2) ID and CSC name of the source (given in Table\,\ref{tbl_observations_short}); %a $\ast$ symbol indicates the cases where the Gaussian fit is not reliable, as discussed in Sect.\,\ref{sec_gaussfit}; 
(3), (7), (11) the classification of each fitted Gaussian component (C$_1$, C$_2$ and C$_3$), into narrow (N), broad (B) or secondary peaks (P2) as discussed in the main text; (4)-(6), (8)-(10), (12)-(14) the central velocity (\vpeak), full width at half maximum (FWHM) and peak temperature (\tpeak) is presented for each component.}
\end{longtab}

\setlength{\tabcolsep}{6pt} % A.3
	
\begin{longtab}
\setcounter{table}{3}
\setlength{\tabcolsep}{4pt}
\captionsetup{width=0.9\textwidth}
\begin{longtable}{cl|lrrr|lrrr|lrrr}
    \caption{\label{tbl_gauss_fitting_co76_fixbeam}Gaussian decomposition of the CO\,(7--6) profiles towards the \atgtop sample.}\\
\hline\hline
ID	&	\multicolumn{1}{c|}{CSC Name}	&	C$_1$	&	\vpeak	&	FWHM	&	\tpeak	&	C$_2$	&	\vpeak	&	FWHM	&	\tpeak	&	C$_3$	&	\vpeak	&	FWHM	&	\tpeak	 \\ 
	&	&	&	(km\,s$^{-1}$)	&	(km\,s$^{-1}$)	&	(K)	&		&	(km\,s$^{-1}$)	&	(km\,s$^{-1}$)	&	(K)	&		&	(km\,s$^{-1}$)	&	(km\,s$^{-1}$)	&	(K)	 \\ 
\hline
\endfirsthead
\caption{continued.} \\
\hline\hline
ID	&	\multicolumn{1}{c|}{CSC Name}	&	C$_1$	&	\vpeak	&	FWHM	&	\tpeak	&	C$_2$	&	\vpeak	&	FWHM	&	\tpeak	&	C$_3$	&	\vpeak	&	FWHM	&	\tpeak	 \\ 
	&	&	&	(km\,s$^{-1}$)	&	(km\,s$^{-1}$)	&	(K)	&		&	(km\,s$^{-1}$)	&	(km\,s$^{-1}$)	&	(K)	&		&	(km\,s$^{-1}$)	&	(km\,s$^{-1}$)	&	(K)	 \\ 
\hline
\endhead
\hline
\endfoot
1	&	AGAL008.684$-$00.367	&	B1	&	38.3	&	9.4	&	15.1	&	B2	&	50.2	&	27.1	&	1.8	&	--	&	--	&	--	&	--	\\
2	&	AGAL008.706$-$00.414	&	N	&	39.8	&	6.8	&	2.4	&	--	&	--	&	--	&	--	&	--	&	--	&	--	&	--	\\
3	&	AGAL010.444$-$00.017	&	N	&	75.0	&	7.0	&	3.8	&	P2	&	66.6	&	3.7	&	1.8	&	--	&	--	&	--	&	--	\\
4	&	AGAL010.472$+$00.027	&	N	&	66.1	&	7.2	&	63.4	&	B	&	70.6	&	20.1	&	8.2	&	--	&	--	&	--	&	--	\\
5	&	AGAL010.624$-$00.384	&	B1	&	$-$2.4	&	9.7	&	88.4	&	B2	&	$-$1.6	&	17.4	&	30.2	&	--	&	--	&	--	&	--	\\
6	&	AGAL012.804$-$00.199	&	B1	&	35.0	&	10.0	&	107.1	&	B2	&	39.8	&	17.4	&	28.4	&	--	&	--	&	--	&	--	\\
7	&	AGAL013.178$+$00.059	&	B	&	49.2	&	11.0	&	16.0	&	--	&	--	&	--	&	--	&	--	&	--	&	--	&	--	\\
8	&	AGAL013.658$-$00.599	&	N	&	48.3	&	6.5	&	10.0	&	B	&	43.6	&	69.9	&	2.9	&	--	&	--	&	--	&	--	\\
9	&	AGAL014.114$-$00.574	&	N	&	19.4	&	4.2	&	20.4	&	B	&	19.9	&	14.3	&	6.3	&	--	&	--	&	--	&	--	\\
10	&	AGAL014.194$-$00.194	&	B1	&	40.1	&	7.7	&	11.1	&	B2	&	41.3	&	29.2	&	3.6	&	--	&	--	&	--	&	--	\\
11	&	AGAL014.492$-$00.139	&	B	&	39.8	&	10.3	&	5.5	&	--	&	--	&	--	&	--	&	--	&	--	&	--	&	--	\\
12	&	AGAL014.632$-$00.577	&	N	&	17.9	&	7.1	&	26.3	&	B	&	21.3	&	20.2	&	1.8	&	--	&	--	&	--	&	--	\\
13	&	AGAL015.029$-$00.669	&	B1	&	20.7	&	7.7	&	104.6	&	B2	&	18.8	&	13.7	&	25.0	&	--	&	--	&	--	&	--	\\
14	&	AGAL018.606$-$00.074	&	B	&	45.2	&	7.5	&	10.2	&	--	&	--	&	--	&	--	&	--	&	--	&	--	&	--	\\
15	&	AGAL018.734$-$00.226	&	N	&	40.4	&	6.7	&	10.1	&	B	&	44.9	&	46.0	&	2.6	&	--	&	--	&	--	&	--	\\
16	&	AGAL018.888$-$00.474	&	N	&	65.8	&	5.7	&	13.2	&	B	&	66.0	&	18.0	&	6.2	&	--	&	--	&	--	&	--	\\
17	&	AGAL019.882$-$00.534	&	N	&	45.1	&	6.6	&	13.1	&	B1	&	45.0	&	17.2	&	14.9	&	B2	&	65.7	&	96.9	&	2.3	\\
18	&	AGAL022.376$+$00.447	&	N	&	53.3	&	6.3	&	7.6	&	--	&	--	&	--	&	--	&	--	&	--	&	--	&	--	\\
19	&	AGAL023.206$-$00.377	&	B1	&	78.0	&	11.6	&	18.4	&	B2	&	78.2	&	50.0	&	3.6	&	--	&	--	&	--	&	--	\\
20	&	AGAL024.629$+$00.172	&	B	&	115.8	&	7.8	&	3.7	&	--	&	--	&	--	&	--	&	--	&	--	&	--	&	--	\\
21	&	AGAL028.564$-$00.236	&	N	&	86.4	&	4.0	&	3.3	&	--	&	--	&	--	&	--	&	--	&	--	&	--	&	--	\\
22	&	AGAL028.861$+$00.066	&	B1	&	104.6	&	12.2	&	21.5	&	B2	&	80.0	&	81.9	&	3.4	&	--	&	--	&	--	&	--	\\
23	&	AGAL030.818$-$00.056	&	N	&	97.9	&	7.4	&	48.2	&	B	&	96.6	&	50.5	&	5.6	&	--	&	--	&	--	&	--	\\
24	&	AGAL030.848$-$00.081	&	B	&	96.2	&	10.9	&	9.5	&	--	&	--	&	--	&	--	&	--	&	--	&	--	&	--	\\
25	&	AGAL030.893$+$00.139	&	N	&	106.3	&	2.0	&	7.2	&	P2	&	95.0	&	3.5	&	3.9	&	--	&	--	&	--	&	--	\\
26	&	AGAL031.412$+$00.307	&	B1	&	97.6	&	7.5	&	40.5	&	B2	&	97.4	&	85.6	&	2.7	&	--	&	--	&	--	&	--	\\
27	&	AGAL034.258$+$00.154	&	B1	&	57.7	&	8.9	&	71.7	&	B2	&	63.8	&	37.4	&	8.7	&	--	&	--	&	--	&	--	\\
28	&	AGAL034.401$+$00.226	&	B1	&	58.0	&	10.3	&	29.6	&	B2	&	63.5	&	64.3	&	2.0	&	--	&	--	&	--	&	--	\\
29	&	AGAL034.411$+$00.234	&	B1	&	58.2	&	10.4	&	16.3	&	B2	&	61.9	&	65.6	&	2.3	&	--	&	--	&	--	&	--	\\
30	&	AGAL034.821$+$00.351	&	N	&	57.7	&	6.9	&	11.4	&	B	&	59.2	&	28.2	&	5.0	&	--	&	--	&	--	&	--	\\
31	&	AGAL035.197$-$00.742	&	B1	&	34.6	&	8.9	&	38.8	&	B2	&	32.8	&	20.1	&	19.6	&	--	&	--	&	--	&	--	\\
32	&	AGAL037.554$+$00.201	&	B1	&	85.5	&	7.7	&	12.8	&	B2	&	88.0	&	52.3	&	3.1	&	--	&	--	&	--	&	--	\\
33	&	AGAL043.166$+$00.011	&	B1	&	15.9	&	12.1	&	30.9	&	B2	&	0.2	&	14.1	&	49.8	&	B3	&	$-$1.9	&	70.9	&	6.5	\\
%34	&	AGAL049.489$-$00.389$^\ast$	&	N	&	51.1	&	5.8	&	36.3	&	B	&	59.2	&	27.1	&	17.3	&	--	&	--	&	--	&	--	\\
%34	&	AGAL049.489$-$00.389$^\ast$	&	--	&	--	&	--	&	--	&	--	&	--	&	--	&	--	&	--	&	--	&	--	&	--	\\
35	&	AGAL053.141$+$00.069	&	N	&	22.4	&	6.3	&	30.5	&	B	&	21.1	&	16.0	&	11.8	&	--	&	--	&	--	&	--	\\
36	&	AGAL059.782$+$00.066	&	B1	&	22.7	&	8.7	&	29.6	&	B2	&	21.6	&	28.0	&	4.0	&	--	&	--	&	--	&	--	\\
37	&	AGAL301.136$-$00.226	&	B1	&	$-$38.4	&	9.5	&	37.8	&	B2	&	$-$36.8	&	32.7	&	14.8	&	B3	&	$-$22.9	&	70.3	&	7.8	\\
38	&	AGAL305.192$-$00.006	&	B1	&	$-$33.9	&	7.9	&	15.2	&	B2	&	$-$34.1	&	24.5	&	3.0	&	--	&	--	&	--	&	--	\\
39	&	AGAL305.209$+$00.206	&	B1	&	$-$42.5	&	10.7	&	45.0	&	B2	&	$-$37.5	&	43.1	&	6.8	&	--	&	--	&	--	&	--	\\
40	&	AGAL305.562$+$00.014	&	N	&	$-$38.7	&	5.9	&	50.2	&	B	&	$-$42.7	&	17.8	&	11.3	&	--	&	--	&	--	&	--	\\
41	&	AGAL305.794$-$00.096	&	B	&	$-$42.5	&	11.1	&	5.7	&	--	&	--	&	--	&	--	&	--	&	--	&	--	&	--	\\
42	&	AGAL309.384$-$00.134	&	B	&	$-$51.3	&	10.4	&	17.4	&	--	&	--	&	--	&	--	&	--	&	--	&	--	&	--	\\
43	&	AGAL310.014$+$00.387	&	B1	&	$-$42.7	&	8.7	&	15.5	&	B2	&	$-$43.1	&	35.1	&	4.7	&	--	&	--	&	--	&	--	\\
44	&	AGAL313.576$+$00.324	&	N	&	$-$46.8	&	6.4	&	19.4	&	B	&	$-$43.0	&	25.2	&	3.4	&	--	&	--	&	--	&	--	\\
45	&	AGAL316.641$-$00.087	&	B	&	$-$17.8	&	17.1	&	5.9	&	--	&	--	&	--	&	--	&	--	&	--	&	--	&	--	\\
46	&	AGAL317.867$-$00.151	&	B	&	$-$39.7	&	8.2	&	17.3	&	--	&	--	&	--	&	--	&	--	&	--	&	--	&	--	\\
47	&	AGAL318.779$-$00.137	&	B	&	$-$38.3	&	7.9	&	10.3	&	--	&	--	&	--	&	--	&	--	&	--	&	--	&	--	\\
48	&	AGAL320.881$-$00.397	&	N	&	$-$45.4	&	4.9	&	8.6	&	--	&	--	&	--	&	--	&	--	&	--	&	--	&	--	\\
49	&	AGAL326.661$+$00.519	&	N	&	$-$39.4	&	4.0	&	69.0	&	--	&	--	&	--	&	--	&	--	&	--	&	--	&	--	\\
50	&	AGAL326.987$-$00.032	&	B	&	$-$57.4	&	8.1	&	12.1	&	--	&	--	&	--	&	--	&	--	&	--	&	--	&	--	\\
51	&	AGAL327.119$+$00.509	&	B	&	$-$84.0	&	9.9	&	11.4	&	--	&	--	&	--	&	--	&	--	&	--	&	--	&	--	\\
52	&	AGAL327.393$+$00.199	&	B	&	$-$88.9	&	9.4	&	14.8	&	--	&	--	&	--	&	--	&	--	&	--	&	--	&	--	\\
53	&	AGAL329.029$-$00.206	&	N	&	$-$43.8	&	7.1	&	15.8	&	B	&	$-$46.4	&	16.2	&	11.0	&	--	&	--	&	--	&	--	\\
54	&	AGAL329.066$-$00.307	&	N	&	$-$42.0	&	6.7	&	12.4	&	B	&	$-$47.5	&	29.0	&	3.1	&	--	&	--	&	--	&	--	\\
55	&	AGAL330.879$-$00.367	&	B1	&	$-$61.3	&	12.7	&	40.6	&	B2	&	$-$69.5	&	22.2	&	19.3	&	B3	&	$-$78.8	&	88.3	&	3.6	\\
56	&	AGAL330.954$-$00.182	&	B1	&	$-$92.4	&	10.0	&	51.6	&	B2	&	$-$94.1	&	26.8	&	24.4	&	--	&	--	&	--	&	--	\\
57	&	AGAL331.709$+$00.582	&	B1	&	$-$65.9	&	8.9	&	11.6	&	B2	&	$-$66.5	&	29.4	&	5.5	&	--	&	--	&	--	&	--	\\
58	&	AGAL332.094$-$00.421	&	B1	&	$-$57.4	&	7.6	&	19.4	&	B2	&	$-$55.2	&	26.4	&	9.9	&	--	&	--	&	--	&	--	\\
%59	&	AGAL332.826$-$00.549$^\ast$	&	B1	&	$-$57.8	&	12.0	&	95.4	&	B2	&	$-$59.6	&	33.9	&	7.0	&	B3	&	$-$64.9	&	118.0	&	0.9	\\
%59	&	AGAL332.826$-$00.549$^\ast$	&	--	&	--	&	--	&	--	&	--	&	--	&	--	&	--	&	--	&	--	&	--	&	--	\\
%60	&	AGAL333.134$-$00.431$^\ast$	&	B1	&	$-$50.2	&	8.9	&	40.0	&	B2	&	$-$55.2	&	22.0	&	49.8	&	--	&	--	&	--	&	--	\\
%60	&	AGAL333.134$-$00.431$^\ast$	&	--	&	--	&	--	&	--	&	--	&	--	&	--	&	--	&	--	&	--	&	--	&	--	\\
61	&	AGAL333.284$-$00.387	&	B	&	$-$51.2	&	7.6	&	55.9	&	--	&	--	&	--	&	--	&	--	&	--	&	--	&	--	\\
62	&	AGAL333.314$+$00.106	&	B1	&	$-$45.9	&	10.7	&	16.9	&	B2	&	$-$58.7	&	90.9	&	2.2	&	--	&	--	&	--	&	--	\\
63	&	AGAL333.604$-$00.212	&	B1	&	$-$44.1	&	11.9	&	94.3	&	B2	&	$-$49.0	&	24.1	&	40.8	&	--	&	--	&	--	&	--	\\
64	&	AGAL333.656$+$00.059	&	B	&	$-$85.7	&	9.9	&	5.5	&	--	&	--	&	--	&	--	&	--	&	--	&	--	&	--	\\
65	&	AGAL335.789$+$00.174	&	N	&	$-$49.7	&	5.2	&	20.1	&	B1	&	$-$50.6	&	19.5	&	13.3	&	B2	&	$-$72.3	&	116.4	&	1.2	\\
66	&	AGAL336.958$-$00.224	&	N	&	$-$71.7	&	7.3	&	6.3	&	--	&	--	&	--	&	--	&	--	&	--	&	--	&	--	\\
67	&	AGAL337.176$-$00.032	&	B1	&	$-$68.6	&	8.2	&	11.2	&	B2	&	$-$66.7	&	32.7	&	2.4	&	P2	&	$-$79.2	&	4.7	&	5.2	\\
68	&	AGAL337.258$-$00.101	&	B	&	$-$68.1	&	10.1	&	8.6	&	--	&	--	&	--	&	--	&	--	&	--	&	--	&	--	\\
69	&	AGAL337.286$+$00.007	&	B	&	$-$105.6	&	8.3	&	3.0	&	--	&	--	&	--	&	--	&	--	&	--	&	--	&	--	\\
70	&	AGAL337.406$-$00.402	&	N	&	$-$40.9	&	5.4	&	75.5	&	B	&	$-$40.5	&	16.2	&	13.2	&	B	&	$-$24.9	&	61.1	&	2.0	\\
71	&	AGAL337.704$-$00.054	&	B1	&	$-$47.2	&	11.1	&	24.7	&	B2	&	$-$40.4	&	68.9	&	3.0	&	--	&	--	&	--	&	--	\\
72	&	AGAL337.916$-$00.477	&	B1	&	$-$40.3	&	10.6	&	40.8	&	B2	&	$-$46.1	&	33.1	&	18.4	&	B3	&	$-$46.9	&	113.4	&	3.0	\\
73	&	AGAL338.066$+$00.044	&	B	&	$-$69.9	&	13.3	&	3.3	&	P2	&	$-$13.5	&	10.5	&	2.8	&	--	&	--	&	--	&	--	\\
74	&	AGAL338.786$+$00.476	&	B	&	$-$63.7	&	8.9	&	5.7	&	--	&	--	&	--	&	--	&	--	&	--	&	--	&	--	\\
75	&	AGAL338.926$+$00.554	&	B1	&	$-$62.3	&	8.6	&	43.2	&	B2	&	$-$60.9	&	26.9	&	6.5	&	--	&	--	&	--	&	--	\\
76	&	AGAL339.623$-$00.122	&	N	&	$-$33.8	&	7.0	&	14.3	&	B	&	$-$32.2	&	21.6	&	9.7	&	--	&	--	&	--	&	--	\\
77	&	AGAL340.374$-$00.391	&	B1	&	$-$44.3	&	8.2	&	6.6	&	B2	&	$-$43.5	&	61.0	&	1.6	&	--	&	--	&	--	&	--	\\
78	&	AGAL340.746$-$01.001	&	B1	&	$-$28.9	&	7.5	&	15.6	&	B2	&	$-$18.6	&	30.2	&	3.4	&	--	&	--	&	--	&	--	\\
79	&	AGAL340.784$-$00.097	&	B	&	$-$101.6	&	10.1	&	8.0	&	--	&	--	&	--	&	--	&	--	&	--	&	--	&	--	\\
80	&	AGAL341.217$-$00.212	&	N	&	$-$43.1	&	5.6	&	28.2	&	B	&	$-$46.2	&	68.9	&	3.7	&	--	&	--	&	--	&	--	\\
81	&	AGAL342.484$+$00.182	&	B1	&	$-$42.4	&	9.6	&	9.6	&	B2	&	$-$40.8	&	67.9	&	2.3	&	--	&	--	&	--	&	--	\\
82	&	AGAL343.128$-$00.062	&	B1	&	$-$29.1	&	18.4	&	47.8	&	B2	&	$-$23.3	&	52.3	&	8.2	&	--	&	--	&	--	&	--	\\
83	&	AGAL343.756$-$00.164	&	B	&	$-$27.1	&	7.8	&	22.2	&	--	&	--	&	--	&	--	&	--	&	--	&	--	&	--	\\
84	&	AGAL344.227$-$00.569	&	B1	&	$-$22.1	&	7.7	&	16.2	&	B2	&	$-$25.3	&	30.3	&	8.0	&	--	&	--	&	--	&	--	\\
85	&	AGAL345.003$-$00.224	&	B1	&	$-$26.6	&	8.7	&	38.9	&	B2	&	$-$27.3	&	21.4	&	17.1	&	B3	&	$-$13.3	&	108.3	&	3.3	\\
%86	&	AGAL345.488$+$00.314$^\ast$	&	N	&	$-$18.7	&	5.9	&	122.9	&	B	&	$-$17.5	&	11.9	&	24.7	&	--	&	--	&	--	&	--	\\
%86	&	AGAL345.488$+$00.314$^\ast$	&	--	&	--	&	--	&	--	&	--	&	--	&	--	&	--	&	--	&	--	&	--	&	--	\\
87	&	AGAL345.504$+$00.347	&	B1	&	$-$16.7	&	9.4	&	37.4	&	B2	&	$-$14.1	&	24.9	&	9.2	&	--	&	--	&	--	&	--	\\
88	&	AGAL345.718$+$00.817	&	N	&	$-$11.5	&	5.5	&	15.3	&	--	&	--	&	--	&	--	&	--	&	--	&	--	&	--	\\
89	&	AGAL351.131$+$00.771	&	N	&	$-$5.4	&	4.7	&	13.7	&	--	&	--	&	--	&	--	&	--	&	--	&	--	&	--	\\
90	&	AGAL351.161$+$00.697	&	B1	&	$-$6.4	&	11.0	&	79.2	&	B2	&	$-$8.9	&	28.0	&	11.5	&	--	&	--	&	--	&	--	\\
91	&	AGAL351.244$+$00.669	&	B1	&	$-$2.9	&	8.4	&	108.6	&	B2	&	0.8	&	25.7	&	13.7	&	B3	&	19.9	&	120.2	&	2.0	\\
92	&	AGAL351.416$+$00.646	&	B1	&	$-$7.0	&	8.5	&	59.0	&	B2	&	$-$11.7	&	31.7	&	17.0	&	B3	&	$-$19.5	&	76.3	&	5.4	\\
93	&	AGAL351.444$+$00.659	&	B1	&	$-$4.3	&	11.0	&	45.6	&	B2	&	$-$6.9	&	28.0	&	12.2	&	--	&	--	&	--	&	--	\\
94	&	AGAL351.571$+$00.762	&	N	&	$-$4.2	&	4.4	&	11.9	&	--	&	--	&	--	&	--	&	--	&	--	&	--	&	--	\\
95	&	AGAL351.581$-$00.352	&	B1	&	$-$96.0	&	14.6	&	18.4	&	B2	&	$-$101.1	&	58.1	&	4.4	&	--	&	--	&	--	&	--	\\
96	&	AGAL351.774$-$00.537	&	B1	&	$-$1.6	&	16.4	&	40.8	&	B2	&	$-$6.1	&	57.8	&	20.2	&	--	&	--	&	--	&	--	\\
97	&	AGAL353.066$+$00.452	&	N	&	$-$1.2	&	6.6	&	10.6	&	--	&	--	&	--	&	--	&	--	&	--	&	--	&	--	\\
98	&	AGAL353.417$-$00.079	&	N	&	$-$55.4	&	5.1	&	6.1	&	--	&	--	&	--	&	--	&	--	&	--	&	--	&	--	\\
99	&	AGAL354.944$-$00.537	&	N	&	$-$5.2	&	6.4	&	5.1	&	--	&	--	&	--	&	--	&	--	&	--	&	--	&	--	\\
\end{longtable}
\tablefoot{The columns are as follows: (1)-(2) ID and CSC name of the source (given in Table\,\ref{tbl_observations_short}); %a $\ast$ symbol indicates the cases where the Gaussian fit is not reliable, as discussed in Sect.\,\ref{sec_gaussfit}; 
(3), (7), (11) the classification of each fitted Gaussian component (C$_1$, C$_2$ and C$_3$), into narrow (N), broad (B) or secondary peaks (P2) as discussed in the main text; (4)-(6), (8)-(10), (12)-(14) the central velocity (\vpeak), full width at half maximum (FWHM) and peak temperature (\tpeak) is presented for each component.}
\end{longtab}

\setlength{\tabcolsep}{6pt} % A.4

\begin{longtab}	
\setcounter{table}{4}	
\captionsetup{width=0.9\textwidth}	
\setlength{\tabcolsep}{3pt}	
\begin{longtable}{rc|ccccc|ccccc}	
    \caption{\label{table_co_extension}Extension of the CO\,(6--5) and the \textit{Herschel}-PACS 70\,\um emission towards the \atgtop clumps.}\\	
\hline\hline	
      &	 	&	\multicolumn{5}{c|}{CO\,(6--5)}	&	\multicolumn{5}{c}{PACS 70\,\um}	\\	
ID	&	CSC Name	&	$\Delta \theta_{\rm{max}}$	&	$\Delta \theta_{\rm{min}}$	&	$\Delta \theta_{\rm{avg}}$	&	$\Delta s_{\rm{avg}}$	&	$\sigma_{\rm{s}}$	&	$\Delta \theta_{\rm{max}}$	&	$\Delta \theta_{\rm{min}}$	&	$\Delta \theta_{\rm{avg}}$	&	$\Delta s_{\rm{avg}}$	&	$\sigma_{\rm{s}}$	\\
	&	&	(\arcsec)	&	(\arcsec)	&	(\arcsec)	&	(pc)	&	(pc)	&	(\arcsec)	&	(\arcsec)	&	(\arcsec)	&	(pc)	&	(pc)	\\
\hline	
\endfirsthead	
\caption{continued.} \\	
\hline\hline	
      &	 	&	\multicolumn{5}{c|}{CO\,(6--5)}	&	\multicolumn{5}{c}{PACS 70\,\um}	\\	
ID	&	CSC Name	&	$\Delta \theta_{\rm{max}}$	&	$\Delta \theta_{\rm{min}}$	&	$\Delta \theta_{\rm{avg}}$	&	$\Delta s_{\rm{avg}}$	&	$\sigma_{\rm{s}}$	&	$\Delta \theta_{\rm{max}}$	&	$\Delta \theta_{\rm{min}}$	&	$\Delta \theta_{\rm{avg}}$	&	$\Delta s_{\rm{avg}}$	&	$\sigma_{\rm{s}}$	\\
	&	&	(\arcsec)	&	(\arcsec)	&	(\arcsec)	&	(pc)	&	(pc)	&	(\arcsec)	&	(\arcsec)	&	(\arcsec)	&	(pc)	&	(pc)	\\
\hline	
\endhead	
\hline	
\endfoot	
1	&	AGAL008.684$-$00.367	&	16.8	&	11.7	&	14.2	&	0.331	&	0.085	&	12.9	&	15.2	&	14.1	&	0.326	&	0.038	\\
2	&	AGAL008.706$-$00.414	&	30.5	&	20.3	&	25.4	&	0.589	&	0.130	&	--	&	--	&	--	&	--	&	--	\\
3	&	AGAL010.444$-$00.017	&	38.1	&	17.2	&	27.6	&	1.146	&	0.256	&	10.3	&	9.4	&	9.9	&	0.409	&	0.025	\\
4	&	AGAL010.472$+$00.027	&	41.2	&	21.1	&	31.2	&	1.292	&	0.280	&	15.2	&	14.1	&	14.7	&	0.607	&	0.032	\\
5	&	AGAL010.624$-$00.384	&	47.3	&	29.7	&	38.5	&	0.924	&	0.193	&	15.2	&	17.3	&	16.2	&	0.389	&	0.036	\\
6	&	AGAL012.804$-$00.199	&	88.5	&	57.9	&	73.2	&	0.852	&	0.173	&	13.3	&	17.1	&	15.2	&	0.177	&	0.031	\\
7	&	AGAL013.178$+$00.059	&	47.3	&	29.7	&	38.5	&	0.448	&	0.094	&	--	&	--	&	--	&	--	&	--	\\
8	&	AGAL013.658$-$00.599	&	25.9	&	12.5	&	19.2	&	0.416	&	0.101	&	13.8	&	12.2	&	13.0	&	0.281	&	0.024	\\
9	&	AGAL014.114$-$00.574	&	45.8	&	19.6	&	32.7	&	0.407	&	0.089	&	9.0	&	8.3	&	8.6	&	0.108	&	0.006	\\
10	&	AGAL014.194$-$00.194	&	27.5	&	18.8	&	23.1	&	0.438	&	0.098	&	10.9	&	9.8	&	10.4	&	0.196	&	0.014	\\
11	&	AGAL014.492$-$00.139	&	36.6	&	21.9	&	29.2	&	0.549	&	0.119	&	--	&	--	&	--	&	--	&	--	\\
12	&	AGAL014.632$-$00.577	&	36.6	&	25.8	&	31.2	&	0.277	&	0.059	&	11.3	&	13.2	&	12.3	&	0.109	&	0.012	\\
13	&	AGAL015.029$-$00.669	&	91.5	&	68.9	&	80.2	&	0.770	&	0.155	&	34.6	&	20.4	&	27.5	&	0.264	&	0.097	\\
14	&	AGAL018.606$-$00.074	&	24.4	&	20.3	&	22.3	&	0.470	&	0.105	&	10.5	&	9.4	&	9.9	&	0.209	&	0.018	\\
15	&	AGAL018.734$-$00.226	&	24.4	&	12.5	&	18.5	&	1.117	&	0.272	&	14.0	&	11.3	&	12.7	&	0.765	&	0.112	\\
16	&	AGAL018.888$-$00.474	&	62.5	&	36.8	&	49.7	&	1.141	&	0.235	&	15.2	&	10.0	&	12.6	&	0.290	&	0.085	\\
17	&	AGAL019.882$-$00.534	&	35.1	&	20.3	&	27.7	&	0.492	&	0.107	&	13.7	&	11.4	&	12.5	&	0.222	&	0.029	\\
18	&	AGAL022.376$+$00.447	&	13.7	&	11.7	&	12.7	&	0.244	&	0.065	&	10.0	&	8.2	&	9.1	&	0.175	&	0.025	\\
19	&	AGAL023.206$-$00.377	&	13.7	&	12.5	&	13.1	&	0.292	&	0.076	&	12.5	&	11.2	&	11.9	&	0.265	&	0.021	\\
20	&	AGAL024.629$+$00.172	&	13.7	&	12.5	&	13.1	&	0.491	&	0.128	&	10.0	&	8.8	&	9.4	&	0.352	&	0.033	\\
21	&	AGAL028.564$-$00.236	&	38.1	&	27.4	&	32.8	&	0.865	&	0.183	&	--	&	--	&	--	&	--	&	--	\\
22	&	AGAL028.861$+$00.066	&	27.5	&	15.7	&	21.6	&	0.775	&	0.178	&	14.5	&	12.1	&	13.3	&	0.478	&	0.059	\\
23	&	AGAL030.818$-$00.056	&	51.9	&	30.5	&	41.2	&	0.979	&	0.204	&	14.6	&	13.2	&	13.9	&	0.330	&	0.024	\\
24	&	AGAL030.848$-$00.081	&	91.5	&	64.2	&	77.8	&	1.849	&	0.374	&	--	&	--	&	--	&	--	&	--	\\
25	&	AGAL030.893$+$00.139	&	38.1	&	20.3	&	29.2	&	0.694	&	0.151	&	--	&	--	&	--	&	--	&	--	\\
26	&	AGAL031.412$+$00.307	&	19.8	&	18.8	&	19.3	&	0.458	&	0.105	&	15.1	&	13.7	&	14.4	&	0.343	&	0.024	\\
27	&	AGAL034.258$+$00.154	&	41.2	&	29.7	&	35.5	&	0.268	&	0.056	&	26.5	&	20.1	&	23.3	&	0.176	&	0.034	\\
28	&	AGAL034.401$+$00.226	&	51.9	&	24.3	&	38.1	&	0.288	&	0.061	&	13.9	&	12.7	&	13.3	&	0.101	&	0.006	\\
29	&	AGAL034.411$+$00.234	&	19.8	&	18.0	&	18.9	&	0.143	&	0.033	&	12.7	&	11.4	&	12.0	&	0.091	&	0.007	\\
30	&	AGAL034.821$+$00.351	&	27.5	&	14.1	&	20.8	&	0.157	&	0.037	&	13.0	&	11.9	&	12.5	&	0.094	&	0.006	\\
31	&	AGAL035.197$-$00.742	&	88.5	&	31.3	&	59.9	&	0.636	&	0.132	&	15.2	&	14.1	&	14.7	&	0.156	&	0.008	\\
32	&	AGAL037.554$+$00.201	&	18.3	&	12.5	&	15.4	&	0.501	&	0.125	&	12.3	&	11.1	&	11.7	&	0.382	&	0.027	\\
33	&	AGAL043.166$+$00.011	&	61.0	&	29.0	&	45.0	&	2.423	&	0.505	&	22.5	&	13.2	&	17.8	&	0.961	&	0.354	\\
34	&	AGAL049.489$-$00.389	&	114.5	&	43.1	&	78.8	&	2.066	&	0.440	&	18.9	&	15.9	&	17.4	&	0.457	&	0.055	\\
35	&	AGAL053.141$+$00.069	&	61.0	&	18.8	&	39.9	&	0.309	&	0.067	&	13.6	&	12.4	&	13.0	&	0.101	&	0.006	\\
36	&	AGAL059.782$+$00.066	&	62.5	&	21.9	&	42.2	&	0.442	&	0.094	&	15.3	&	11.9	&	13.6	&	0.142	&	0.026	\\
37	&	AGAL301.136$-$00.226	&	18.3	&	12.5	&	15.4	&	0.329	&	0.082	&	18.7	&	17.1	&	17.9	&	0.381	&	0.024	\\
38	&	AGAL305.192$-$00.006	&	35.1	&	25.8	&	30.4	&	0.561	&	0.120	&	12.2	&	9.9	&	11.1	&	0.204	&	0.029	\\
39	&	AGAL305.209$+$00.206	&	45.8	&	23.5	&	34.7	&	0.638	&	0.136	&	10.7	&	12.6	&	11.7	&	0.215	&	0.024	\\
40	&	AGAL305.562$+$00.014	&	48.8	&	26.6	&	37.7	&	0.695	&	0.146	&	15.2	&	13.6	&	14.4	&	0.265	&	0.021	\\
41	&	AGAL305.794$-$00.096	&	77.8	&	29.0	&	53.4	&	0.983	&	0.205	&	--	&	--	&	--	&	--	&	--	\\
42	&	AGAL309.384$-$00.134	&	38.1	&	19.6	&	28.8	&	0.746	&	0.163	&	8.2	&	11.6	&	9.9	&	0.256	&	0.062	\\
43	&	AGAL310.014$+$00.387	&	30.5	&	26.6	&	28.5	&	0.500	&	0.107	&	13.0	&	12.6	&	12.8	&	0.224	&	0.005	\\
44	&	AGAL313.576$+$00.324	&	18.3	&	14.1	&	16.2	&	0.297	&	0.072	&	11.0	&	12.2	&	11.6	&	0.212	&	0.015	\\
45	&	AGAL316.641$-$00.087	&	19.8	&	12.5	&	16.1	&	0.093	&	0.023	&	12.3	&	10.3	&	11.3	&	0.065	&	0.008	\\
46	&	AGAL317.867$-$00.151	&	41.2	&	19.6	&	30.4	&	0.435	&	0.095	&	8.2	&	11.2	&	9.7	&	0.139	&	0.031	\\
47	&	AGAL318.779$-$00.137	&	25.9	&	15.7	&	20.8	&	0.280	&	0.065	&	--	&	--	&	--	&	--	&	--	\\
48	&	AGAL320.881$-$00.397	&	38.1	&	34.4	&	36.2	&	1.753	&	0.366	&	--	&	--	&	--	&	--	&	--	\\
49	&	AGAL326.661$+$00.519	&	44.2	&	33.7	&	39.0	&	0.343	&	0.071	&	13.0	&	11.7	&	12.4	&	0.109	&	0.008	\\
50	&	AGAL326.987$-$00.032	&	33.6	&	19.6	&	26.6	&	0.509	&	0.112	&	10.3	&	9.1	&	9.7	&	0.186	&	0.016	\\
51	&	AGAL327.119$+$00.509	&	36.6	&	25.0	&	30.8	&	0.823	&	0.176	&	11.5	&	12.4	&	11.9	&	0.319	&	0.017	\\
52	&	AGAL327.393$+$00.199	&	29.0	&	21.1	&	25.0	&	0.719	&	0.158	&	12.8	&	11.3	&	12.1	&	0.346	&	0.032	\\
53	&	AGAL329.029$-$00.206	&	50.3	&	19.6	&	35.0	&	1.946	&	0.423	&	21.1	&	12.4	&	16.7	&	0.932	&	0.345	\\
54	&	AGAL329.066$-$00.307	&	35.1	&	26.6	&	30.8	&	1.732	&	0.369	&	14.5	&	8.2	&	11.4	&	0.637	&	0.249	\\
55	&	AGAL330.879$-$00.367	&	25.9	&	21.9	&	23.9	&	0.482	&	0.106	&	16.7	&	15.1	&	15.9	&	0.320	&	0.024	\\
56	&	AGAL330.954$-$00.182	&	29.0	&	22.7	&	25.9	&	1.168	&	0.254	&	19.3	&	20.1	&	19.7	&	0.890	&	0.024	\\
57	&	AGAL331.709$+$00.582	&	21.4	&	19.6	&	20.5	&	1.046	&	0.237	&	11.7	&	12.5	&	12.1	&	0.619	&	0.029	\\
58	&	AGAL332.094$-$00.421	&	22.9	&	20.3	&	21.6	&	0.377	&	0.085	&	13.2	&	11.6	&	12.4	&	0.217	&	0.020	\\
59	&	AGAL332.826$-$00.549	&	29.0	&	21.1	&	25.0	&	0.437	&	0.096	&	29.4	&	20.1	&	24.7	&	0.431	&	0.115	\\
60	&	AGAL333.134$-$00.431	&	45.8	&	34.4	&	40.1	&	0.700	&	0.145	&	22.7	&	17.3	&	20.0	&	0.349	&	0.066	\\
61	&	AGAL333.284$-$00.387	&	86.9	&	61.0	&	73.9	&	1.291	&	0.261	&	21.9	&	12.4	&	17.2	&	0.299	&	0.118	\\
62	&	AGAL333.314$+$00.106	&	16.8	&	15.7	&	16.2	&	0.283	&	0.068	&	13.4	&	11.9	&	12.6	&	0.221	&	0.018	\\
63	&	AGAL333.604$-$00.212	&	73.2	&	53.2	&	63.2	&	1.103	&	0.224	&	24.7	&	17.9	&	21.3	&	0.372	&	0.085	\\
64	&	AGAL333.656$+$00.059	&	12.2	&	11.0	&	11.6	&	0.297	&	0.082	&	--	&	--	&	--	&	--	&	--	\\
65	&	AGAL335.789$+$00.174	&	29.0	&	23.5	&	26.2	&	0.467	&	0.101	&	15.9	&	13.0	&	14.4	&	0.257	&	0.036	\\
66	&	AGAL336.958$-$00.224	&	12.2	&	11.0	&	11.6	&	0.612	&	0.168	&	9.5	&	8.7	&	9.1	&	0.482	&	0.031	\\
67	&	AGAL337.176$-$00.032	&	41.2	&	20.3	&	30.8	&	1.641	&	0.357	&	9.1	&	12.4	&	10.7	&	0.572	&	0.125	\\
68	&	AGAL337.258$-$00.101	&	18.3	&	14.1	&	16.2	&	0.864	&	0.210	&	11.0	&	10.1	&	10.5	&	0.562	&	0.034	\\
69	&	AGAL337.286$+$00.007	&	33.6	&	14.9	&	24.2	&	1.109	&	0.256	&	--	&	--	&	--	&	--	&	--	\\
70	&	AGAL337.406$-$00.402	&	15.3	&	14.1	&	14.7	&	0.232	&	0.058	&	17.1	&	15.9	&	16.5	&	0.261	&	0.013	\\
71	&	AGAL337.704$-$00.054	&	24.4	&	21.9	&	23.1	&	1.376	&	0.304	&	15.6	&	10.4	&	13.0	&	0.772	&	0.220	\\
72	&	AGAL337.916$-$00.477	&	30.5	&	22.7	&	26.6	&	0.413	&	0.090	&	17.0	&	16.7	&	16.8	&	0.261	&	0.004	\\
73	&	AGAL338.066$+$00.044	&	13.7	&	12.5	&	13.1	&	0.298	&	0.078	&	--	&	--	&	--	&	--	&	--	\\
74	&	AGAL338.786$+$00.476	&	33.6	&	32.1	&	32.8	&	0.715	&	0.151	&	--	&	--	&	--	&	--	&	--	\\
75	&	AGAL338.926$+$00.554	&	61.0	&	26.6	&	43.8	&	0.934	&	0.196	&	12.7	&	15.6	&	14.2	&	0.302	&	0.044	\\
76	&	AGAL339.623$-$00.122	&	35.1	&	21.1	&	28.1	&	0.410	&	0.089	&	12.9	&	11.3	&	12.1	&	0.177	&	0.017	\\
77	&	AGAL340.374$-$00.391	&	22.9	&	15.7	&	19.3	&	0.336	&	0.078	&	--	&	--	&	--	&	--	&	--	\\
78	&	AGAL340.746$-$01.001	&	36.6	&	25.0	&	30.8	&	0.412	&	0.088	&	11.9	&	10.5	&	11.2	&	0.150	&	0.014	\\
79	&	AGAL340.784$-$00.097	&	16.8	&	14.9	&	15.8	&	0.767	&	0.186	&	11.2	&	11.5	&	11.4	&	0.550	&	0.013	\\
80	&	AGAL341.217$-$00.212	&	18.3	&	14.9	&	16.6	&	0.295	&	0.071	&	17.3	&	17.3	&	17.3	&	0.307	&	0.000	\\
81	&	AGAL342.484$+$00.182	&	35.1	&	22.7	&	28.9	&	1.758	&	0.379	&	12.8	&	10.8	&	11.8	&	0.718	&	0.082	\\
82	&	AGAL343.128$-$00.062	&	42.7	&	18.0	&	30.4	&	0.447	&	0.099	&	18.4	&	16.0	&	17.2	&	0.253	&	0.025	\\
83	&	AGAL343.756$-$00.164	&	29.0	&	20.3	&	24.6	&	0.347	&	0.076	&	12.6	&	12.0	&	12.3	&	0.173	&	0.006	\\
84	&	AGAL344.227$-$00.569	&	27.5	&	13.3	&	20.4	&	0.249	&	0.059	&	13.8	&	11.9	&	12.8	&	0.157	&	0.017	\\
85	&	AGAL345.003$-$00.224	&	39.7	&	18.0	&	28.9	&	0.422	&	0.093	&	16.7	&	15.2	&	16.0	&	0.234	&	0.016	\\
86	&	AGAL345.488$+$00.314	&	61.0	&	43.8	&	52.4	&	0.564	&	0.115	&	18.5	&	16.9	&	17.7	&	0.190	&	0.012	\\
87	&	AGAL345.504$+$00.347	&	68.6	&	37.6	&	53.1	&	0.579	&	0.119	&	14.9	&	14.0	&	14.4	&	0.158	&	0.007	\\
88	&	AGAL345.718$+$00.817	&	38.1	&	27.4	&	32.8	&	0.248	&	0.052	&	14.6	&	17.6	&	16.1	&	0.122	&	0.016	\\
89	&	AGAL351.131$+$00.771	&	91.5	&	86.1	&	88.8	&	0.783	&	0.158	&	--	&	--	&	--	&	--	&	--	\\
90	&	AGAL351.161$+$00.697	&	42.7	&	34.4	&	38.6	&	0.340	&	0.071	&	25.1	&	18.1	&	21.6	&	0.191	&	0.044	\\
91	&	AGAL351.244$+$00.669	&	53.4	&	35.2	&	44.3	&	0.391	&	0.081	&	33.3	&	18.3	&	25.8	&	0.227	&	0.094	\\
92	&	AGAL351.416$+$00.646	&	48.8	&	25.0	&	36.9	&	0.240	&	0.051	&	23.0	&	20.9	&	22.0	&	0.143	&	0.010	\\
93	&	AGAL351.444$+$00.659	&	73.2	&	47.7	&	60.4	&	0.393	&	0.080	&	17.3	&	12.6	&	15.0	&	0.097	&	0.022	\\
94	&	AGAL351.571$+$00.762	&	53.4	&	27.4	&	40.4	&	0.262	&	0.055	&	--	&	--	&	--	&	--	&	--	\\
95	&	AGAL351.581$-$00.352	&	13.7	&	13.3	&	13.5	&	0.446	&	0.114	&	14.2	&	13.2	&	13.7	&	0.452	&	0.021	\\
96	&	AGAL351.774$-$00.537	&	33.6	&	15.7	&	24.6	&	0.119	&	0.027	&	21.8	&	20.0	&	20.9	&	0.101	&	0.006	\\
97	&	AGAL353.066$+$00.452	&	83.9	&	23.5	&	53.7	&	0.224	&	0.047	&	--	&	--	&	--	&	--	&	--	\\
98	&	AGAL353.417$-$00.079	&	--	&	--	&	--	&	--	&	--	&	--	&	--	&	--	&	--	&	--	\\
99	&	AGAL354.944$-$00.537	&	22.9	&	14.9	&	18.9	&	0.175	&	0.041	&	--	&	--	&	--	&	--	&	--	\\
\end{longtable}	
\tablefoot{The extension of the CO emission is measured from the 50\% peak contour on the maps presented in Appendix\,\ref{appendix_co_fixbeam}. The columns are as follows:	
(1) ID of the source;	
(2) the CSC name of the ATLASGAL clump;	
(3) the maximum elongation of the CO emission (in arcseconds);	
(4) the minimum elongation of the CO emission (in arcseconds);	
(5) the average size of the CO emission (in arcseconds);	
(6) the linear size of the CO emission (in parsecs), considering the average between the data presented in columns (3)-(4) and taking into account the distance of the source from Table\,\ref{tbl_observations_short};	
(7)  error of the linear extent of the CO emission, considering an uncertainty of 1.5\arcsec on the angular sizes presented in columns (3)-(4);	
(8)-(12) same as columns (3)-(7) but for the extension of the \textit{Herschel}-PACS 70\,\um emission towards the 70\,\um-bright clumps.
Columns (8) and (9) are extracted from \citet{Molinari16}.}	
\end{longtab}	
\setlength{\tabcolsep}{6pt}			% A.5

\begin{landscape}
\begin{longtab}
\setcounter{table}{5}
\captionsetup{width=0.9\textwidth}
\setlength{\tabcolsep}{4pt}
\begin{longtable}{rr|rrrrrc|rrrrrc|rrrrrc}
\caption{\label{tbl_intpropCO_fixbeam}Integrated properties of the CO emission profiles convolved to 13\farcs4.}\\
\hline\hline
	&	&	\multicolumn{6}{c|}{CO\,(4--3)}	&	\multicolumn{6}{c|}{CO\,(6--5)}	&	\multicolumn{6}{c}{CO\,(7--6)}	 \\ 
\multicolumn{1}{c}{ID}	&	\multicolumn{1}{c|}{CSC Name}	&	\multicolumn{1}{c}{rms}	&	\multicolumn{1}{c}{FWZP}	&	\sint	&	\lco	&	\multicolumn{1}{c}{$\sigma_{\rm L}$}	&	&	\multicolumn{1}{c}{rms}	&	\multicolumn{1}{c}{FWZP}	&	\sint	&	\lco	&	\multicolumn{1}{c}{$\sigma_{\rm L}$}	&	&	\multicolumn{1}{c}{rms}	&	\multicolumn{1}{c}{FWZP}	&	\sint	&	\lco	&	\multicolumn{1}{c}{$\sigma_{\rm L}$}	&	 \\ 
\hline
\endfirsthead
\caption{continued.} \\
\hline\hline
	&	&	\multicolumn{6}{c|}{CO\,(4--3)}	&	\multicolumn{6}{c|}{CO\,(6--5)}	&	\multicolumn{6}{c}{CO\,(7--6)}	 \\ 
\multicolumn{1}{c}{ID}	&	\multicolumn{1}{c|}{CSC Name}	&	\multicolumn{1}{c}{rms}	&	\multicolumn{1}{c}{FWZP}	&	\sint	&	\lco	&	\multicolumn{1}{c}{$\sigma_{\rm L}$}	&	&	\multicolumn{1}{c}{rms}	&	\multicolumn{1}{c}{FWZP}	&	\sint	&	\lco	&	\multicolumn{1}{c}{$\sigma_{\rm L}$}	&	&	\multicolumn{1}{c}{rms}	&	\multicolumn{1}{c}{FWZP}	&	\sint	&	\lco	&	\multicolumn{1}{c}{$\sigma_{\rm L}$}	&	 \\ 
\hline
\endhead
\hline
\endfoot
1	&	AGAL008.684$-$00.367	&	0.16 	&	32	&	79.30 	&	17.38 	&	6.02 	&	$\ast$	&	0.23 	&	38	&	69.86 	&	15.26 	&	5.29 	&	$\ast$	&	1.28 	&	24 	&	71.81 	&	15.69 	&	5.43 	&	$\ast$	\\
2	&	AGAL008.706$-$00.414	&	0.12 	&	18	&	37.48 	&	8.21 	&	2.85 	&		&	0.18 	&	20	&	12.89 	&	2.82 	&	0.98 	&		&	0.82 	&	10 	&	7.88 	&	1.72 	&	0.60 	&		\\
3	&	AGAL010.444$-$00.017	&	0.22 	&	28	&	43.70 	&	30.65 	&	10.62 	&		&	0.16 	&	26	&	24.39 	&	17.05 	&	5.91 	&		&	0.37 	&	24 	&	16.95 	&	11.85 	&	4.10 	&		\\
4	&	AGAL010.472$+$00.027	&	0.28 	&	42	&	228.67 	&	160.37 	&	55.55 	&	$\ast$	&	0.43 	&	44	&	197.14 	&	137.81 	&	47.74 	&	$\ast$	&	0.91 	&	32 	&	170.60 	&	119.26 	&	41.31 	&	$\ast$	\\
5	&	AGAL010.624$-$00.384	&	0.19 	&	44	&	535.01 	&	125.76 	&	43.57 	&	$\ast$	&	0.15 	&	60	&	458.13 	&	107.34 	&	37.18 	&	$\ast$	&	0.87 	&	38 	&	421.05 	&	98.65 	&	34.17 	&	$\ast$	\\
6	&	AGAL012.804$-$00.199	&	0.15 	&	52	&	452.52 	&	25.01 	&	8.66 	&	$\ast$	&	0.15 	&	60	&	529.02 	&	29.14 	&	10.09 	&	$\ast$	&	0.43 	&	46 	&	532.39 	&	29.32 	&	10.16 	&	$\ast$	\\
7	&	AGAL013.178$+$00.059	&	0.13 	&	66	&	134.91 	&	7.45 	&	2.58 	&		&	0.22 	&	40	&	93.44 	&	5.15 	&	1.78 	&		&	0.90 	&	24 	&	92.03 	&	5.07 	&	1.76 	&		\\
8	&	AGAL013.658$-$00.599	&	0.15 	&	94	&	151.85 	&	29.11 	&	10.08 	&	$\ast$	&	0.16 	&	98	&	109.50 	&	20.92 	&	7.25 	&	$\ast$	&	0.92 	&	80 	&	120.92 	&	23.10 	&	8.00 	&	$\ast$	\\
9	&	AGAL014.114$-$00.574	&	0.14 	&	40	&	46.68 	&	2.96 	&	1.02 	&	$\ast$	&	0.12 	&	38	&	59.85 	&	3.78 	&	1.31 	&	$\ast$	&	0.40 	&	24 	&	73.35 	&	4.63 	&	1.60 	&	$\ast$	\\
10	&	AGAL014.194$-$00.194	&	0.17 	&	48	&	189.30 	&	27.62 	&	9.57 	&		&	0.14 	&	50	&	113.51 	&	16.51 	&	5.72 	&		&	0.52 	&	40 	&	90.43 	&	13.15 	&	4.56 	&		\\
11	&	AGAL014.492$-$00.139	&	0.19 	&	44	&	64.86 	&	9.32 	&	3.23 	&	$\ast$	&	0.18 	&	28	&	29.37 	&	4.21 	&	1.46 	&	$\ast$	&	1.09 	&	10 	&	17.26 	&	2.47 	&	0.86 	&	$\ast$	\\
12	&	AGAL014.632$-$00.577	&	0.17 	&	44	&	110.21 	&	3.54 	&	1.23 	&	$\ast$	&	0.19 	&	40	&	96.73 	&	3.10 	&	1.07 	&	$\ast$	&	0.40 	&	26 	&	93.95 	&	3.01 	&	1.04 	&	$\ast$	\\
13	&	AGAL015.029$-$00.669	&	0.17 	&	36	&	564.08 	&	21.22 	&	7.35 	&	$\ast$	&	0.21 	&	38	&	580.80 	&	21.77 	&	7.54 	&		&	1.14 	&	26 	&	588.89 	&	22.08 	&	7.65 	&		\\
14	&	AGAL018.606$-$00.074	&	0.16 	&	22	&	65.14 	&	93.93 	&	32.54 	&	$\ast$	&	0.10 	&	28	&	45.94 	&	8.24 	&	2.85 	&	$\ast$	&	0.41 	&	18 	&	35.66 	&	6.39 	&	2.21 	&		\\
15	&	AGAL018.734$-$00.226	&	0.19 	&	48	&	99.23 	&	148.27 	&	51.36 	&		&	0.17 	&	50	&	81.83 	&	121.87 	&	42.22 	&		&	0.86 	&	28 	&	60.64 	&	90.32 	&	31.29 	&		\\
16	&	AGAL018.888$-$00.474	&	0.17 	&	38	&	188.78 	&	40.69 	&	14.10 	&		&	0.20 	&	52	&	128.67 	&	27.64 	&	9.58 	&		&	1.05 	&	22 	&	77.99 	&	16.76 	&	5.80 	&		\\
17	&	AGAL019.882$-$00.534	&	0.19 	&	88	&	389.08 	&	50.00 	&	17.32 	&		&	0.50 	&	94	&	267.32 	&	34.24 	&	11.86 	&		&	0.64 	&	84 	&	229.71 	&	29.42 	&	10.19 	&		\\
18	&	AGAL022.376$+$00.447	&	0.14 	&	62	&	84.15 	&	12.66 	&	4.39 	&	$\ast$	&	0.28 	&	22	&	28.50 	&	4.27 	&	1.48 	&	$\ast$	&	1.24 	&	14 	&	21.57 	&	3.23 	&	1.12 	&	$\ast$	\\
19	&	AGAL023.206$-$00.377	&	0.20 	&	98	&	253.96 	&	51.33 	&	17.78 	&	$\ast$	&	0.11 	&	82	&	162.87 	&	32.81 	&	11.37 	&	$\ast$	&	0.37 	&	88 	&	202.84 	&	40.87 	&	14.16 	&	$\ast$	\\
20	&	AGAL024.629$+$00.172	&	0.29 	&	26	&	41.95 	&	23.99 	&	8.31 	&	$\ast$	&	0.21 	&	24	&	17.99 	&	10.25 	&	3.55 	&	$\ast$	&	0.91 	&	10 	&	11.49 	&	6.55 	&	2.27 	&	$\ast$	\\
21	&	AGAL028.564$-$00.236	&	0.25 	&	18	&	43.03 	&	12.26 	&	4.25 	&	$\ast$	&	0.22 	&	28	&	21.99 	&	6.24 	&	2.16 	&	$\ast$	&	1.15 	&	4 	&	4.35 	&	1.23 	&	0.43 	&	$\ast$	\\
22	&	AGAL028.861$+$00.066	&	0.37 	&	42	&	206.51 	&	108.78 	&	37.68 	&	$\ast$	&	0.44 	&	64	&	230.64 	&	121.10 	&	41.95 	&	$\ast$	&	1.27 	&	50 	&	205.30 	&	107.79 	&	37.34 	&	\\	
23	&	AGAL030.818$-$00.056	&	0.27 	&	48	&	207.05 	&	47.69 	&	16.52 	&	$\ast$	&	0.22 	&	108	&	185.22 	&	42.53 	&	14.73 	&	$\ast$	&	0.90 	&	66 	&	209.78 	&	48.17 	&	16.69 	&	$\ast$	\\
24	&	AGAL030.848$-$00.081	&	0.14 	&	24	&	99.20 	&	22.85 	&	7.92 	&	$\ast$	&	0.21 	&	32	&	60.99 	&	14.00 	&	4.85 	&	$\ast$	&	0.99 	&	22 	&	45.63 	&	10.48 	&	3.63 	&	$\ast$	\\
25	&	AGAL030.893$+$00.139	&	0.20 	&	22	&	65.85 	&	15.17 	&	5.25 	&		&	0.21 	&	28	&	31.09 	&	7.14 	&	2.47 	&		&	1.45 	&	6 	&	7.86 	&	1.80 	&	0.63 	&		\\
26	&	AGAL031.412$+$00.307	&	0.32 	&	44	&	118.49 	&	27.29 	&	9.45 	&	$\ast$	&	0.24 	&	66	&	154.61 	&	35.50 	&	12.30 	&	$\ast$	&	0.37 	&	54 	&	166.62 	&	38.25 	&	13.25 	&	$\ast$	\\
27	&	AGAL034.258$+$00.154	&	0.47 	&	58	&	267.05 	&	6.23 	&	2.16 	&	$\ast$	&	0.52 	&	106	&	371.07 	&	8.64 	&	2.99 	&	$\ast$	&	0.47 	&	80 	&	383.48 	&	8.92 	&	3.09 	&	$\ast$	\\
28	&	AGAL034.401$+$00.226	&	0.14 	&	62	&	227.22 	&	5.30 	&	1.84 	&	$\ast$	&	0.13 	&	94	&	217.43 	&	5.06 	&	1.75 	&	$\ast$	&	0.41 	&	50 	&	183.55 	&	4.27 	&	1.48 	&	$\ast$	\\
29	&	AGAL034.411$+$00.234	&	0.18 	&	92	&	210.08 	&	4.90 	&	1.70 	&	$\ast$	&	0.28 	&	86	&	136.45 	&	3.18 	&	1.10 	&	$\ast$	&	1.24 	&	24 	&	99.08 	&	2.31 	&	0.80 	&		\\
30	&	AGAL034.821$+$00.351	&	0.18 	&	104	&	156.90 	&	3.66 	&	1.27 	&	$\ast$	&	0.19 	&	50	&	101.50 	&	2.36 	&	0.82 	&	$\ast$	&	1.34 	&	22 	&	77.53 	&	1.80 	&	0.63 	&		\\
31	&	AGAL035.197$-$00.742	&	0.26 	&	42	&	269.51 	&	12.40 	&	4.30 	&	$\ast$	&	0.09 	&	76	&	348.76 	&	16.00 	&	5.54 	&	$\ast$	&	0.34 	&	46 	&	295.51 	&	13.55 	&	4.69 	&	$\ast$	\\
32	&	AGAL037.554$+$00.201	&	0.15 	&	44	&	142.11 	&	61.38 	&	21.26 	&	$\ast$	&	0.26 	&	38	&	81.59 	&	35.13 	&	12.17 	&	$\ast$	&	1.56 	&	26 	&	70.58 	&	30.39 	&	10.53 	&	$\ast$	\\
33	&	AGAL043.166$+$00.011	&	0.52 	&	66	&	1128.22 	&	1335.97 	&	462.80 	&	$\ast$	&	0.25 	&	132	&	904.50 	&	1067.60 	&	369.83 	&		&	1.59 	&	84 	&	771.20 	&	910.26 	&	315.32 	&		\\
34	&	AGAL049.489$-$00.389	&	0.63 	&	58	&	542.23 	&	152.25 	&	52.74 	&	$\ast$	&	0.39 	&	82	&	264.61 	&	74.06 	&	25.65 	&	$\ast$	&	0.91 	&	70 	&	327.19 	&	91.57 	&	31.72 	&	$\ast$	\\
35	&	AGAL053.141$+$00.069	&	0.15 	&	76	&	211.85 	&	5.20 	&	1.80 	&	$\ast$	&	0.14 	&	72	&	179.15 	&	4.39 	&	1.52 	&	$\ast$	&	0.35 	&	38 	&	159.77 	&	3.91 	&	1.35 	&	$\ast$	\\
36	&	AGAL059.782$+$00.066	&	0.24 	&	56	&	246.98 	&	11.05 	&	3.83 	&	$\ast$	&	0.12 	&	66	&	204.38 	&	9.12 	&	3.16 	&	$\ast$	&	0.43 	&	40 	&	170.32 	&	7.60 	&	2.63 	&	$\ast$	\\
37	&	AGAL301.136$-$00.226	&	0.34 	&	134	&	730.69 	&	135.71 	&	47.01 	&		&	0.88 	&	130	&	705.32 	&	130.58 	&	45.23 	&	$\ast$	&	1.62 	&	126 	&	780.76 	&	144.54 	&	50.07 	&		\\
38	&	AGAL305.192$-$00.006	&	1.57 	&	12	&	50.73 	&	7.03 	&	2.43 	&		&	0.66 	&	30	&	115.42 	&	15.94 	&	5.52 	&		&	1.06 	&	26 	&	95.41 	&	13.18 	&	4.56 	&		\\
39	&	AGAL305.209$+$00.206	&	1.60 	&	36	&	433.43 	&	60.04 	&	20.80 	&	$\ast$	&	0.22 	&	96	&	403.76 	&	55.75 	&	19.31 	&	$\ast$	&	0.83 	&	70 	&	396.50 	&	54.75 	&	18.97 	&	$\ast$	\\
40	&	AGAL305.562$+$00.014	&	0.40 	&	32	&	376.91 	&	52.21 	&	18.09 	&	$\ast$	&	0.23 	&	38	&	271.66 	&	37.51 	&	12.99 	&	$\ast$	&	1.65 	&	26 	&	248.49 	&	34.31 	&	11.89 	&	$\ast$	\\
41	&	AGAL305.794$-$00.096	&	0.34 	&	22	&	13.80 	&	1.91 	&	0.66 	&	$\ast$	&	0.22 	&	24	&	21.01 	&	2.90 	&	1.00 	&	$\ast$	&	1.62 	&	16 	&	34.13 	&	4.71 	&	1.63 	&	$\ast$	\\
42	&	AGAL309.384$-$00.134	&	0.53 	&	26	&	133.61 	&	36.55 	&	12.66 	&	$\ast$	&	0.21 	&	44	&	95.86 	&	26.14 	&	9.05 	&	$\ast$	&	1.58 	&	26 	&	99.76 	&	27.20 	&	9.42 	&	$\ast$	\\
43	&	AGAL310.014$+$00.387	&	0.45 	&	48	&	182.21 	&	22.78 	&	7.89 	&	$\ast$	&	0.21 	&	74	&	120.32 	&	14.99 	&	5.19 	&	$\ast$	&	1.27 	&	30 	&	104.86 	&	13.07 	&	4.53 	&		\\
44	&	AGAL313.576$+$00.324	&	0.35 	&	42	&	174.54 	&	23.93 	&	8.29 	&	$\ast$	&	0.21 	&	60	&	131.00 	&	17.90 	&	6.20 	&	$\ast$	&	1.18 	&	28 	&	104.31 	&	14.25 	&	4.94 	&	$\ast$	\\
45	&	AGAL316.641$-$00.087	&	0.35 	&	54	&	122.67 	&	1.67 	&	0.58 	&	$\ast$	&	0.22 	&	40	&	64.08 	&	0.87 	&	0.30 	&		&	1.42 	&	30 	&	55.15 	&	0.75 	&	0.26 	&		\\
46	&	AGAL317.867$-$00.151	&	0.35 	&	22	&	110.70 	&	9.24 	&	3.20 	&	$\ast$	&	0.23 	&	42	&	77.81 	&	6.48 	&	2.24 	&	$\ast$	&	1.49 	&	20 	&	73.94 	&	6.15 	&	2.13 	&	$\ast$	\\
47	&	AGAL318.779$-$00.137	&	0.33 	&	36	&	70.07 	&	5.19 	&	1.80 	&	$\ast$	&	0.21 	&	30	&	39.33 	&	2.91 	&	1.01 	&	$\ast$	&	1.18 	&	16 	&	43.92 	&	3.25 	&	1.12 	&		\\
48	&	AGAL320.881$-$00.397	&	0.43 	&	28	&	90.61 	&	86.40 	&	29.93 	&		&	0.16 	&	26	&	36.39 	&	34.59 	&	14.26 	&		&	0.91 	&	10 	&	20.54 	&	19.53 	&	8.05 	&		\\
49	&	AGAL326.661$+$00.519	&	0.36 	&	100	&	334.51 	&	10.63 	&	3.68 	&		&	0.17 	&	24	&	169.20 	&	5.36 	&	2.21 	&		&	0.95 	&	18 	&	151.69 	&	4.80 	&	1.98 	&		\\
50	&	AGAL326.987$-$00.032	&	0.28 	&	28	&	61.97 	&	9.28 	&	3.21 	&	$\ast$	&	0.18 	&	54	&	63.15 	&	9.42 	&	3.26 	&	$\ast$	&	0.79 	&	20 	&	45.92 	&	6.85 	&	2.37 	&	$\ast$	\\
51	&	AGAL327.119$+$00.509	&	0.24 	&	42	&	90.43 	&	26.34 	&	9.12 	&	$\ast$	&	0.24 	&	38	&	69.34 	&	20.13 	&	8.30 	&	$\ast$	&	1.00 	&	22 	&	49.12 	&	14.26 	&	5.88 	&	$\ast$	\\
52	&	AGAL327.393$+$00.199	&	0.32 	&	54	&	140.73 	&	47.32 	&	16.39 	&		&	0.14 	&	54	&	84.02 	&	28.16 	&	9.75 	&		&	0.77 	&	22 	&	67.46 	&	22.61 	&	7.83 	&		\\
53	&	AGAL329.029$-$00.206	&	0.25 	&	42	&	179.16 	&	226.91 	&	78.61 	&	$\ast$	&	0.23 	&	94	&	172.95 	&	218.34 	&	75.64 	&	$\ast$	&	0.55 	&	38 	&	145.34 	&	183.49 	&	63.56 	&		\\
54	&	AGAL329.066$-$00.307	&	0.24 	&	42	&	111.89 	&	143.95 	&	49.86 	&	$\ast$	&	0.17 	&	46	&	86.23 	&	110.58 	&	63.52 	&	$\ast$	&	0.77 	&	24 	&	65.68 	&	84.22 	&	48.38 	&		\\
55	&	AGAL330.879$-$00.367	&	0.38 	&	76	&	716.58 	&	118.97 	&	41.21 	&	$\ast$	&	0.33 	&	110	&	570.27 	&	94.37 	&	32.69 	&	$\ast$	&	0.66 	&	92 	&	583.59 	&	96.58 	&	33.45 	&	$\ast$	\\
56	&	AGAL330.954$-$00.182	&	0.57 	&	54	&	519.36 	&	432.79 	&	149.92 	&	$\ast$	&	0.22 	&	102	&	487.98 	&	405.33 	&	140.41 	&	$\ast$	&	0.43 	&	70 	&	462.96 	&	384.55 	&	133.21 	&	$\ast$	\\
57	&	AGAL331.709$+$00.582	&	0.44 	&	48	&	275.53 	&	293.09 	&	101.53 	&	$\ast$	&	0.16 	&	76	&	163.27 	&	173.12 	&	59.97 	&	$\ast$	&	0.85 	&	40 	&	125.29 	&	132.84 	&	46.02 	&	$\ast$	\\
58	&	AGAL332.094$-$00.421	&	0.38 	&	56	&	257.43 	&	32.01 	&	11.09 	&	$\ast$	&	0.18 	&	68	&	183.51 	&	22.74 	&	7.88 	&	$\ast$	&	0.76 	&	44 	&	185.02 	&	22.93 	&	7.94 	&	$\ast$	\\
59	&	AGAL332.826$-$00.549	&	0.46 	&	48	&	416.12 	&	51.74 	&	17.92 	&	$\ast$	&	0.26 	&	94	&	400.36 	&	49.62 	&	17.19 	&	$\ast$	&	0.55 	&	76 	&	438.98 	&	54.40 	&	18.85 	&	$\ast$	\\
60	&	AGAL333.134$-$00.431	&	0.47 	&	52	&	711.47 	&	88.46 	&	30.64 	&	$\ast$	&	0.15 	&	68	&	686.75 	&	85.11 	&	29.48 	&	$\ast$	&	0.79 	&	50 	&	738.38 	&	91.51 	&	31.70 	&	$\ast$	\\
61	&	AGAL333.284$-$00.387	&	0.29 	&	28	&	250.92 	&	31.20 	&	10.81 	&	$\ast$	&	0.19 	&	76	&	208.57 	&	25.85 	&	8.95 	&	$\ast$	&	0.76 	&	30 	&	193.79 	&	24.02 	&	8.32 	&	$\ast$	\\
62	&	AGAL333.314$+$00.106	&	0.37 	&	82	&	318.10 	&	39.55 	&	13.70 	&	$\ast$	&	0.18 	&	88	&	156.50 	&	19.40 	&	6.72 	&	$\ast$	&	0.83 	&	32 	&	112.52 	&	13.94 	&	4.83 	&	$\ast$	\\
63	&	AGAL333.604$-$00.212	&	0.37 	&	52	&	892.79 	&	111.00 	&	38.45 	&		&	0.18 	&	68	&	1018.59 	&	126.23 	&	43.73 	&		&	0.79 	&	50 	&	1109.47 	&	137.50 	&	47.63 	&		\\
64	&	AGAL333.656$+$00.059	&	0.29 	&	26	&	80.47 	&	21.60 	&	7.48 	&		&	0.15 	&	28	&	37.24 	&	9.96 	&	3.45 	&		&	0.87 	&	30 	&	31.73 	&	8.49 	&	2.94 	&		\\
65	&	AGAL335.789$+$00.174	&	0.28 	&	54	&	350.01 	&	45.23 	&	15.67 	&	$\ast$	&	0.15 	&	94	&	206.53 	&	26.60 	&	9.21 	&	$\ast$	&	0.82 	&	46 	&	189.92 	&	24.46 	&	8.47 	&	$\ast$	\\
66	&	AGAL336.958$-$00.224	&	0.35 	&	40	&	86.22 	&	98.45 	&	34.10 	&	$\ast$	&	0.23 	&	60	&	50.64 	&	57.63 	&	19.97 	&	$\ast$	&	0.86 	&	22 	&	28.95 	&	32.95 	&	11.42 	&		\\
67	&	AGAL337.176$-$00.032	&	0.37 	&	34	&	97.82 	&	113.55 	&	39.34 	&	$\ast$	&	0.18 	&	30	&	62.38 	&	72.18 	&	25.00 	&	$\ast$	&	0.70 	&	32 	&	61.29 	&	70.92 	&	24.57 	&	$\ast$	\\
68	&	AGAL337.258$-$00.101	&	0.51 	&	20	&	39.65 	&	8.08 	&	2.80 	&	$\ast$	&	0.20 	&	30	&	46.90 	&	54.26 	&	18.80 	&	$\ast$	&	0.93 	&	26 	&	44.80 	&	51.84 	&	17.96 	&	$\ast$	\\
69	&	AGAL337.286$+$00.007	&	0.38 	&	32	&	40.76 	&	34.84 	&	12.07 	&	$\ast$	&	0.24 	&	24	&	16.26 	&	13.85 	&	4.80 	&	$\ast$	&	0.97 	&	12 	&	9.80 	&	8.35 	&	2.89 	&	$\ast$	\\
70	&	AGAL337.406$-$00.402	&	0.58 	&	56	&	243.43 	&	24.82 	&	8.60 	&	$\ast$	&	0.58 	&	116	&	309.66 	&	31.47 	&	10.90 	&	$\ast$	&	1.31 	&	66 	&	256.30 	&	26.05 	&	9.02 	&	$\ast$	\\
71	&	AGAL337.704$-$00.054	&	0.43 	&	32	&	101.60 	&	146.50 	&	50.75 	&	$\ast$	&	0.27 	&	60	&	167.49 	&	240.73 	&	83.39 	&	$\ast$	&	0.97 	&	36 	&	168.65 	&	242.41 	&	83.97 	&	$\ast$	\\
72	&	AGAL337.916$-$00.477	&	0.97 	&	140	&	897.61 	&	88.18 	&	30.55 	&	$\ast$	&	0.42 	&	110	&	638.22 	&	62.49 	&	21.65 	&		&	0.70 	&	88 	&	607.40 	&	59.48 	&	20.60 	&		\\
73	&	AGAL338.066$+$00.044	&	0.44 	&	58	&	124.04 	&	26.18 	&	9.07 	&	$\ast$	&	0.22 	&	40	&	35.35 	&	7.44 	&	2.58 	&	$\ast$	&	1.15 	&	12 	&	15.34 	&	3.23 	&	1.12 	&	$\ast$	\\
74	&	AGAL338.786$+$00.476	&	0.33 	&	18	&	50.57 	&	9.78 	&	3.39 	&	$\ast$	&	0.16 	&	20	&	18.99 	&	3.66 	&	1.27 	&	$\ast$	&	0.84 	&	14 	&	17.82 	&	3.43 	&	1.19 	&	$\ast$	\\
75	&	AGAL338.926$+$00.554	&	0.49 	&	58	&	244.95 	&	45.50 	&	15.76 	&	$\ast$	&	0.20 	&	86	&	244.47 	&	45.26 	&	15.68 	&	$\ast$	&	1.00 	&	50 	&	242.78 	&	44.95 	&	15.57 	&	$\ast$	\\
76	&	AGAL339.623$-$00.122	&	0.45 	&	62	&	242.43 	&	21.07 	&	7.30 	&	$\ast$	&	0.25 	&	78	&	197.35 	&	17.10 	&	5.92 	&	$\ast$	&	0.89 	&	42 	&	147.88 	&	12.81 	&	4.44 	&	$\ast$	\\
77	&	AGAL340.374$-$00.391	&	0.35 	&	48	&	63.05 	&	7.84 	&	2.72 	&	$\ast$	&	0.20 	&	56	&	45.55 	&	5.64 	&	1.96 	&	$\ast$	&	0.94 	&	22 	&	40.20 	&	4.98 	&	1.73 	&	$\ast$	\\
78	&	AGAL340.746$-$01.001	&	0.29 	&	32	&	164.06 	&	11.99 	&	4.15 	&	$\ast$	&	0.16 	&	68	&	97.74 	&	7.12 	&	2.47 	&	$\ast$	&	1.22 	&	22 	&	71.16 	&	5.18 	&	1.80 	&	$\ast$	\\
79	&	AGAL340.784$-$00.097	&	0.35 	&	40	&	114.61 	&	109.73 	&	38.01 	&	$\ast$	&	0.17 	&	42	&	47.41 	&	45.25 	&	15.67 	&	$\ast$	&	0.74 	&	24 	&	38.09 	&	36.35 	&	12.59 	&	$\ast$	\\
80	&	AGAL341.217$-$00.212	&	0.36 	&	68	&	176.05 	&	22.75 	&	7.88 	&	$\ast$	&	0.17 	&	86	&	160.76 	&	20.71 	&	7.17 	&	$\ast$	&	0.91 	&	58 	&	159.15 	&	20.50 	&	7.10 	&	$\ast$	\\
81	&	AGAL342.484$+$00.182	&	0.43 	&	34	&	136.93 	&	206.90 	&	71.67 	&	$\ast$	&	0.14 	&	54	&	81.07 	&	122.10 	&	70.14 	&	$\ast$	&	0.70 	&	30 	&	64.75 	&	97.53 	&	56.02 	&	$\ast$	\\
82	&	AGAL343.128$-$00.062	&	0.37 	&	82	&	680.59 	&	60.34 	&	20.90 	&	$\ast$	&	0.24 	&	104	&	571.99 	&	50.55 	&	17.51 	&	$\ast$	&	1.20 	&	76 	&	552.46 	&	48.82 	&	16.91 	&	$\ast$	\\
83	&	AGAL343.756$-$00.164	&	0.51 	&	34	&	99.60 	&	8.04 	&	2.78 	&	$\ast$	&	0.19 	&	42	&	77.30 	&	6.22 	&	2.15 	&	$\ast$	&	1.10 	&	20 	&	91.39 	&	7.35 	&	2.55 	&	$\ast$	\\
84	&	AGAL344.227$-$00.569	&	0.52 	&	104	&	321.68 	&	19.60 	&	6.79 	&	$\ast$	&	0.27 	&	96	&	164.03 	&	9.96 	&	3.45 	&	$\ast$	&	0.91 	&	40 	&	137.89 	&	8.37 	&	2.90 	&	$\ast$	\\
85	&	AGAL345.003$-$00.224	&	0.55 	&	58	&	236.55 	&	20.70 	&	7.17 	&	$\ast$	&	0.94 	&	108	&	381.10 	&	33.24 	&	11.51 	&	$\ast$	&	1.27 	&	96 	&	454.00 	&	39.59 	&	13.72 	&	$\ast$	\\
86	&	AGAL345.488$+$00.314	&	0.34 	&	28	&	278.89 	&	13.19 	&	4.57 	&	$\ast$	&	0.24 	&	56	&	232.56 	&	10.96 	&	3.80 	&	$\ast$	&	1.13 	&	34 	&	232.16 	&	10.94 	&	3.79 	&	$\ast$	\\
87	&	AGAL345.504$+$00.347	&	0.42 	&	52	&	388.73 	&	18.88 	&	6.54 	&	$\ast$	&	0.31 	&	72	&	297.81 	&	14.42 	&	4.99 	&	$\ast$	&	1.03 	&	40 	&	251.73 	&	12.19 	&	4.22 	&	$\ast$	\\
88	&	AGAL345.718$+$00.817	&	0.37 	&	24	&	77.99 	&	1.82 	&	0.63 	&	$\ast$	&	0.20 	&	24	&	50.60 	&	1.18 	&	0.41 	&	$\ast$	&	0.96 	&	12 	&	34.86 	&	0.81 	&	0.28 	&	$\ast$	\\
89	&	AGAL351.131$+$00.771	&	0.31 	&	12	&	71.39 	&	2.27 	&	0.79 	&	$\ast$	&	0.17 	&	16	&	37.63 	&	1.19 	&	0.41 	&	$\ast$	&	0.91 	&	12 	&	27.49 	&	0.87 	&	0.30 	&	$\ast$	\\
90	&	AGAL351.161$+$00.697	&	0.38 	&	70	&	567.62 	&	18.04 	&	6.25 	&	$\ast$	&	0.17 	&	84	&	461.74 	&	14.63 	&	5.07 	&	$\ast$	&	0.62 	&	64 	&	468.44 	&	14.84 	&	5.14 	&	$\ast$	\\
91	&	AGAL351.244$+$00.669	&	0.34 	&	50	&	544.55 	&	17.30 	&	5.99 	&	$\ast$	&	0.22 	&	146	&	604.24 	&	19.14 	&	6.63 	&	$\ast$	&	0.81 	&	66 	&	641.90 	&	20.33 	&	7.04 	&	$\ast$	\\
92	&	AGAL351.416$+$00.646	&	1.02 	&	72	&	785.21 	&	13.53 	&	4.69 	&	$\ast$	&	0.99 	&	130	&	699.69 	&	12.01 	&	4.16 	&	$\ast$	&	0.93 	&	142 	&	706.14 	&	12.12 	&	4.20 	&	$\ast$	\\
93	&	AGAL351.444$+$00.659	&	0.55 	&	56	&	418.17 	&	7.20 	&	2.50 	&	$\ast$	&	0.20 	&	94	&	371.22 	&	6.37 	&	2.21 	&	$\ast$	&	1.25 	&	66 	&	408.63 	&	7.02 	&	2.43 	&	$\ast$	\\
94	&	AGAL351.571$+$00.762	&	0.41 	&	18	&	52.86 	&	0.91 	&	0.32 	&	$\ast$	&	0.30 	&	14	&	16.09 	&	0.28 	&	0.10 	&	$\ast$	&	2.85 	&	12 	&	25.21 	&	0.43 	&	0.15 	&	$\ast$	\\
95	&	AGAL351.581$-$00.352	&	0.55 	&	60	&	144.40 	&	64.24 	&	22.25 	&	$\ast$	&	0.35 	&	84	&	228.92 	&	101.52 	&	35.17 	&	$\ast$	&	1.07 	&	54 	&	206.39 	&	91.53 	&	31.71 	&	$\ast$	\\
96	&	AGAL351.774$-$00.537	&	0.85 	&	100	&	740.71 	&	7.11 	&	2.46 	&		&	0.88 	&	162	&	816.24 	&	7.81 	&	2.70 	&		&	1.37 	&	96 	&	777.98 	&	7.44 	&	2.58 	&		\\
97	&	AGAL353.066$+$00.452	&	0.58 	&	10	&	29.88 	&	0.21 	&	0.07 	&	$\ast$	&	0.19 	&	26	&	63.88 	&	0.45 	&	0.16 	&	$\ast$	&	1.39 	&	10 	&	28.05 	&	0.20 	&	0.07 	&	$\ast$	\\
98	&	AGAL353.417$-$00.079	&	0.63 	&	24	&	42.10 	&	14.83 	&	5.14 	&	$\ast$	&	0.22 	&	18	&	11.63 	&	4.08 	&	1.42 	&	$\ast$	&	1.84 	&	10 	&	16.82 	&	5.91 	&	2.05 	&	$\ast$	\\
99	&	AGAL354.944$-$00.537	&	0.62 	&	12	&	36.75 	&	1.29 	&	0.45 	&		&	0.23 	&	22	&	18.97 	&	0.66 	&	0.23 	&		&	1.42 	&	8 	&	10.72 	&	0.37 	&	0.13 	&		\\
\end{longtable}
\tablefoot{The columns are as follows: (1)-(2) ID and CSC name of the source; (3) rms of the CO\,(4--3) data (in K), (4) the full width at zero power (FWZP, in km\,s$^{-1}$); (5) integrated intensity of the line (\sint, in K\,km\,s$^{-1}$); (6)-(7) the line luminosity and its associated uncertainty (\lco and $\sigma_{\rm L}$, in K\,km\,s$^{-1}$\,pc$^{2}$) of the CO\,(4--3) spectra; (8) an asterisk mark indicates if the spectrum is contaminated by a self-absorption feature;	
the same properties computed for the CO\,(6--5) and CO\,(7--6) data are presented in columns (9)-(14) and (15)-(20), respectively.}
\end{longtab}
\setlength{\tabcolsep}{6pt}
\end{landscape} % A.6

\begin{longtab}	
\setcounter{table}{6}	
\setlength{\tabcolsep}{4pt}	
\captionsetup{width=0.9\textwidth}	
\begin{longtable}{cc|rr|rrrl}	
    \caption{\label{tbl_excitation_temperature}Excitation temperature derived from the CO\,(6--5) spectra convolved to 13\farcs4.}\\	
\hline\hline	
ID	&	CSC Name	&	$T_{\rm ex}$	&	$\sigma_T$	&	$T_{\rm ex,gauss}$	&	$\sigma_{T+}$	&	$\sigma_{T-}$	&	\\
\hline	
\endfirsthead	
\caption{continued.} \\	
\hline\hline	
ID	&	CSC Name	&	$T_{\rm ex}$	&	$\sigma_T$	&	$T_{\rm ex,gauss}$	&	$\sigma_{T+}$	&	$\sigma_{T-}$	&	\\
\hline	
\endhead	
\hline	
\endfoot	
1	&	AGAL008.684$-$00.367	&	28.0 	&	0.3 	&	43.2 	&	5.9 	&	5.2 	&	$\ast$	\\
2	&	AGAL008.706$-$00.414	&	14.2 	&	0.3 	&	13.9 	&	0.3 	&	0.3 	&	\\
3	&	AGAL010.444$-$00.017	&	17.5 	&	0.2 	&	39.9 	&	21.2 	&	13.9 	&	\\
4	&	AGAL010.472$+$00.027	&	52.3 	&	0.4 	&	97.0 	&	0.6 	&	0.6 	&	\\
5	&	AGAL010.624$-$00.384	&	84.9 	&	0.2 	&	131.5 	&	27.9 	&	28.0 	&	\\
6	&	AGAL012.804$-$00.199	&	97.1 	&	0.2 	&	143.0 	&	11.5 	&	12.4 	&	\\
7	&	AGAL013.178$+$00.059	&	31.0 	&	0.2 	&	32.1 	&	7.4 	&	7.7 	&	\\
8	&	AGAL013.658$-$00.599	&	29.7 	&	0.2 	&	30.8 	&	18.6 	&	16.2 	&	\\
9	&	AGAL014.114$-$00.574	&	24.4 	&	0.1 	&	35.3 	&	4.2 	&	3.8 	&	$\ast$	\\
10	&	AGAL014.194$-$00.194	&	34.0 	&	0.2 	&	37.1 	&	17.0 	&	18.6 	&	\\
11	&	AGAL014.492$-$00.139	&	20.5 	&	0.2 	&	22.8 	&	6.1 	&	6.8 	&	\\
12	&	AGAL014.632$-$00.577	&	38.1 	&	0.2 	&	41.6 	&	10.9 	&	11.3 	&	\\
13	&	AGAL015.029$-$00.669	&	146.9 	&	0.2 	&	67.2 	&	41.2 	&	25.5 	&	\\
14	&	AGAL018.606$-$00.074	&	26.1 	&	0.1 	&	33.8 	&	6.1 	&	6.3 	&	\\
15	&	AGAL018.734$-$00.226	&	29.4 	&	0.2 	&	29.2 	&	2.7 	&	2.8 	&	\\
16	&	AGAL018.888$-$00.474	&	36.8 	&	0.2 	&	36.8 	&	6.6 	&	6.8 	&	\\
17	&	AGAL019.882$-$00.534	&	49.7 	&	0.5 	&	50.0 	&	2.8 	&	2.8 	&	\\
18	&	AGAL022.376$+$00.447	&	20.4 	&	0.4 	&	29.0 	&	13.1 	&	15.2 	&	\\
19	&	AGAL023.206$-$00.377	&	32.4 	&	0.1 	&	34.9 	&	0.5 	&	0.5 	&	\\
20	&	AGAL024.629$+$00.172	&	15.8 	&	0.3 	&	15.8 	&	7.8 	&	11.3 	&	\\
21	&	AGAL028.564$-$00.236	&	15.3 	&	0.3 	&	17.7 	&	4.8 	&	5.8 	&	\\
22	&	AGAL028.861$+$00.066	&	46.1 	&	0.5 	&	45.4 	&	10.7 	&	11.0 	&	\\
23	&	AGAL030.818$-$00.056	&	44.1 	&	0.2 	&	66.7 	&	43.0 	&	45.2 	&	\\
24	&	AGAL030.848$-$00.081	&	25.8 	&	0.2 	&	30.4 	&	14.9 	&	10.0 	&	\\
25	&	AGAL030.893$+$00.139	&	18.8 	&	0.3 	&	18.5 	&	6.8 	&	9.1 	&	\\
26	&	AGAL031.412$+$00.307	&	42.9 	&	0.2 	&	61.5 	&	10.4 	&	10.5 	&	\\
27	&	AGAL034.258$+$00.154	&	87.4 	&	0.5 	&	59.9 	&	9.8 	&	8.4 	&	$\ast$	\\
28	&	AGAL034.401$+$00.226	&	52.5 	&	0.1 	&	59.5 	&	47.8 	&	23.8 	&	\\
29	&	AGAL034.411$+$00.234	&	34.0 	&	0.3 	&	35.7 	&	1.0 	&	1.0 	&	\\
30	&	AGAL034.821$+$00.351	&	29.3 	&	0.2 	&	32.4 	&	2.2 	&	2.2 	&	\\
31	&	AGAL035.197$-$00.742	&	65.3 	&	0.1 	&	84.1 	&	0.1 	&	0.1 	&	\\
32	&	AGAL037.554$+$00.201	&	28.3 	&	0.3 	&	32.8 	&	3.4 	&	3.5 	&	\\
33	&	AGAL043.166$+$00.011	&	84.9 	&	0.3 	&	135.3 	&	98.8 	&	57.1 	&	\\
34	&	AGAL049.489$-$00.389	&	58.3 	&	0.4 	&	56.3 	&	8.9 	&	7.7 	&	$\ast$	\\
35	&	AGAL053.141$+$00.069	&	49.2 	&	0.1 	&	61.7 	&	7.0 	&	7.0 	&	\\
36	&	AGAL059.782$+$00.066	&	51.7 	&	0.1 	&	57.4 	&	0.4 	&	0.4 	&	\\
37	&	AGAL301.136$-$00.226	&	59.7 	&	0.9 	&	73.6 	&	0.1 	&	0.1 	&	\\
38	&	AGAL305.192$-$00.006	&	33.5 	&	0.7 	&	40.6 	&	23.9 	&	26.9 	&	\\
39	&	AGAL305.209$+$00.206	&	69.7 	&	0.2 	&	68.9 	&	2.2 	&	2.2 	&	\\
40	&	AGAL305.562$+$00.014	&	74.5 	&	0.2 	&	74.8 	&	2.9 	&	3.0 	&	\\
41	&	AGAL305.794$-$00.096	&	15.9 	&	0.3 	&	15.5 	&	1.0 	&	1.0 	&	\\
42	&	AGAL309.384$-$00.134	&	32.1 	&	0.2 	&	35.1 	&	5.8 	&	5.9 	&	\\
43	&	AGAL310.014$+$00.387	&	30.4 	&	0.2 	&	30.0 	&	2.4 	&	2.4 	&	\\
44	&	AGAL313.576$+$00.324	&	35.5 	&	0.2 	&	37.7 	&	3.7 	&	3.8 	&	\\
45	&	AGAL316.641$-$00.087	&	21.0 	&	0.3 	&	20.4 	&	3.0 	&	3.3 	&	\\
46	&	AGAL317.867$-$00.151	&	28.2 	&	0.3 	&	29.2 	&	0.4 	&	0.4 	&	\\
47	&	AGAL318.779$-$00.137	&	20.6 	&	0.3 	&	22.3 	&	4.9 	&	5.3 	&	\\
48	&	AGAL320.881$-$00.397	&	24.0 	&	0.2 	&	24.5 	&	4.5 	&	4.8 	&	\\
49	&	AGAL326.661$+$00.519	&	89.7 	&	0.2 	&	89.7 	&	13.6 	&	13.7 	&	\\
50	&	AGAL326.987$-$00.032	&	23.8 	&	0.2 	&	27.0 	&	14.4 	&	19.1 	&	\\
51	&	AGAL327.119$+$00.509	&	28.2 	&	0.3 	&	33.8 	&	6.1 	&	6.3 	&	\\
52	&	AGAL327.393$+$00.199	&	28.8 	&	0.2 	&	29.9 	&	4.9 	&	5.1 	&	\\
53	&	AGAL329.029$-$00.206	&	36.0 	&	0.3 	&	40.3 	&	7.8 	&	7.9 	&	\\
54	&	AGAL329.066$-$00.307	&	28.8 	&	0.2 	&	29.5 	&	3.9 	&	4.0 	&	\\
55	&	AGAL330.879$-$00.367	&	64.9 	&	0.3 	&	75.0 	&	0.5 	&	0.5 	&	\\
56	&	AGAL330.954$-$00.182	&	74.8 	&	0.2 	&	94.9 	&	17.0 	&	17.1 	&	\\
57	&	AGAL331.709$+$00.582	&	37.3 	&	0.2 	&	51.3 	&	29.3 	&	18.6 	&	\\
58	&	AGAL332.094$-$00.421	&	46.5 	&	0.2 	&	35.3 	&	1.1 	&	1.1 	&	\\
59	&	AGAL332.826$-$00.549	&	85.2 	&	0.3 	&	73.6 	&	13.3 	&	11.2 	&	$\ast$	\\
60	&	AGAL333.134$-$00.431	&	91.3 	&	0.2 	&	76.8 	&	14.1 	&	11.9 	&	$\ast$	\\
61	&	AGAL333.284$-$00.387	&	67.4 	&	0.2 	&	72.8 	&	0.7 	&	0.7 	&	\\
62	&	AGAL333.314$+$00.106	&	35.8 	&	0.2 	&	38.1 	&	7.8 	&	8.0 	&	\\
63	&	AGAL333.604$-$00.212	&	140.5 	&	0.2 	&	136.3 	&	1.2 	&	1.2 	&	\\
64	&	AGAL333.656$+$00.059	&	21.9 	&	0.2 	&	22.4 	&	7.6 	&	8.9 	&	\\
65	&	AGAL335.789$+$00.174	&	40.3 	&	0.2 	&	49.3 	&	27.8 	&	29.8 	&	\\
66	&	AGAL336.958$-$00.224	&	21.7 	&	0.3 	&	20.9 	&	3.5 	&	3.7 	&	\\
67	&	AGAL337.176$-$00.032	&	30.0 	&	0.2 	&	29.9 	&	5.2 	&	5.4 	&	\\
68	&	AGAL337.258$-$00.101	&	22.4 	&	0.2 	&	21.8 	&	0.3 	&	0.3 	&	\\
69	&	AGAL337.286$+$00.007	&	15.2 	&	0.3 	&	14.2 	&	0.9 	&	1.0 	&	\\
70	&	AGAL337.406$-$00.402	&	61.6 	&	0.6 	&	95.4 	&	31.0 	&	31.3 	&	\\
71	&	AGAL337.704$-$00.054	&	41.7 	&	0.3 	&	45.5 	&	6.4 	&	5.6 	&	$\ast$	\\
72	&	AGAL337.916$-$00.477	&	68.9 	&	0.4 	&	76.4 	&	0.3 	&	0.3 	&	\\
73	&	AGAL338.066$+$00.044	&	17.4 	&	0.3 	&	18.6 	&	2.1 	&	2.3 	&	\\
74	&	AGAL338.786$+$00.476	&	17.1 	&	0.2 	&	17.4 	&	0.3 	&	0.3 	&	\\
75	&	AGAL338.926$+$00.554	&	51.4 	&	0.2 	&	62.3 	&	37.5 	&	23.4 	&	\\
76	&	AGAL339.623$-$00.122	&	42.5 	&	0.3 	&	42.6 	&	12.2 	&	12.6 	&	\\
77	&	AGAL340.374$-$00.391	&	17.1 	&	0.3 	&	18.3 	&	4.7 	&	5.6 	&	\\
78	&	AGAL340.746$-$01.001	&	32.9 	&	0.2 	&	34.1 	&	17.6 	&	20.1 	&	\\
79	&	AGAL340.784$-$00.097	&	21.4 	&	0.2 	&	23.5 	&	2.5 	&	2.6 	&	\\
80	&	AGAL341.217$-$00.212	&	42.8 	&	0.2 	&	51.9 	&	11.2 	&	11.4 	&	\\
81	&	AGAL342.484$+$00.182	&	27.4 	&	0.2 	&	27.1 	&	1.1 	&	1.1 	&	\\
82	&	AGAL343.128$-$00.062	&	68.0 	&	0.2 	&	75.3 	&	0.6 	&	0.6 	&	\\
83	&	AGAL343.756$-$00.164	&	31.5 	&	0.2 	&	34.1 	&	14.1 	&	15.3 	&	\\
84	&	AGAL344.227$-$00.569	&	34.1 	&	0.3 	&	37.4 	&	2.2 	&	2.3 	&	\\
85	&	AGAL345.003$-$00.224	&	63.4 	&	1.0 	&	86.9 	&	30.5 	&	30.8 	&	\\
86	&	AGAL345.488$+$00.314	&	59.6 	&	0.3 	&	61.6 	&	10.2 	&	8.7 	&	$\ast$	\\
87	&	AGAL345.504$+$00.347	&	63.4 	&	0.3 	&	67.0 	&	145.2 	&	141.7 	&	\\
88	&	AGAL345.718$+$00.817	&	26.3 	&	0.2 	&	27.3 	&	3.1 	&	3.2 	&	\\
89	&	AGAL351.131$+$00.771	&	26.5 	&	0.2 	&	30.6 	&	0.2 	&	0.2 	&	\\
90	&	AGAL351.161$+$00.697	&	89.9 	&	0.2 	&	99.1 	&	0.1 	&	0.1 	&	\\
91	&	AGAL351.244$+$00.669	&	112.8 	&	0.2 	&	126.5 	&	16.3 	&	16.3 	&	\\
92	&	AGAL351.416$+$00.646	&	84.5 	&	1.0 	&	105.7 	&	19.2 	&	19.3 	&	\\
93	&	AGAL351.444$+$00.659	&	62.2 	&	0.2 	&	81.2 	&	45.1 	&	46.1 	&	\\
94	&	AGAL351.571$+$00.762	&	19.7 	&	0.4 	&	19.6 	&	0.5 	&	0.5 	&	\\
95	&	AGAL351.581$-$00.352	&	40.4 	&	0.4 	&	49.0 	&	4.3 	&	4.3 	&	\\
96	&	AGAL351.774$-$00.537	&	74.1 	&	0.9 	&	88.0 	&	9.1 	&	9.1 	&	\\
97	&	AGAL353.066$+$00.452	&	28.7 	&	0.2 	&	13.2 	&	4.9 	&	3.6 	&	\\
98	&	AGAL353.417$-$00.079	&	15.5 	&	0.3 	&	32.9 	&	16.5 	&	11.0 	&	\\
99	&	AGAL354.944$-$00.537	&	16.6 	&	0.3 	&	20.6 	&	8.9 	&	6.2 	&	\\
\end{longtable}	
\tablefoot{The columns are as follows: (1)-(2) ID and CSC name of the source (given in Table\,\ref{tbl_observations_short}); (3)-(4) excitation temperature (in K) and its uncertainty derived from the peak intensity of the CO\,(6--5) spectra; (5)-(7) excitation temperature (in K) and its upper and lower uncertainty derived from the peak intensity of the Gaussian fit of the CO\,(6--5) spectra; (8) an asterisks indicates the cases where the Gaussian fit is dubious and, therefore, the excitation temperature was obtained from the relation ($\log(T_{\rm ex})=(0.75\pm0.10)+(0.21\pm0.02)\log(L_{\rm bol})$ (see Sect.\,\ref{sec_texc} for further details).}	
\end{longtab}	
\setlength{\tabcolsep}{6pt}	
		% A.7

  \clearpage

\section{CO spectra and CO\,(6--5) maps}
\label{appendix_co_fixbeam}

In Fig.\,\ref{fig_fixbeam_gaussian_fit} we show the integrated CO\,(6--5) maps towards the \atgtop sample together with the CO\,(4--3) spectra from \flash observations, the convolved \champ mid-$J$ CO spectra using a fixed beam size of 13\farcs4 and the isotopologue C$^{17}$O\,(3--2) or C$^{18}$O\,(2--1) spectra from \citet{Giannetti14}. The CO profiles that are not overlaid by the Gaussian fit correspond to those that were not properly fitted.

\begin{figure*}[ht!]
    \centering
\includegraphics[height=0.33\linewidth]{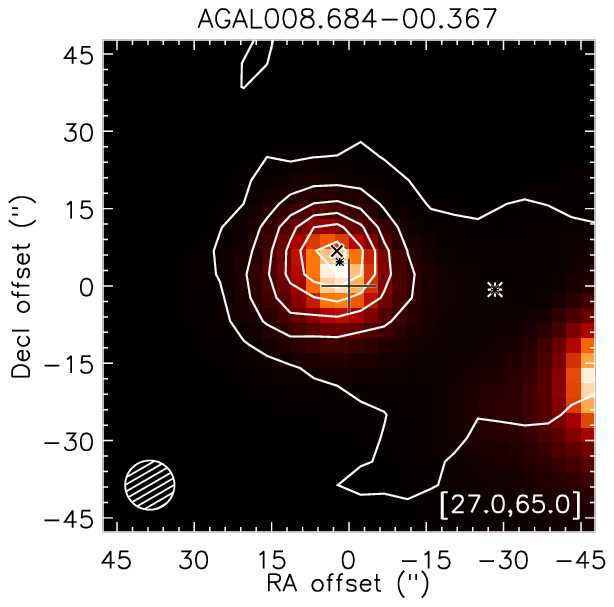}
\includegraphics[height=0.33\linewidth]{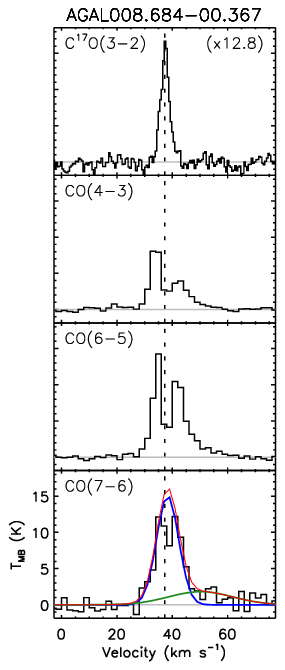}
\includegraphics[height=0.33\linewidth]{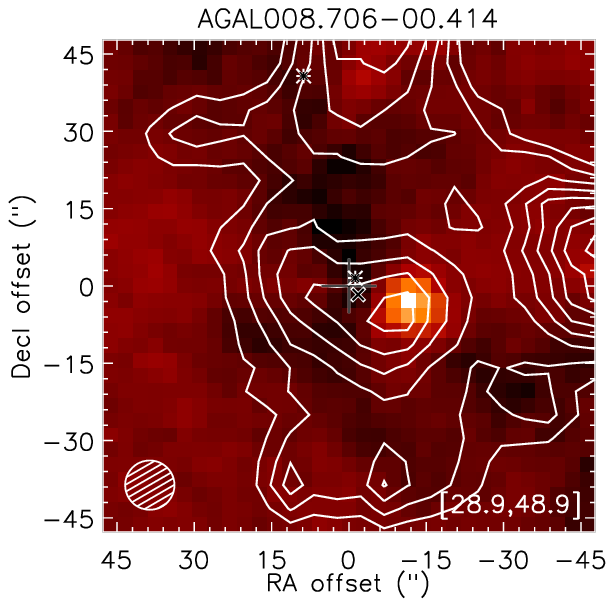}
\includegraphics[height=0.33\linewidth]{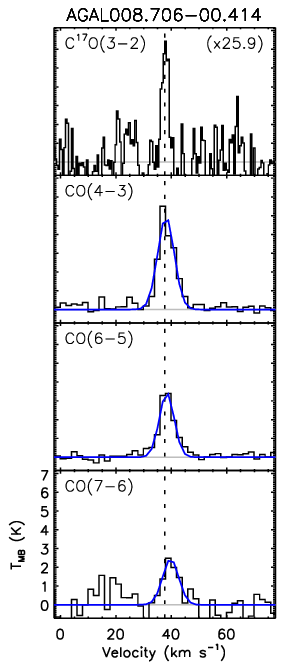}

\includegraphics[height=0.33\linewidth]{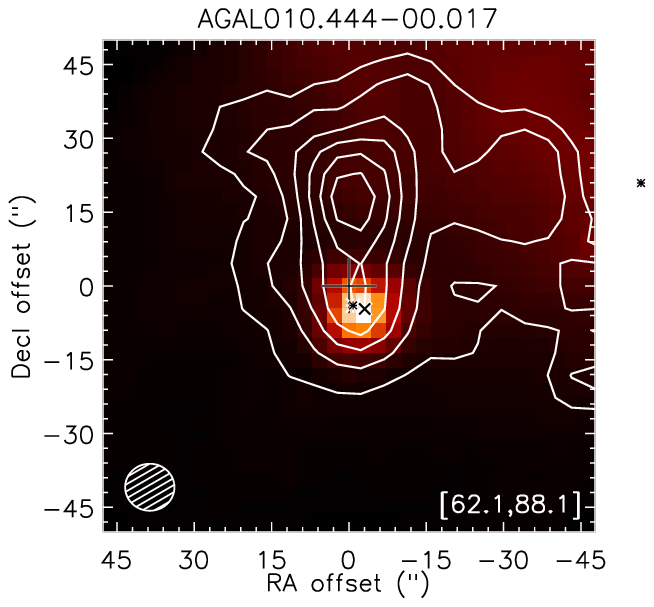}
\includegraphics[height=0.33\linewidth]{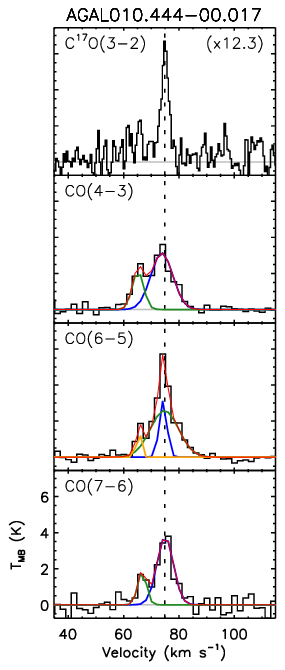}
\includegraphics[height=0.33\linewidth]{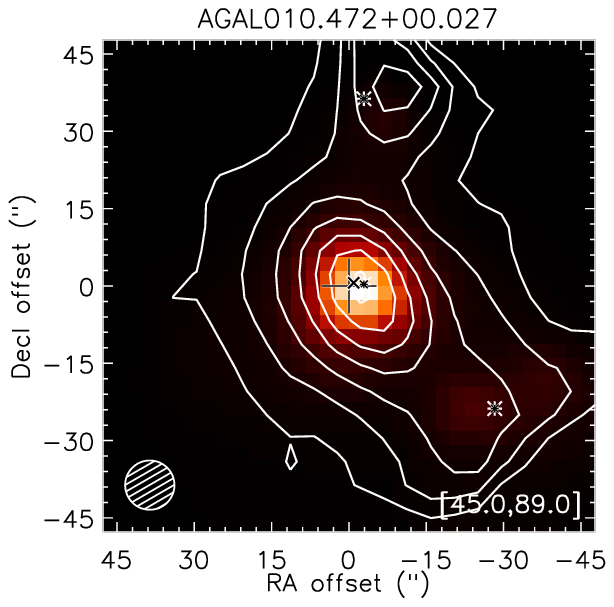}
\includegraphics[height=0.33\linewidth]{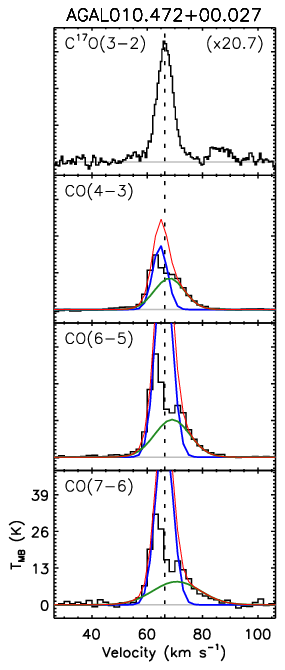}
    \caption{Left: False-colour {\it Herschel}/PACS images at 70\,$\mu$m overlaid by the CO\,(6--5) emission contours towards the \atgtop source.
    The CO contours correspond to the emission integrated over the full-width at zero power (FWZP) of the CO\,(6--5) profile, the velocity range shown in the bottom right side of the image, and the contour levels are shown from 20\% to 90\% of the peak emission of each map, in steps of 10\%.
    The (0,0) position of the map is shown as a $+$ symbol, the position of the CSC source from \citet{Contreras13} is shown as a $\times$ symbol and the dust continuum emission peaks from \citet{Csengeri14} are shown as asterisks.
    Right, from top to bottom panel: C$^{17}$O\,(2--1) or C$^{18}$O\,(1--0) from \citet{Giannetti14}, CO\,(4--3), CO\,(6--5) and CO\,(7--6) profiles towards the \atgtop sample, convolved into a fixed beam size of 13\farcs4\xspace (shown in black) and their fitted Gaussian components.
    The narrower Gaussian component fitted to the data is shown in blue, the second and broader component is shown in green and the third component is shown in yellow. The sum of all Gaussian components is shown in red, except for the cases where a single component was fitted.
    The vertical dashed black line is placed at the rest velocity (\vlsr) of each source.
    The horizontal filled grey line displays the baseline of the data.
    \textit{ArXiV only:} the full Fig.\,\ref{fig_fixbeam_gaussian_fit} is available in the A\&A version.}
    \label{fig_fixbeam_gaussian_fit}
\end{figure*}

\section{Additional material}
\label{appendix_additional_plots}

\subsection{Analysis of the mid-$J$ CO emission in the distance-limited sub-sample}
\label{appendix_distlim}

For completeness, in this section we present the analysis of the  CO emission for the distance-limited sample (defined in Sect.\,\ref{sec_CO_convolution}).

    Figure\,\ref{fig_avgspc_fixscale} shows the average CO spectra per class integrated over a linear scale ($\sim$\,0.24\,pc) for the distance-limited sub-sample.
    When compared to Fig.\,\ref{fig_avgspc_fixbeam}, which shows the same kind of spectra but for the full sample (using the spectra convolved to 13\farcs4, see Sect.\,\ref{sec_CO_convolution}), we found that the \irq and \irb classes are much better separated in the distance-limited sample than in the full dataset smoothed to 13\farcs4. In fact, the \irq and \irb gets less distinguishable when including the outlier \irq clumps located at $d$\,$\geq$\,12\,kpc (AGAL018.606$-$00.074, AGAL018.734$-$00.226, AGAL342.484+00.182).

\begin{figure*}
 \centering
 \includegraphics[width=0.425\linewidth]{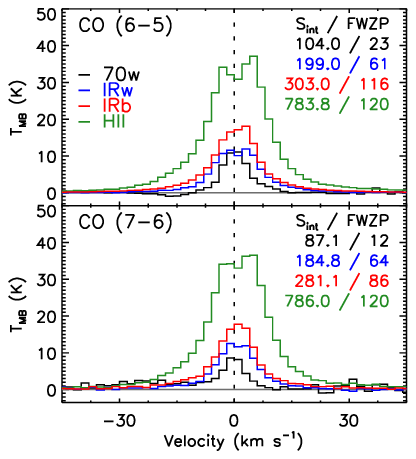}
 \includegraphics[width=0.425\linewidth]{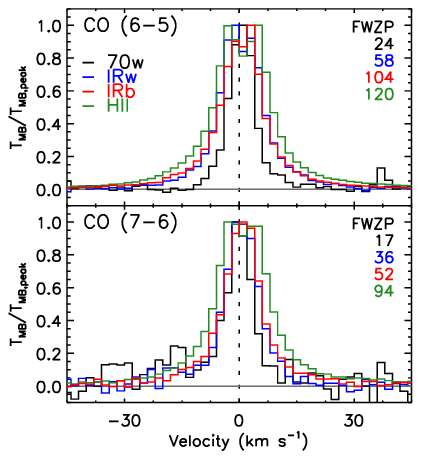} \\[-0.5ex]
 \caption{Left: Average CO\,(6--5) and CO\,(7--6) spectra of each ATLASGAL class scaled to the median distance of the distance-limited sub-sample ($d$\,=\,3.26\,kpc). Right: Same plot, but the average CO spectra were normalised by their peak intensity.
 The baseline level is indicated by the solid grey line. The black dashed line marks a velocity of 0\,\kms.
 The FWZP of the profiles are shown in the upper right side of the panels (in \kms units), together with the integrated intensity (\sint, in K\,\kms units) of the CO profiles shown in the left panels.}
 \label{fig_avgspc_fixscale}
\end{figure*}

    Figure\,\ref{fig_fixscl_cdf} presents the CDF for  CO\,(6--5) and CO\,(7--6) line luminosity. \lco  ranges from 2.7 to 284.0\,K\,km\,s$^{-1}$\,pc$^2$ for the CO\,(6--5) transition, and 1.2 to 276.5\,K\,km\,s$^{-1}$\,pc$^2$ for the CO\,(7--6) line.
    The KS tests indicate that the evolutionary classes are relatively more distinguishable based on the distance-limited sub-sample than on the full sample (excluding the comparison between \irq and \irb, all tests indicated ranks KS\,$\geq$\,0.51, with $p$\,$\leq$\,0.001 for both $J$ transitions, see Table\,\ref{table_lco_classes_linear}).

\begin{table}
 \centering
\setlength{\tabcolsep}{5pt}	
 \caption{\label{table_lco_classes_linear}Kolmogorov-Smirnov statistics of the mid-$J$ CO line luminosity for the distance-limited sub-sample as a function of the evolutionary class of the clumps.}
\begin{tabular}{c|ccccccc}
 \hline
 \hline
Transition	& \fiq-\irq 			&	\fiq-\irb 			& \fiq-\hii & \irq-\irb & \irq-\hii & \irb-\hii \\
\hline
CO\,(6--5)	&	0.60, $p$\,$<$\,0.001&0.80, $p$\,$<$\,0.001	& 0.95, $p$\,$<$\,0.001	&0.31, $p$\,=\,0.08		& 0.71, $p$\,$<$=0.001	&	0.51, $p$\,=\,0.001		\\
CO\,(7--6)	&	0.73, $p$\,$<$\,0.001&0.80, $p$\,$<$\,0.001	& 1.00, $p$\,$<$\,0.001	&0.34, $p$\,=\,0.04	&	0.68, $p$\,$<$=0.001	& 0.52, $p$\,=\,0.001\\
\hline 
 \end{tabular}
\tablefoot{The rank KS and its corresponding probability ($p$) are shown for each comparison. A $p$-value of $<$\,0.001 indicate a correlation at 0.001 significance level. $p$-values of 0.05, 0.002 and $<$\,0.001 represent the $\sim$\,2, 3 and $>$\,3\,$\sigma$ confidence levels.}
\end{table}
\setlength{\tabcolsep}{6pt}	

\begin{figure}
 \centering
 \includegraphics[width=0.485\linewidth]{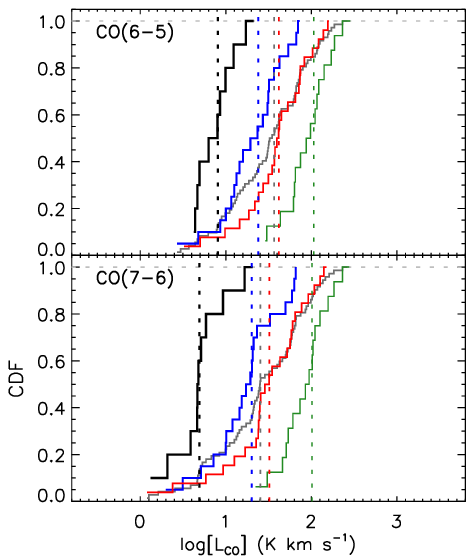}
 \includegraphics[width=0.485\linewidth]{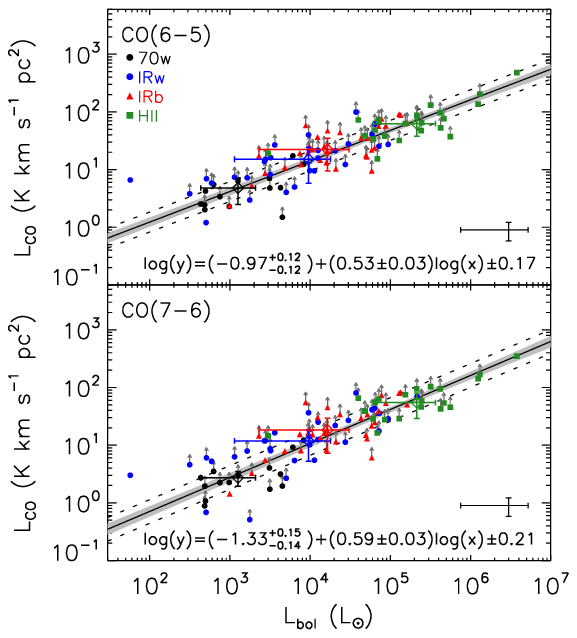}
 \caption{Left panels: Cumulative distribution function (CDF) of the CO line luminosity derived using the spectra convolved to the same linear scale of 0.24\,pc. The median values per class are shown as vertical dashed lines in their corresponding colours. 
 Right: Line luminosity of the same CO lines versus the bolometric luminosity of the \atgtop clumps in distance-limited sub-sample.  The median values for each class are shown as open diamonds and their error bars correspond to the absolute deviation of the data from their median value.
 Points having an upward arrow indicate a self-absorption feature in the spectrum and correspond to a lower limit.
 The typical error bars are shown in the bottom right side of the plots. The black solid line is the best fit, the light grey shaded area indicates the 68\% uncertainty, and the dashed lines show the intrinsic scatter ($\epsilon$) of the relation.}
 \label{fig_fixscl_cdf}
\end{figure}

We also looked at the CO\,(6--5) and CO\,(7--6) line luminosities as function of the bolometric luminosity of the sources (see right panel of Fig.\,\ref{fig_fixscl_cdf}), their mass and their  luminosity-to-mass ratio (See Fig.\,\ref{fig_fixscl_cdf_linscl_full}).   Table\,\ref{table_fixscl_co_correlation} lists the Spearman correlation factor, $\rho$, and its associated probability, $p$, for the CO line luminosity versus the bolometric luminosity, the clump mass and the luminosity-to-mass ratio of the clumps.
    For \lbol and \mclump, the partial Spearman rank, excluding the dependency on the distance, is also provided. The $\rho$ values are similar to those reported on Table\,\ref{table_co_correlation} for the correlation between \lco and \lbol, based on the 13\farcs4 dataset, indicating no significant improvement on the correlation between these quantities.
    The correlation with \mclump, however, is weaker ($\rho$\,$\leq$\,0.44, $p$\,$<$\,0.001) than the one found towards the 13\farcs4 dataset ($\rho$\,$\geq$\,0.72, $p$\,$<$\,0.001, see Table\,\ref{table_co_correlation}). The weaker correlation between \lco and \mclump on the distance-limited sub-sample might arise from the fact that the CO line luminosity is integrated over only a fraction of the beam used for estimating the mass of the clumps. Indeed \citet{Koenig15} used a minimum aperture size of 55\farcs1 for their study, while the minimum beam size adopted for the convolution of the CO was about 10\arcsec (see Sect.\,\ref{sec_CO_convolution}).
    
    We also found that \lco is relatively better correlated with \lmratio for the distance-limited dataset ($\rho$\,$\geq$\,0.67, $p$\,$<$\,0.001 for all lines, see Table\,\ref{table_fixscl_co_correlation}) rather than the 13\farcs4 spectra ($\rho$\,$\leq$\,0.50, $p$\,$\leq$\,0.003 for the mid-$J$ CO lines, see Table\,\ref{table_co_correlation}).
    
   We compared the best fits obtained for the mid-$J$ CO line luminosity convolved to the same linear scale with those derived using the 13\farcs4 data. Table\,\ref{table_fixscl_lco_fit} reports the coefficients of the individual fits.
    We find that \lco vs. \lbol follows a power-law distribution with indices of 0.53$\pm$0.03 and 0.59$\pm$0.03 for the CO\,(6--5) and CO\,(7--6) lines, respectively.
    Such power-law distributions are relatively less steeper than those derived towards the 13\farcs4 dataset, with indices of 0.61$\pm$0.03 and 0.67$\pm$0.03 for the same $J$ transitions (see Table\,\ref{table_lco_fit}).
    The offset of the fits indicates the brightness of the CO emission is roughly 0.4\,dex larger than the \lco values derived using the spectra convolved to a 13\farcs4 beam.
    At least for the closest sources, such an increment in \lco is expected due to the larger size of the beam corresponding to 0.24\,pc.
For example, at $d$\,=\,1.85\,kpc, the linear scale of 0.24\,pc corresponds to a beam of 26\farcs8, which is sampling an area 4 times larger than the 13\farcs4 dataset.

\begin{figure*}[!ht]
\centering
 \includegraphics[width=0.485\linewidth]{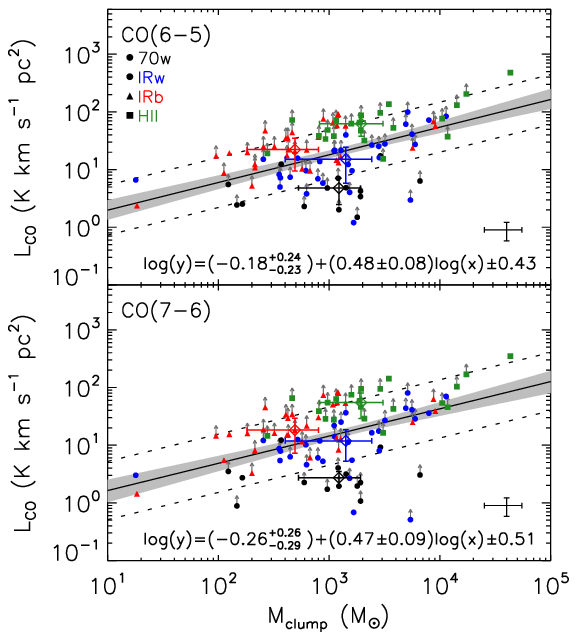}
 \includegraphics[width=0.485\linewidth]{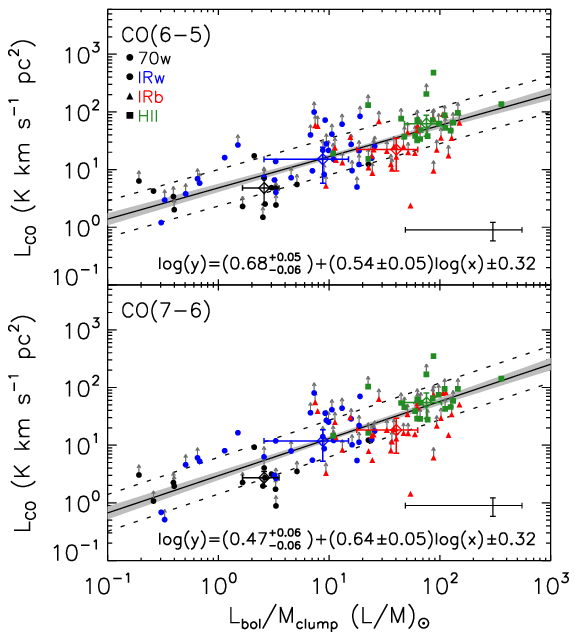} \\[-1.0ex]
\caption{The CO line luminosity derived using the spectra convolved to the same linear scale of 0.24\,pc are shown versus their masses (left panels) and their luminosity-to-mass ratios (right). For a complete description of the plots, see Fig.\,\ref{fig_fixscl_cdf}.}
 \label{fig_fixscl_cdf_linscl_full}
\end{figure*}

\begin{table}[!ht]
\caption{\label{table_fixscl_co_correlation}Spearman rank correlation statistics for the CO line luminosity as a function of the clump properties towards the distance-limited sub-sample.}
\centering
\setlength{\tabcolsep}{4pt}
\begin{tabular}{l|cc}
	\hline\hline
Property				&	CO\,(6--5)				&	CO\,(7--6)				\\
\hline								
\multirow{2}{*}{\lbol}	&	0.86, $p$\,$<$\,0.001; 	&	0.87, $p$\,$<$\,0.001;	\\
						&	$\rho_p$\,=\,0.91		&	 $\rho_p$\,=\,0.88		\\
\multirow{2}{*}{\mclump}&	0.44, $p$\,$<$\,0.001;	&	0.41, $p$\,$<$\,0.001; 	\\
						&	$\rho_p$\,=\,0.85		&	 $\rho_p$\,=\,0.81		\\
\lmratio				&   0.67, $p$\,$<$\,0.001	&   0.79, $p$\,$<$\,0.001	\\
\hline
\end{tabular}
\tablefoot{The rank $\rho$ and its corresponding probability ($p$) are shown for each comparison. A $p$-value of $<$\,0.001 indicate a correlation at 0.001 significance level. $p$-values of 0.05, 0.002 and $<$\,0.001 represent the $\sim$\,2, 3 and $>$\,3\,$\sigma$ confidence levels. For \lbol and \mclump, the partial correlation coefficient, $\rho_p$, is also shown.}
\end{table}
\setlength{\tabcolsep}{6pt}

\begin{table}[!ht]
\caption{\label{table_fixscl_lco_fit}Parameters of the fits of \lco, extracted within a common linear scale, as a function of the clump properties.}
\centering
\setlength{\tabcolsep}{4pt}
\begin{tabular}{cl|rcc}
	\hline\hline
Transition	&	Property	&	\multicolumn{1}{c}{$\alpha$}	&	$\beta$	&	$\epsilon$	\\ 
\hline
			&	\lbol		&	$-$0.97$^{+0.12}_{-0.12}$	&	0.53$\pm$0.03	&	0.17	\\
CO\,(6--5)	&	\mclump		&	$-$0.18$^{+0.24}_{-0.23}$	&	0.48$\pm$0.08	&	0.43	\\
			&	\lmratio	&	   0.68$^{+0.05}_{-0.06}$	&	0.54$\pm$0.05	&	0.32	\\
\hline
			&	\lbol		&	$-$1.33$^{+0.15}_{-0.14}$	&	0.59$\pm$0.03	&	0.21	\\
CO\,(7--6)	&	\mclump		&	$-$0.26$^{+0.26}_{-0.29}$	&	0.47$\pm$0.08	&	0.51	\\
			&	\lmratio	&	   0.47$^{+0.06}_{-0.06}$	&	0.64$\pm$0.05	&	0.32	\\      
\hline
\end{tabular}
\tablefoot{The fits were performed by adjusting a model with three free parameters in the form of $\log(y) = \alpha + \beta \log(x) \pm \epsilon$, where $\alpha$, $\beta$ and $\epsilon$ correspond to  the intercept, the slope and the intrinsic scatter, respectively.}
\end{table}
\setlength{\tabcolsep}{6pt}

    Finally, we further investigated the effects of beam dilution on our results by integrating the CO emission over the full maps for those clumps where the aperture used by \citet{Koenig15} to derive the bolometric luminosity of the clumps, $\Delta\theta_{\rm{ap}}$, was larger or equal to the CO emission extension ($\Delta\theta$\,$\leq$\,$\Delta\theta_{\rm ap}$). Such criterion was satisfied for 92 of the 99 clumps.
    The best fit of the data indicates that \lco increases with \lbol with a power-law index of 0.59\,$\pm$\,0.03 and 0.68\,$^{+0.05}_{-0.06}$ for the CO\,(6--5) and CO\,(7--6) transitions, consistently with the results based on the 13\farcs4 dataset (see Fig.\,\ref{fig_lco_correlation_fixbeam_lbol}).
    Similar results are also found when comparing \lco versus \mclump and \lmratio.
    The overall results  suggests that the analysis of the mid-$J$ CO emission is robust in terms of beam dilution effects.

\begin{table}[!ht]
	\centering
	\caption{\label{table:kstest_integratedCO_fixbeam} Spearman statistics for the CO line luminosity as a function of the clump properties using the spectra integrated on the whole CHAMP$^+$ maps.}
	\begin{tabular}{l|ccc}
\hline\hline
Transition	&	\lbol	&	\mclump	&	\lmratio	\\
\hline
CO\,(6--5)	&	0.84, $p$\,$<$\,0.001	&	0.72, $p$\,$<$\,0.001	&	0.43, $p$\,$<$\,0.001	\\
			&	$\rho_p$\,=\,0.96		&	 $\rho_p$\,=\,0.93		&	\\
CO\,(7--6)	&	0.82, $p$\,$<$\,0.001	&	0.59, $p$\,$<$\,0.001	&	0.49, $p$\,$<$\,0.001	\\
			&	$\rho_p$\,=\,0.97		&	 $\rho_p$\,=\,0.94		&	\\
\hline
        \end{tabular}
        \tablefoot{The rank $\rho$ and it corresponding probability ($p$) are shown for each comparison. A $p$-value of $<$0.001 indicate a correlation at 0.001 significance level. $p$-values of 0.05, 0.002 and $<$0.001 represent the $\sim$2, 3 and $>$3$\sigma$ confidence levels.}
\end{table}

\subsection{Analysis of the CO emission using the Gaussian profiles}
\label{appendix_gaussfit}

We further investigated the effects of self-absorption by computing the CO line luminosities using the integrated flux over the Gaussian fit of the CO profiles (see Sect\,\ref{sec_gaussfit}). Then, we compared the Gaussian CO luminosities with the clump properties and compared the results with those reported in Sect.\,\ref{sec_CO_correlations}.

First, we checked the correlation between the Gaussian \lco values and the clump properties by means of their Spearman rank correlation factor. The results are summarised in Table\,\ref{table_co_gauus_correlation}.
    when compared to the Spearman factors of the observed CO line luminosity against the clump properties (see Table\,\ref{table_co_correlation}), we found a slightly improvement on the correlation between \lco and the bolometric luminosity of the clumps (e.g. for the CO\,(6--5) line, the correlation slightly improves from $\rho$\,=\,0.85 to 0.88), and with their \lmratio ratio (e.g. from $\rho$\,=\,0.46 to 0.49 for the same transition).
    No significant changes in the correlation between \lco and \mclump were found (e.g. from $\rho$\,=\,0.72 to 0.74 for the CO\,(6--5) line), indicating that the observed correlation is likely dependent on the distance rather than the mass of the clumps. 

\begin{table}[h!]
\caption{\label{table_co_gauus_correlation}Spearman rank correlation statistics for the CO line luminosity as a function of the clump properties towards the \atgtop sample for the observed and the Gaussian CO luminosities.}
\centering
\begin{tabular}{l|ccc}			
	\hline\hline
\multirow{2}{*}{Property}	&	\multicolumn{3}{c}{Observed $\rho$ / Gaussian $\rho$}	\\
	&	CO\,(4--3)	&	CO\,(6--5)	&	CO\,(7--6)	\\
\hline
\lbol		&	0.71 / 0.78 &	0.85 / 0.88 &	0.89 / 0.90	\\
\mclump		&	0.75 / 0.77	&	0.72 / 0.73	&	0.69 / 0.69	\\
\lmratio	&   0.29 / 0.34	&   0.46 / 0.49	&   0.50 / 0.54	\\
\hline
\end{tabular}
\tablefoot{The rank $\rho$ are shown for each comparison between the observed and Gaussian CO luminosities ratio (Observed/Gaussian). The corresponding probability ($p$) of most comparisons are $p$\,$<$\,0.001, except for the comparison between the CO\,(4--3) luminosity vs \lmratio ($p$\,=\,0.09). A $p$-value of $<$\,0.001 indicate a correlation at 0.001 significance level. $p$-values of 0.05, 0.002 and $<$\,0.001 represent the $\sim$\,2, 3 and $>$\,3\,$\sigma$ confidence levels. }
\end{table}
\setlength{\tabcolsep}{6pt}

Figure\,\ref{fig_lco_correlation_gauss} presents the distribution of the Gaussian CO line luminosities as a function of the clump properties. The parameters of the fits are summarised in Table\,\ref{table_lco_gauss_fit}.
Although the distribution of the points indicates higher correlation with \lbol and \lmratio, the steepness of the relations are consistent with those reported in Sect.\,\ref{sec_CO_correlations}. For example, the slope of the best fit of \lco against \lbol is 0.63$\pm$0.04 for the observed \lco values, and 0.62$\pm$0.04 for the Gaussian CO luminosities, respectively. These findings suggests that the relations between the CO line luminosities and the clump properties are robust in terms of the self-absorption observed in the CO spectra of the \atgtop.

\begin{figure*}[!]
 \centering

  \includegraphics[width=0.33\linewidth]{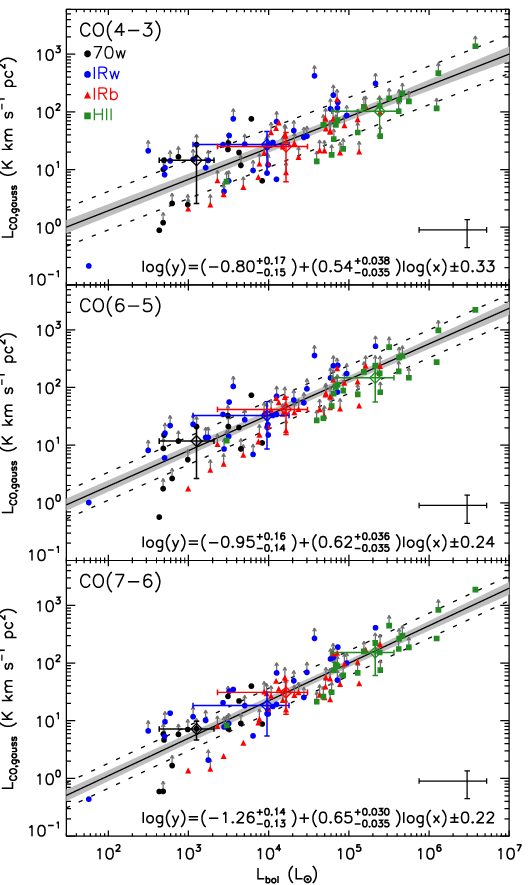}
  \includegraphics[width=0.33\linewidth]{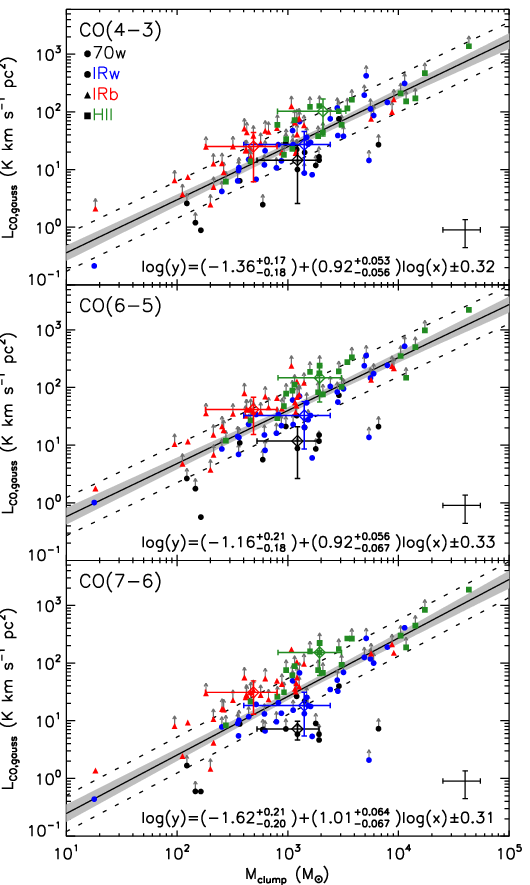}
  \includegraphics[width=0.33\linewidth]{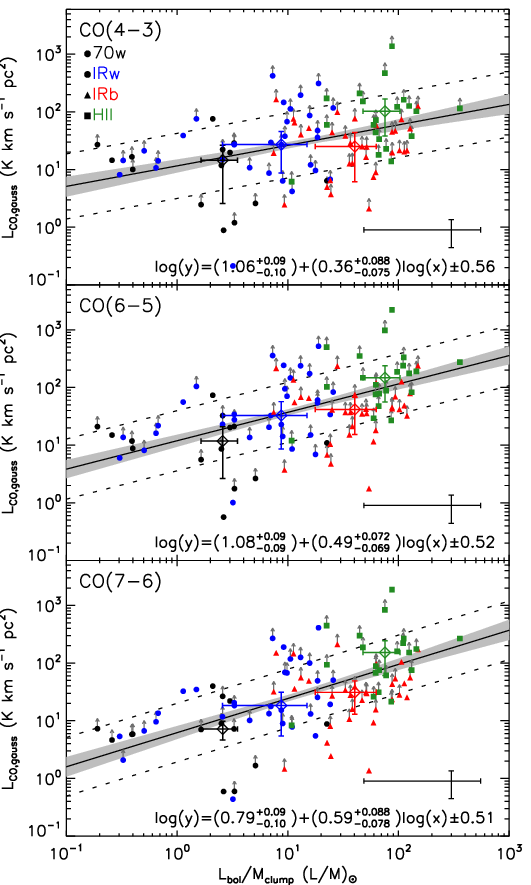}  \\[-1.5ex]
 \caption{Line luminosity of the CO\,(4--3) (upper panels), CO\,(6--5) (middle) and CO\,(7--6) emission (bottom) versus the bolometric luminosity (left panels), the mass of the clumps (middle) and the \lmratio ratio (right) of the \atgtop sources.
 The median values for each class are shown as open diamonds and their error bars correspond to the absolute deviation of the data from their median value.
 Points having an upward arrow indicate a self-absorption feature in the spectrum convolved to 13\farcs4 and correspond to a lower limit.
 The typical error bars are shown at the bottom right side of the plots. The black solid line is the best fit, the light grey shaded area indicates the 68\% uncertainty, and the dashed lines show the intrinsic scatter ($\epsilon$) of the relation.}
 \label{fig_lco_correlation_gauss}
\end{figure*}

\begin{table}[h!]
\caption{\label{table_lco_gauss_fit}Parameters of the fits of \lco as a function of the clump properties for the Gaussian fluxes.}
\centering
\begin{tabular}{cl|ccc}
	\hline\hline
Transition	&	Property	&	$\alpha$	&	$\beta$	&	$\epsilon$	\\
\hline
			&	\lbol	&	$-$0.80$^{+0.17}_{-0.15}$	&	0.54$\pm$0.04	&	0.33	\\
CO\,(4--3)	&	\mclump	&	$-$1.36$^{+0.17}_{-0.18}$	&	0.92$\pm$0.06	&	0.32	\\
			&	\lmratio&	   1.06$^{+0.09}_{-0.10}$	&	0.36$\pm$0.09	&	0.56	\\
\hline
			&	\lbol	&	$-$0.95$^{+0.16}_{-0.14}$	&	0.62$\pm$0.04	&	0.24	\\
CO\,(6--5)	&	\mclump	&	$-$1.16$^{+0.21}_{-0.18}$	&	0.92$\pm$0.07	&	0.33	\\
			&	\lmratio&	   1.08$^{+0.09}_{-0.09}$	&	0.49$\pm$0.07	&	0.52	\\
\hline
			&	\lbol	&	$-$1.26$^{+0.14}_{-0.13}$	&	0.65$\pm$0.03	&	0.22	\\
CO\,(7--6)	&	\mclump	&	$-$1.62$^{+0.21}_{-0.20}$	&	1.01$\pm$0.07	&	0.31	\\
			&	\lmratio&	   0.79$^{+0.09}_{-0.10}$	&	0.59$\pm$0.09	&	0.51	\\
\hline
\end{tabular}
\tablefoot{The fits were performed by adjusting a model with three free parameters in the form of $\log(y) = \alpha + \beta \log(x) \pm \epsilon$, where $\alpha$, $\beta$ and $\epsilon$ correspond to the intercept, the slope and the intrinsic scatter, respectively.}
\end{table}

\subsection{Integrated CO intensity maps of the secondary Gaussian components}
\label{appendix_secondary}

Figure\,\ref{fig_laboca_secondary_components} presents the 870\,\um LABOCA maps towards five \atgtop clumps displaying secondary CO peaks in their spectra. The integrated CO\,(6--5) distribution of the two velocity components clearly shows that the two components trace different structures in the observed field.

\begin{figure*}[!ht]
\includegraphics[height=0.32\linewidth]{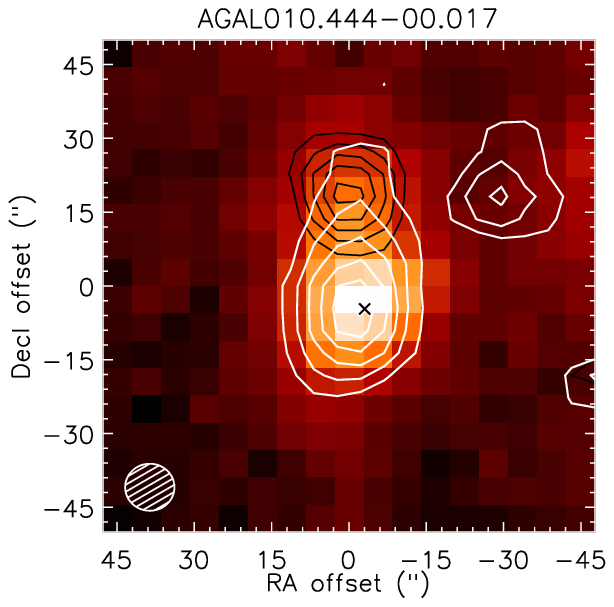}
\includegraphics[height=0.32\linewidth]{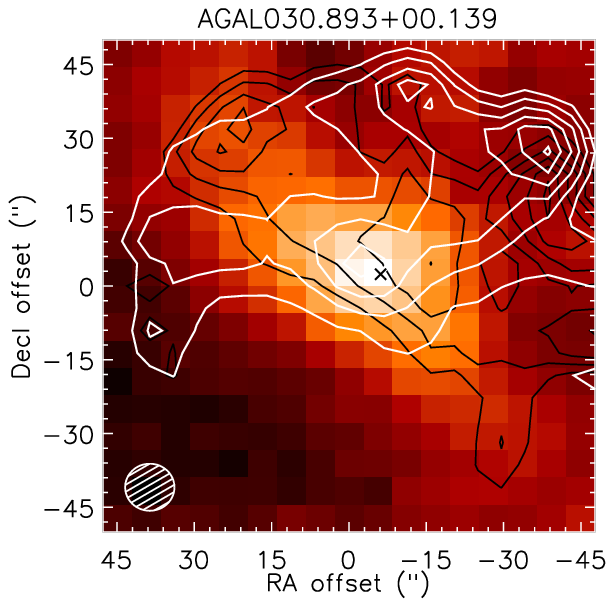}
\includegraphics[height=0.32\linewidth]{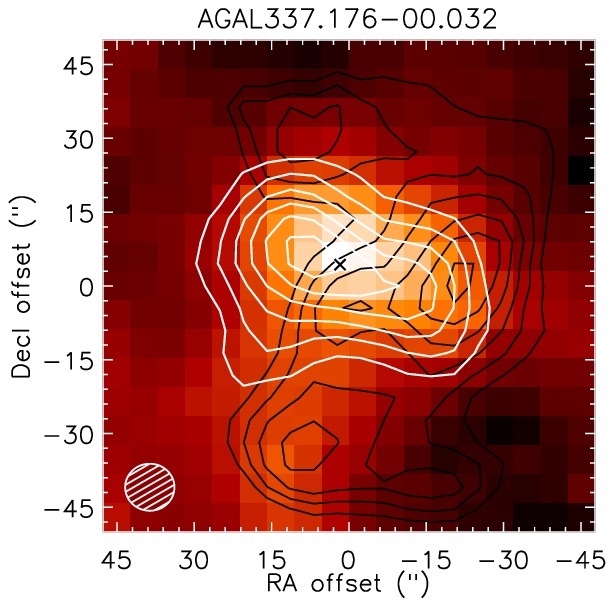} \\
\includegraphics[height=0.32\linewidth]{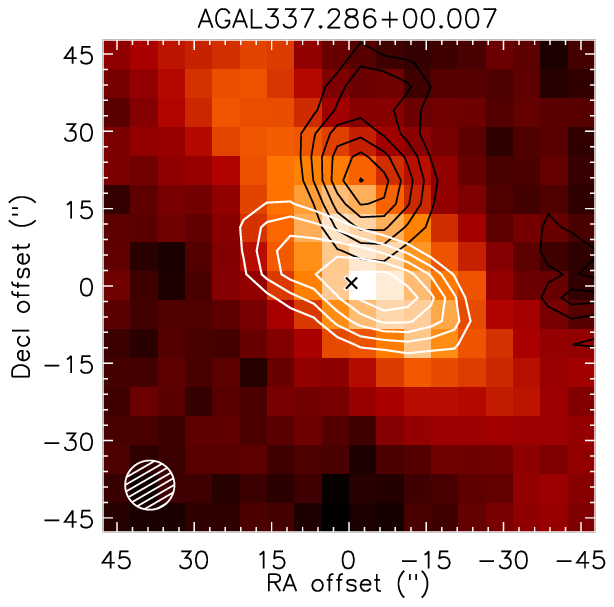}
\includegraphics[height=0.32\linewidth]{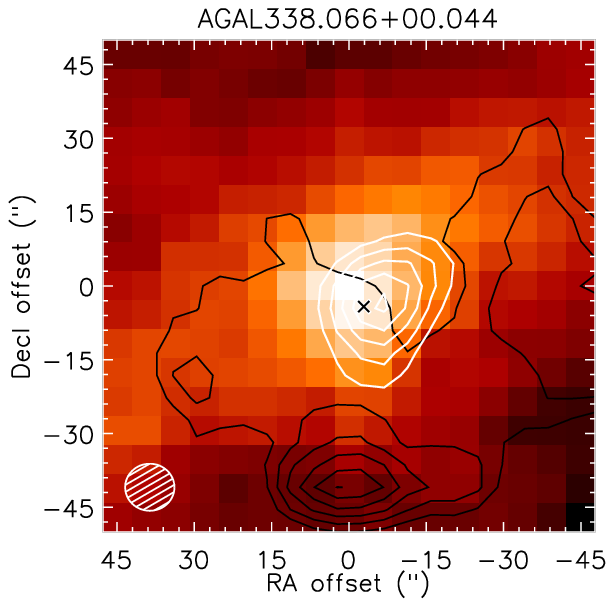} \\[-3.0ex]
    \caption{APEX-LABOCA images at 870\,$\mu$m overlaid by the CO\,(6--5) emission contours (C1 component in white,  C2 component in black; in both cases, the contours are from 30\% to 90\% of the peak emission of the corresponding component in steps of 10\%) towards the \atgtop clumps.
    The position of the CSC source from \citet{Contreras13} is shown as a $\times$ symbol.}
    \label{fig_laboca_secondary_components}
\end{figure*}

  \end{appendix}
\end{onecolumn}

\label{lastpage}

\end{document}